\documentclass[fleqn,usenatbib]{mnras}

\usepackage{newtxtext,newtxmath}
\usepackage{lscape}

\usepackage[T1]{fontenc}

\DeclareRobustCommand{\VAN}[3]{#2}
\let\VANthebibliography\thebibliography
\def\thebibliography{\DeclareRobustCommand{\VAN}[3]{##3}\VANthebibliography}

\usepackage{graphicx}	
\usepackage{amsmath}	
\usepackage{xspace}
\usepackage{longtable}
\usepackage{subfig}
\usepackage{booktabs}
\usepackage{lscape}
\usepackage{siunitx}
\usepackage[symbol]{footmisc}
\usepackage[dvipsnames]{xcolor}
\newcommand{\oxfordastro}{Department of Astrophysics, University of Oxford, Denys Wilkinson Building, Keble Road, Oxford OX1 3RH, UK}
\newcommand{\oxfordmag}{Magdalen College, University of Oxford, Oxford OX1 4AU, UK}
\newcommand{\birmingham}{School of Physics \& Astronomy, University of Birmingham, Edgbaston, Birmingham B15 2TT, United Kingdom}
\newcommand{\princeton}{Department of Astrophysical Sciences, Princeton University, 4 Ivy Lane, Princeton, NJ 08544, USA}
\newcommand{\carnegie}{Carnegie Science Earth and Planets Laboratory, 5241 Broad Branch Road, NW, Washington, DC 20015, USA}
\newcommand{\yale}{Department of Astronomy, Yale University, 219 Prospect Street, New Haven, CT 06511, USA}
\newcommand{\liege}{Astrobiology Research Unit, Université de Liège, 19C Allée du 6 Août, 4000 Liège, Belgium}
\newcommand{\miteaps}{Department of Earth, Atmospheric and Planetary Sciences, Massachusetts Institute of Technology, 77 Massachusetts Avenue, Cambridge,\\ MA 02139, USA }
\newcommand{\mitkavli}{Kavli Institute for Astrophysics and Space Research, Massachusetts Institute of Technology, Cambridge, MA 02139, USA }
\newcommand{\sandiego}{Department of Astronomy \& Astrophysics, UC San Diego, 9500 Gilman Drive, La Jolla, CA 92093, USA}
\newcommand{\iac}{Instituto de Astrof\'{i}sica de Canarias (IAC), Calle Vía Láctea s/n, 38200, La Laguna, Tenerife, Spain}
\newcommand{\cambridge}{Cavendish Laboratory, JJ Thomson Avenue, Cambridge CB3 0HE, UK}
\newcommand{\ethch}{Institute for Particle Physics and Astrophysics , ETH Z\"urich, Wolfgang-Pauli-Strasse 2, 8093 Z\"urich, Switzeland}
\newcommand{\cfa}{Center for Astrophysics \textbar  Harvard \& Smithsonian, 60 Garden Street, Cambridge, MA 02138, USA}
\newcommand{\liegestar}{Space Sciences, Technologies and Astrophysics Research (STAR) Institute, Universit\'e de Li\'ege, All\'ee du 6 Ao\^ut 19C, B-4000 Li\'ege, Belgium}
\newcommand{\komaba}{Komaba Institute for Science, The University of Tokyo, 3-8-1 Komaba, Meguro, Tokyo 153-8902, Japan}
\newcommand{\astrojp}{Astrobiology Center, 2-21-1 Osawa, Mitaka, Tokyo 181-8588, Japan}

\newcommand{\geneva}{Observatoire de Gen\`eve, D\'epartement d’Astronomie, Universit\'e de Gen\`eve, Chemin Pegasi 51b, 1290 Versoix, Switzerland}

\newcommand{\tautenburg}{Thüringer Landessternwarte Tautenburg, Sternwarte 5, D-07778, Tautenburg, Germany}
\newcommand{\grenoble}{Univ. Grenoble Alpes, CNRS, IPAG, F-38000 Grenoble, France}
\newcommand{\cadi}{Cadi Ayyad University, Oukaimeden Observatory, High Energy Physics, Astrophysics and Geoscience Laboratory, FSSM, Morocco}
\newcommand{\sharjah}{Department of Applied Physics and Astronomy, and Sharjah Academy for Astronomy, Space Sciences and Technology, University of Sharjah,\\ United Arab Emirates}

\newcommand{\ull}{Deptartamento de Astrof\'{i}sica, Universidad de La Laguna (ULL), 38206 La Laguna, Tenerife, Spain}
\newcommand{\mex}{Universidad Nacional Aut\'onoma de M\'exico, Instituto de Astronom\'ia, AP 70-264, Ciudad de M\'exico, 04510, M\'exico}
\newcommand{\ohio}{Department of Astronomy, McPherson Laboratory, The Ohio State University, 140 W 18th Ave, Columbus, Ohio 43210, USA}
\newcommand{\carnegieobs}{The Observatories of the Carnegie Institution for Science, 813 Santa Barbara Street, Pasadena, CA, 91101}
\newcommand{\colorado}{Department of Astrophysical and Planetary Sciences, University of Colorado Boulder, Boulder, CO 80309, USA}
\newcommand{\seti}{SETI Institute, Mountain View, CA 94043 USA}
\newcommand{\bern}{Center for Space and Habitability, University of Bern, Gesellschaftsstrasse 6, 3012, Bern, Switzerland}
\newcommand{\mariecurie}{Paris Region Fellow, Marie Sklodowska-Curie Action}
\newcommand{\aim}{AIM, CEA, CNRS, Université Paris-Saclay, Université de Paris, F-91191 Gif-sur-Yvette, France}
\newcommand{\leicester}{School of Physics and Astronomy, University of Leicester, University Road, Leicester LE1 7RH, UK}
\newcommand{\iaa}{Instituto de Astrofísica de Andalucía (IAA-CSIC), Glorieta de la Astronomía s/n, 18008 Granada, Spain}
\newcommand{\zoe}{NSF Graduate Research Fellow, MIT Presidential Fellow, MIT CollamoreRogers Fellow, MIT Teaching Development Fellow}

\newcommand{\teff}{\ensuremath{T_{\textup{eff}}}\xspace}
\newcommand{\feh}{\ensuremath{[\text{Fe/H}]}\xspace}

\title[MANGOS II]{MANGOS II: Five new giant planets orbiting low-mass stars\thanks{This paper includes data gathered with the 6.5 meter Magellan Telescopes located at Las Campanas Observatory, Chile.}}

\author[G. Dransfield \& M. Timmermans et al.]{
G. Dransfield$^{1,2,3}$\thanks{email: george.dransfield@physics.ox.ac.uk}\thanks{These two authors contributed equally to this work and should be considered joint first authors.},
M. Timmermans$^{3,4}$\thanks{email:m.timmermans@bham.ac.uk}\color{blue}$^{\ddagger}$
\color{black},
D. Sebastian$^{5}$,
B. V. Rackham$^{6,7}$,
A. Burgasser$^{8}$,
K. Barkaoui$^{9,4,6}$,
\newauthor
A. H. M. J. Triaud$^{3}$,
M. Gillon$^{4}$,
J. M. Almenara$^{10,11}$,
S. L. Casewell$^{12}$,
K. A. Collins$^{13}$,
A. Fukui$^{14,15}$,
\newauthor
C. Jano-Munoz$^{16}$,
S. Kanodia$^{17}$,
N. Narita$^{14,15,9}$,
E. Palle$^{9,18}$,
M. G. Scott$^{3}$,
A. Soubkiou$^{4}$,
\newauthor
A. Stokholm$^{3}$,
J. Audenaert$^{7}$,
G. \'A. Bakos$^{19}$,
Y. Beletsky$^{17}$,
Z. L. de Beurs$^{6,20}$,
Z. Benkhaldoun$^{21,22}$,
\newauthor
A. Burdanov$^{6}$,
R. P. Butler$^{17}$,
D. Caldwell$^{23}$,
J. D. Crane$^{24}$,
Y. T. Davis$^{3}$,
B.O. Demory$^{25}$,
\newauthor
E. Ducrot$^{26,27}$,
Y. G\'{o}mez Maqueo Chew$^{28}$,
M. Gachaoui$^{4,22}$,
J. D. Hartman$^{19}$,
M. J. Hooton$^{16}$,
E. Jehin$^{29}$,
\newauthor
S. Mercier$^{6}$,
F. Murgas$^{9,18}$,
C. Murray$^{30}$,
P. P. Pedersen$^{16,31}$,
F. J. Pozuelos$^{32}$,
M. Rice$^{33}$,
\newauthor
G. Ross$^{19}$,
S. A. Shectman$^{24}$,
E. Softich$^{8}$,
M. Tala Pinto$^{34}$,
A. M. Vanderburg$^{7}$,
J. Villasenor$^{7}$,
\newauthor
J. de Wit$^{6}$,
S. Z\'{u}\~{n}iga-Fern\'{a}ndez$^{4}$
\\
\\
Affiliations can be found after the acknowledgments. 
}

\date{Accepted XXX. Received YYY; in original form ZZZ}

\pubyear{2015}

\begin{document}
\label{firstpage}
\pagerange{\pageref{firstpage}--\pageref{lastpage}}
\maketitle

\begin{abstract}

Giant planets orbiting low-mass stars on short orbits present a conundrum, as in the most extreme cases their existence cannot be reconciled with current models of core accretion. Therefore, surveys dedicated to finding these rare planets have a key role to play by growing the sample to overcome small number statistics. In this work we present MANGOS, a programme dedicated to the search for giant objects (planets, brown dwarfs, and low-mass stars) orbiting M dwarfs. We report on the discovery of five new giant planets (TOI-3288~Ab, TOI-4666~b, TOI-5007~b, TOI-5292~Ab, TOI-5916~b) first detected by \textit{TESS}, and confirmed using ground-based photometry and spectroscopy. The five planets have radii in the range 0.99-- 1.12~$\mathrm{R_{Jup}}$, masses between 0.49--1.69~$\mathrm{M_{Jup}}$, and orbital periods between 1.43 and 2.91~days. We reveal that TOI-3288 and TOI-5292 are wide binaries, and in the case of TOI-5292 we are able to characterise both stellar components. We demonstrate that the planets presented are suitable for further characterisation of their obliquities and atmospheres. We detect a small but significant eccentricity for TOI-5007~b, although for this to be more robust, more observations are needed to fully sample the orbit. Finally, we reveal a correlation between stellar metallicity and planet bulk density for giant planets orbiting low-mass stars.

\end{abstract}
    



\section{Introduction}

Prior to the detection of the first exoplanet orbiting a Sun-like star \citep{1995_51Peg_Mayor_Queloz}, the nature and locations of all eight known planets orbiting a main sequence star were adequately reproduced by the accepted planetary formation models of the day \citep{Lissauer1993}. 51 Peg b was thus the first planet to spoil the fun.

In the decades since, over 6000 exoplanets have been detected\footnote{According to the NASA Exoplanet Archive, accessed on UTC 2025\,Sep\,25.}, most of which also belong to emerging populations not seen in the solar system. The hot Jupiter population in particular consists of giant planets (defined here as having $R_{\rm p}\geq 7~\mathrm{R_\oplus}$) with orbital periods $P<10~\rm days$ resulting in equilibrium temperatures hotter than 1000~K. Thanks to their large sizes and short orbital periods, hot Jupiters produce easily detectable signals in photometry, radial velocity (RV) and both transmission and emission spectroscopy, making them one of the best studied populations. 

Despite this, the puzzle of their formation remains unsolved. Current formation theory of giant planets indicates that the standard core accretion model \citep[e.g.][]{1996_Pollack_core_accretion,2005_Alibert_giant_formation,2020_Liu_Ji_core_accretion} is the most likely formation pathway. However, there is still no consensus on whether formation takes place `in-situ' \citep[e.g.][]{2016_Batygin_in_situ_formation,2016_Boley_in_situ_formation} or beyond the ice line where there are more solids available \citep{Rafikov2006,Piso2015}. In the latter case, planets would then need to migrate toward their present-day positions \citep[e.g.][]{Ida+Lin2004, 2005_Alibert_giant_formation}, which could happen via disc migration \citep{Lin1996,Baruteau2014} or high eccentricity tidal migration \citep{Rasio1996}.

Large numbers of hot Jupiters were discovered in the early days of the field; however, RV surveys estimated the occurrence rate for this population to be of the order of 1\% for FGK stars \citep[e.g.][]{2012_Wright}. This occurrence rate is consistent with the results from population synthesis models \citep{2021_pop_sythesis_paper2}. 

But what of M dwarfs? They constitute the most abundant stellar population in the galaxy \citep[accounting for 60--70\% of all stars;][]{2006_Henry,2021_Reyle}, and host on average 1 to 2 planets each \citep{2023_Ribas_carmenes_occurrence_rates}. Given that on average the mass of a protoplanetary disc scales with stellar mass \citep{Andrews2013}, and low-mass stars have longer Keplerian timescales \citep{Laughlin2004}, giant planet occurrence rate is expected to decrease with stellar mass. This prediction is supported by the population synthesis models of  \cite{Burn2021}, which also found that no giant planets were formed around stars with $M_\star <0.5 \rm M_\odot$. 
However, recent observation-driven occurrence rates have been calculated based on the \textit{TESS} mission data. \cite{2023_Gan_occurrence} find an occurrence rate for giant planets with $P < 10$ days of $0.27\pm0.09$\% for stellar masses between 0.45 and 0.65 M$_\odot$. This is in agreement with \cite{Bryant2023}, who found that giant planets occur at a rate of $0.29\pm0.15$\%, $0.108\pm0.083$\%, and $0.137\pm0.097$\%, for stellar hosts with $M_{\star}$=0.42--0.71 M$_\odot$,  0.26-0.42 M$_\odot$, and 0.088-0.26 M$_\odot$, respectively. \cite{2025_Glusman_occurrence} extended this work to include more recent data, finding slightly lower occurrence rates of $0.067\pm0.047$\%, $0.139\pm0.069$\%, and $0.032\pm0.032$\% for a median stellar mass of 0.54M$_\odot$, 0.35M$_\odot$, and 0.18M$_\odot$, respectively. Given the model predictions of \cite{Burn2021} that no giant planets should form around hosts with $M_\star <0.5 \rm M_\odot$, these small but non-zero measured occurrence rates around the lowest-mass stars present a conundrum.

Thus far, 32 short period giant planets with $P < 10$ days have been discovered orbiting M dwarfs, with hosts ranging from 0.2-0.7 M$_\odot$ \citep[e.g.][]{2022_TOI-3714b_TOI-3629b,2025_TOI-7149b}. Among those, six orbit stars around mid to late M dwarfs with $M_\star<0.4\mathrm{M}_\odot$: TOI-3235 b \citep{2023_TOI-3235b}, TOI-5205 b \citep{2023_TOI-5205b}, TOI-7149 b \citep{2025_TOI-7149b}, TOI-4860 b \citep{2023_TOI-4860b}, TOI-519 b \citep{2021_TOI-519b,2023_TOI-519b_mass}, and TOI-6894 b \citep{2025_TOI-6894b}. These discoveries showcase the need to re-evaluate formation theories for these objects. One alternative pathway to core accretion to have been proposed for low-mass stars is gravitational instability \citep{1997_Boss,2006_Boss, Boss2023}. However, \cite{2020_Mercer_Stamatellos} have shown that this mechanism, where planets form from the collapse of a gravitationally unstable disc, should produce planets of masses larger than 2$\mathrm{M}_{\rm Jup}$. Only six super-Jupiter mass planets in our parameter space of interest have been found, which are excellent test cases to reconcile theory and observations: TOI-4201 b \citep{2024_TOI-4201b}, TOI-2379 b, TOI-2384 b \citep{2024_TOI-2379}, TIC 46432937 b \citep{2024_TIC46432937b}, TOI-6303 b, and TOI-6330 b \citep{6330}. The latter is the most massive planet to have been found orbiting an M star with a mass of $10.0\pm0.3\mathrm{M}_\mathrm{Jup}$.

While the mass of TOI-6330 b is remarkably large, it is still below the deuterium burning limit that defines brown dwarfs (BDs), canonically found in the mass range $13~\mathrm{M_{Jup}}\lesssim M_\mathrm{BD} \lesssim 80~\mathrm{M_{Jup}}$ \citep{Chabrier2014}. The occurrence rate of BDs as companions to FGK stars on short orbits is low \citep[e.g.][]{Grether2006}, with values as low as 0.56\% estimated from radial velocity surveys \citep{Grieves2017}, and this paucity inside 10~AU is known as the `brown dwarf desert' \citep[e.g.][]{Sahlmann2011}. Brown dwarf companions orbiting around low-mass stars are even rarer than giant planets, with only 12 transiting BDs confirmed thus far \citep[e.g.][]{Maldonado2023,Gan2025,Barkaoui2025a}. Studying BDs around low-mass stars is also crucial to understand the formation environment around these stars. It is also unclear whether the `dryness' and `width' of the so-called brown dwarf desert are determined by companion mass or mass ratio. Thus discovering and characterising BD companions to M dwarfs on short orbits adds critical pieces to the puzzle of giant planet formation and low-mass stars as planet hosts.

In all cases, the characterisation of planets and BDs is fundamentally constrained by the accuracy of the stellar parameters of their host stars. Stellar properties such as mass and radius provide the basis for deriving the orbital and physical parameters of the associated planets. For single stars, we use evolutionary stellar models such as \cite{Baraffe2015} to derive stellar parameters. However, several works have shown that observed radii of M dwarfs were inflated compared to those expected from benchmark models \citep[e.g.][]{2008_Casagrande_Mdwarf,2013_Spada_Mdwarf}. Multiple hypotheses are considered to explain this phenomenon, such as stellar magnetic activity inhibiting convection and producing stellar spots which can cause an under- or overestimation of the eclipse depth of the binary \citep{2010_Morales,2013_Feiden}, or stellar metallicity which affects the opacity of the outer layers of the star \citep{2006_Berger,2019_von_Boetticher}. This radius inflation can significantly impact inferences of key planetary parameters. In order to increase the statistical significance of empirical scaling relations such as the ones of \cite{Mann2015} and \cite{Mann2019}, it is critical to add to the sample of low-mass stars with absolute mass and radius measurements, i.e. low-mass double-lined eclipsing binaries. These objects can mimic the photometric behavior of giant planet candidates when looking at M dwarfs since the companion star would be of equal or lower mass. Notably, the eclipsing binaries with low mass (EBLM) project \citep{2013_Triaud} has provided absolute physical parameters of dozens of low-mass binaries \citep[e.g.][]{2023_Sebastian,2024_Swayne}, and applied high-resolution cross-correlation techniques to deal with extreme flux ratios, effectively turning single-lined eclipsing binaries into double-lined ones \citep{2025_Sebastian}.

In this context, we present MANGOS (\textbf{M} dwarfs \textbf{A}ccompanied by close-i\textbf{N} \textbf{G}iant \textbf{O}rbiters with \textbf{S}PECULOOS), a comprehensive program designed to detect, catalogue, and characterise giant companions orbiting low-mass stars. The MANGOS initiative encompasses three main scientific cases: giant planets, brown dwarfs, and double-lined eclipsing binaries. This project was first presented in \cite{2023_TOI-4860b}, where we reported the discovery of TOI-4860~b, a Jupiter-sized planet orbiting an M4.5 host. In the present work, we report on the discovery, confirmation and characterisation of five new MANGOS planets, orbiting hosts with masses $0.48-0.67~\rm M_\odot$.

Our paper is structured as follows: we begin in Section \ref{sec:strategy} by describing the MANGOS programme strategy for planets. We then describe our full photometric and spectroscopic observing campaigns in Section \ref{sec:observations}. In Section \ref{sec:stellar_char} we characterise our host stars, followed by a global analysis of all available data in Section \ref{sec:global}. We present a summary of key resulting parameters in Section \ref{sec:results}, which we discuss in the context of the field in Section~\ref{sec:discuss}. Finally, we conclude in Section~\ref{sec:we_conclude}.

\section{The MANGOS strategy}
\label{sec:strategy}

As described above, MANGOS is a holistic program of discovery and characterisation of giant objects hosted by low-mass stars. While all MANGOS targets begin as planetary candidates, many will be identified as brown dwarf-M dwarf or M dwarf-M dwarf binaries. They are then reclassified and switched to the binary programme. In this section, we describe the steps taken to validate and confirm MANGOS planets, as well as what sets this project apart from others. 

\subsection{MANGOS Planet Candidate Selection and Validation Workflow}

All targets in the MANGOS sample come from the list of \textit{TESS} Objects of Interest \citep[TOIs,][]{2021_Guerrero_TOI}. We begin our cuts by selecting candidates orbiting hosts with $\teff \leq 4000 \rm K$ to ensure all low-mass stars are captured up to M0. We then rule out any TOIs with radii $R_{\rm p}<7 \rm R_{\oplus}$ and periods $P>7\rm days$ to select only close-in giant planet candidates. 

Due to the faintness of many of the hosts, the photometric precision on the \textit{TESS} light curves is often limited, leading to low signal-to-noise ratio (SNR) detections of candidate events. Additionally, the large pixel scale of \textit{TESS} of 21$\arcsec$ often leads to multiple sources in the apertures used to extract the photometry. In the case of crowded stellar fields, the probability of false positives increases. To ensure the transit event is indeed on the target star, a first vetting step is to observe a transit with the TRAPPIST (TRAnsiting Planets and PlanetesImals Small Telescope) telescopes. TRAPPIST-South \citep[TS,][]{Jehin2011,Gillon2011} is located at ESO La Silla Observatory in Chile. It is a 0.6-m telescope equipped with a FLI ProLine PL3041-BB camera and a back-illuminated CCD. The pixel size is $0\farcs64$, providing a total field of view of $22\arcmin\,\times\,22\arcmin$ for an array of $2048\,\times\,2048$ pixels. TRAPPIST-North  \citep[TN,][]{Barkaoui2019_TN} is a twin of its Southern counterpart located at Oukaïmeden Observatory in Morocco. It is equipped with a thermoelectrically cooled 2K$\times$2K Andor iKon-L BEX2-DD CCD camera with a pixel scale of 0\farcs6 and a  field of view of $20\arcmin\times20\arcmin$. Both TS and TN are Ritchey-Chr\'etien telescopes with F/8 equipped with German equatorial mounts. For the purpose of the MANGOS programme, our observations with TRAPPIST are conducted in the \textit{I+z} filter, a custom filter optimised for low-mass stars with a transmittance >90\% from 750 nm to beyond 1000 nm. 

At the end of this first step, the transit is confirmed on target and its ephemeris is corrected if needed. The candidates are then added to our programmes for reconnaissance spectroscopy for host characterisation using optical and infrared spectrographs (see Section \ref{sec:spec}). This step allows us to refine stellar parameters, as well as check for spectroscopic indicators of youth or binarity. 

Further transit observations are then scheduled on the SPECULOOS (Search for habitable Planets EClipsing ULtra-cOOl Stars) telescopes to carry out multi-colour photometric validation. The SPECULOOS-South Observatory \citep[SSO,][]{SPC_Laeti,SPC_Daniel} is located at ESO Paranal Observatory in Chile. It is composed of four identical 1.0m F/8 Ritchey-Chrétien telescopes, named after the Galilean moons Io, Europa, Ganymede, and Callisto. All telescopes are equipped with a deep-depletion Andor iKon-L $2k \times 2k$ CCD camera, resulting in a total field of view of $12'$ for a pixel scale of $0\farcs35$ \citep{SPC_Artem}. Recently, SSO/Callisto was equipped with the SPECULOOS Infra-Red photometric Imager for Transits (SPIRIT). It is an InGaAs CMOS-based instrument with a custom wide-pass filter, \textit{zYJ}, covering the wavelength range 0.81-1.33$\mu$m to minimize the effect of precipitable water vapor on differential light curves \citep{2024SPIE_SPIRIT_Pedersen,2025_TOI-2407_SPIRIT}. SPIRIT has a $1024 \times 1280$ detector with 12-$\micron$ pixels, yielding a $5.3\arcmin\,\times\,6.6\arcmin$ field of view. 

Two facilities cover the northern hemisphere: the SPECULOOS-North Observatory \citep[SNO,][]{Burdanov2022}, which hosts one telescope named Artemis, and SAINT-EX \citep[Search And characterIsatioN of Transiting EXoplanets,][]{Demory_AA_SAINTEX_2020}. Both are twins of the SSO telescopes equipped with identical cameras. SNO is located at Teide Observatory in Tenerife while SAINT-EX is located at the National Astronomical Observatory of Mexico, in San Pedro Mártir. In total, six telescopes may be used to check for a chromatic signature of the transit depths of the MANGOS candidates, a telltale sign of binarity. False positive scenarios could be either an unresolved background blend with our target or another smaller, redder star (or BD) orbiting the target. The chromaticity investigation towards the bluer bands identifies any additional dilution of the transit signal due to a non-opaque orbiting body in the considered wavelength ranges. We also schedule observations at phase 0.5 in our reddest filter available to check for hints of a secondary eclipse. 

We then carry out statistical validation using the \texttt{TRICERATOPS} \citep{Giacalone_2021AJ_triceratops} package, a Bayesian framework to evaluate the flux of nearby stars and assess the false positive probabilities (FPPs). While this is a standard step in our workflow, MANGOS planet candidates will often have high FPPs as many of them present a grazing configuration. In the case of small planets orbiting larger stars, a grazing transit is very often an indicator of binarity, but due to the much larger radius ratios in our sample, we expect to find a higher proportion of truly grazing planets. 

The most promising targets will then be proposed for RV measurements on suitable facilities. Given the faintness of the targets and the high expected RV signals, ESPRESSO and MAROON-X allow us to achieve excellent precision on mass measurements with only a handful of spectra. In this paper, we present planets found in our successful ESPRESSO programmes during P112 and P114 (P112.25ZF, PI: G. Dransfield; P114.27JF, PI: M. Timmermans). Through our membership of the TFOP (\textit{TESS} Follow-up Observing Programme) collaboration, our data are supplemented by observations taken by LCO, MuSCAT2, ExTrA, and PFS.

\subsection{The MANGOS advantage}
The TRAPPIST and SPECULOOS facilities are the primary photometric instruments used for the MANGOS programme. Together, they total three instruments in the northern hemisphere (TN, SNO, and SAINT-EX), and five in the southern one (the four SSO telescopes, and TS). All are optimised in the near-infrared for low-mass stars but extend into the optical down to 350 nm. Having multiple identical telescopes observing simultaneously a transit event generates a more robust detection of chromatic depth variations than at different epochs using different facilities. Notably, the SPIRIT camera allows us to explore up to 1.3$\mu$m, a regime where secondary eclipses could be detected for hot orbiting bodies. When that is the case, we can determine the brightness temperature of the hottest MANGOS candidates or reveal a binary. Finally, our simultaneous multi-color photometry approach also makes it possible to capture the chromaticity of any flares and stellar spots, in turn permitting a direct characterisation of these activity indicators. This is particularly valuable in the case of M dwarf stars as they tend to be very active.

\section{Observations}
\label{sec:observations}

In this section we describe the observations collected to characterise the five host stars, as well as to confirm the planetary nature of the giant planets orbiting them. We begin by describing the photometric data collected from space and the ground, followed by the spectroscopic datasets from six different optical and near infrared instruments. A journal of photometric observations can be found in Tables \ref{tab:photometry_summary} and \ref{tab:photometry_summary2}.

\subsection{Photometry}

In this section we describe the photometric data collected for the five targets presented in this work. In Section \ref{sec:tess} we describe the initial target identification by \textit{TESS}, as well as the available data. The remaining sections describe ground-based photometric data collection.

We note that our scrutiny of archival imaging and \textit{Gaia}, described in Section \ref{sec:blends}, revealed both TOI-3288 and TOI-5292 to have bound companions. We therefore refer to them as TOI-3288~AB and TOI-5292~AB where observations are unable to separate the components, and TOI-3288~A or B and TOI-5292~A or B where observations resolve the individual stars.

\subsubsection{TESS}
\label{sec:tess}
All five targets considered in this work were initially found by the \textit{TESS} mission. For all, threshold crossing events (TCE) were detected and Data Validation (DV) reports were created from the Science Processing Operations Center \citep[SPOC,][]{SPOC} pipeline or Quick Look Pipeline \citep[QLP,][]{QLP} depending on the cadence of the data taken. The DV reports a series of validation and diagnostic tests to evaluate the confidence in the transit events from the TCE alert, they are described in \cite{Twicken:DVdiagnostics2018} and \cite{Li:DVmodelFit2019} for SPOC detections and \cite{2021_Guerrero_TOI} for QLP detections\footnote{The DV reports are available at https://tev.mit.edu/.}. For all candidates, the DV reports associated with the initial detections show the signals pass all the vetting steps apart from TOI-5007.01, which showed some odd/even transit depth variations in the initial QLP detection DV report. This variation can be attributed to the crowded field with many neighboring stars contaminating the \textit{TESS} aperture, resulting in poor data quality for the Sector 39 initial detection. A subsequent SPOC detection in Sector 65 did not show such odd/even variations. The typical apertures used to extract the photometry in the \textit{TESS} pixel files are given in Fig~\ref{fig:tpfs}, and the \textit{TESS} data is described below. 

TOI-5007 (TIC 424196109) was observed in Sectors 12, 39, and 65 with full-frame image (FFI) data (1800s, 600s, and 200s, respectively) and also with 2-minute cadence data in Sector 65. TOI-5007.01 was initially detected in Sector 39 by the QLP faint-star search \citep{2022_Kunimoto} on UTC 2022\,Jan\,6. The planet candidate had a radius of $10.62\pm0.96\, \mathrm{R_\oplus}$, for a transit depth of $21300\pm157$ ppm and a period of $2.5433\pm0.0007$ days. We obtained the Presearch Data Conditioning Simple Aperture Photometry \citep[PDCSAP][]{Stumpe2012,Smith2012,Stumpe2014} flux available for the 2-minute cadence data of Sectors 65 from the NASA Mikulski Archive for Space Telescopes (MAST) using the \texttt{Lightkurve} package \citep{lightkurve}. The PDCSAP flux is produced by the SPOC pipeline where it is corrected for instrumental systematics as well as dilution of the signal due to crowding of the aperture. It offers a much better precision than the QLP data which is why we selected this data product for our transit analysis. 

TOI-5292~AB (TIC 33397739) was observed in Sectors 42, 70 and 92 with FFI data (600s and 200s), and 2-minute cadence data were also taken in Sectors 70 and 92. TOI-5292.01 was detected in Sector 42 by the QLP faint-star search on UTC 2022\,Aug\,22. A depth of $44030\pm913$ ppm was reported for a $2.0231\pm0.0007$ day period, corresponding to a planet with a radius of $13.43\pm2.00\, \mathrm{R_\oplus}$. For the global analysis, similarly to TOI-5007, we used the PDCSAP flux of Sectors 70 and 92 obtained from the SPOC pipeline. 

TOI-5916 (TIC 305506996) was observed in Sector 55 with an exposure of 600s and in Sector 82 with exposures of 200s and 120s. The candidate TOI-5916.01 was also first detected by the QLP faint-star search in Sector 55 on UTC 2022\,Nov\,8. The DVR reports a planet candidate of a size of $10.92\pm0.60\, \mathrm{R_\oplus}$ and a period of $2.3682\pm0.0007$ days, for a transit depth of $48700\pm3699$ ppm. 

\begin{figure*}
    \centering
    \includegraphics[width=1\linewidth]{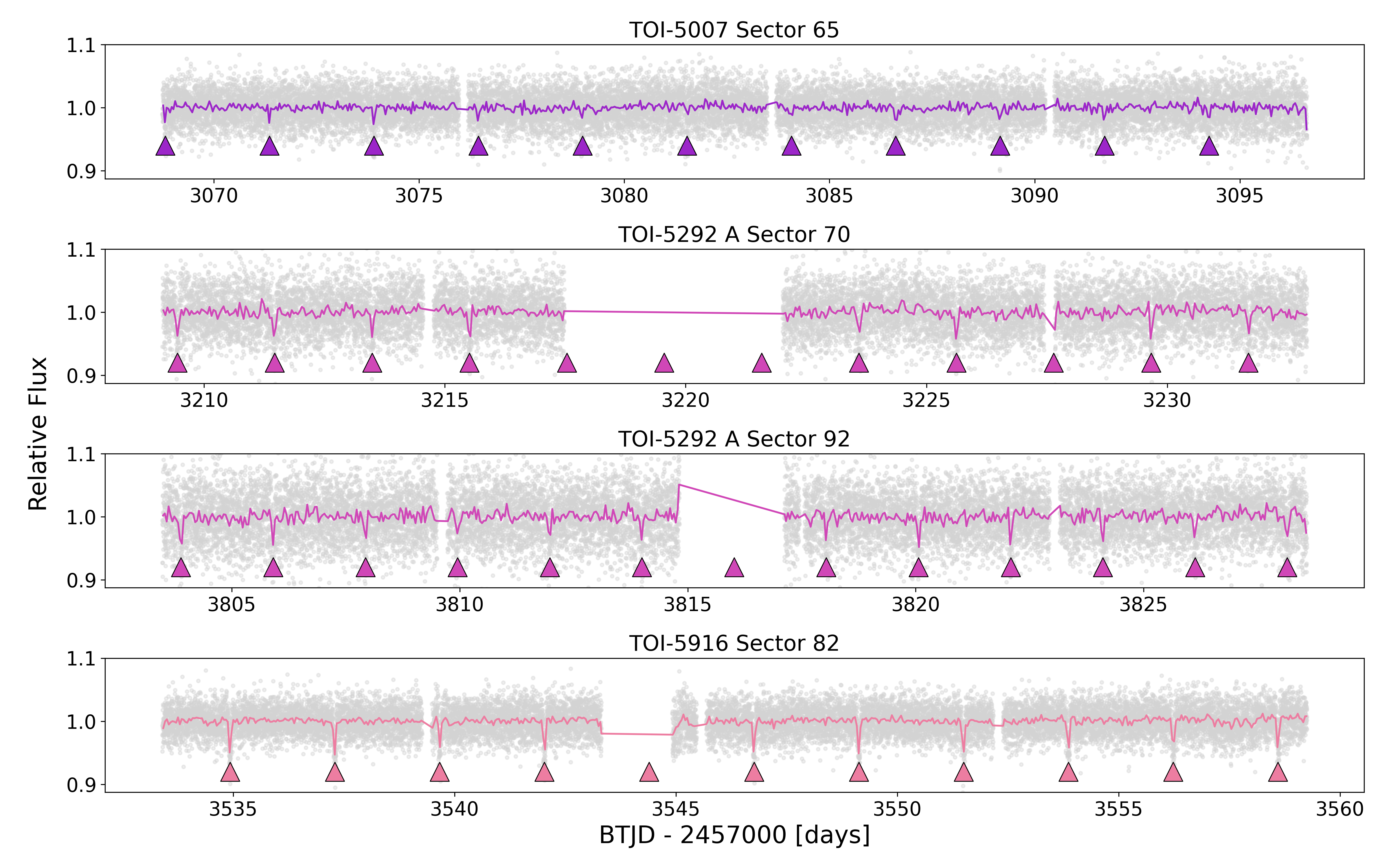}
    \caption{TESS 120 second cadence data of TOI-5007 (top), TOI-5292~A (middle) and TOI-5916 (bottom). In this cadence, TOI-5007 was observed in Sector 65, TOI-5292~A in Sectors 70 and 92, and TOI-5916 in Sector 82. Light grey points show the 120 second exposures, and the coloured lines show the flux in one-hour bins. The transit events of each planet candidate are indicated by coloured arrows.}
    \label{fig:TESS_data}
\end{figure*}

TOI-3288~AB (TIC 79920467) was observed in Sectors 13 and 27 of \textit{TESS} with FFI data (1800 and 600 seconds, respectively), and 67 with both 2-minute cadence and FFI data (120 and 200s). The planet candidate TOI-3288.01 detection was made on UTC 2021\,Jun\,4 by the QLP pipeline faint-star search in Sector 13 and 27. It reported a $31360\pm8$ ppm deep transit at a period of $1.4339\pm0.0001$ days, corresponding to a $12.38\pm0.42\, \mathrm{R_\oplus}$ planet. Finally, TOI-4666 (TIC 165202476) was only observed in Sector 31 with FFI data data (600s). The detection of the $2.9094\pm 0.0002$ day period and $45160\pm16$ ppm deep signal was made on UTC 2021\,Dec\,21, corresponding to a planet radius of $12.87\pm0.47 \mathrm{R_\oplus}$. This signal was also found through QLP's faint-star search. Both of these last two systems are also confirmed by another team (Frensch et al., submitted). Through coordination with them, we agreed not to use the \textit{TESS} data in our analyses in order to report fully independent results. 
The 120\,s cadence \textit{TESS} data used in the analyses for TOI-5007, TOI-5292~A and TOI-5916 are shown in Fig.~\ref{fig:TESS_data}.

\subsubsection{TRAPPIST} \label{sec:trappist}

We scheduled the \textit{TESS} follow-up observations using the \textit{TESS} Transit Finder based on the \texttt{Tapir} interface \citep{2013_Jensen}. The data is then processed using a custom pipeline built with the \texttt{prose} package, a publicly available \texttt{python} framework for processing astronomical images\footnote{\url{https://github.com/lgrcia/prose}} \citep{garcia2021,garcia_prose_2022}. This pipeline provides standard data reduction as well as photometric extraction. 

We observed a full transit TOI-5916.01 with TN on UTC 2022\,Nov\,28 in the \textit{I+z} band and with an exposure time of 120s. FWHM for this observation was $2\farcs27$. This was observed simultaneously with the SNO/Artemis (described above).

A full transit of TOI-3288.01 was observed with TS on UTC 2023\,Sep\,12 in the \textit{I+z} band with an exposure time of 60s and a FWHM of $2\farcs87$. 

We observed a full transit of TOI-4666.01 with TS on UTC 2023\,Jan\,17 (simultaneous observation with SSO, described above) in \textit{Sloan-z'} with an exposure of 130s and a FWHM of $2\farcs52$. A second full transit event was observed with TS on UTC 2024\,Dec\,18 in \textit{Sloan-z'} with an exposure of 120s and a FWHM of $2\farcs32$.

TOI-5292.01 was observed for the first time with TN on UTC 2023\,Dec\,09 in \textit{I+z}, where only the transit egress was detected. The observation was taken with 120s exposures and had a FWHM of $2\farcs13$. On UTC 2023\,Dec\,11, a full transit was observed with TN in \textit{Sloan-z'} with exposures of 150s and a FWHM of $2\farcs43$.

We observed a full transit of TOI-5007.01 with TS on UTC 2022\,May\,02 in \textit{I+z} with an exposure of 60s and FWHM of $1\farcs93$. Second and third full transits were observed on UTC 2022\,Jul\,02 and UTC 2022\,Feb\,03 with exposure=110s and in bands \textit{I+z} and \textit{Sloan-z'} respectively.

\subsubsection{SPECULOOS}

All SPECULOOS observations are scheduled using SPECULOOS's custom scheduler \textsc{SPOCK}\footnote{\url{https://github.com/educrot22/SPOCK_v2/tree/main/SPOCK}} \citep{SPC_Daniel}, and initially processed by an automatic data reduction pipeline, described in \cite{SSOpipeline}. Successful observations of \textit{TESS} targets are then reprocessed manually using a custom pipeline built with the \texttt{prose} package. This allows for a better control over the various parameters of the differential photometry process, namely free control over the background annulus size as well as photometric apertures and choice of comparison stars. This is critical in the case of TOI-5007 b where the field is very crowded.

We collected 17 observations of TOI-3288.01 with SSO on nine different nights, between UTC\,2022\,Jun\,30 and UTC\,2025\,Sep\,15. Exposure times ranged from 18~s for observations in \textit{zYJ}, to 120~s for those collected in \textit{Sloan-g'}. We collected one observation at phase 0.5 on UTC\,2022\,Aug\,07 wtih SSO in \textit{I+z} and did not observe a secondary eclipse.

We collected 11 observations of TOI-4666.01 across 6 nights with SSO between UTC\,2022\,Oct\,10 and UTC\,2024\,Dec\,18, using exposure times of 16--160s. We observed at phase 0.5 on UTC\,2023\,Sep\,08 and UTC\,2023\,Oct\,13 in \textit{I+z} and \textit{Sloan-i'} and did not detect a secondary eclipse.

We collected four transit observations TOI-5007.01 across three nights with SSO, between UTC\,2022\,Aug\,27 and UTC\,2025\,Mar\,12 using 23--120~s exposures.

We obtained three transit observations of TOI-5292.01 with SNO over three nights, between UTC\,2024\,Sep\,27 and UTC\,2024\,Dec\,19 using exposure times between 60~s (for \textit{Sloan-z'}) and 160~s (\textit{Sloan-g'}). We collected one further observation at phase 0.5 on UTC\,2024\,Dec\,18 and ruled out a secondary eclipse event.

We observed TOI-5916.01 on the night of UTC\,2022\,Nov\,28 with SNO using the \textit{Sloan-g'} filter with a 14~s exposure.

\subsubsection{MuSCAT2}

TOI-5916 was observed on the night of UTC 2024\,Aug\,16 using the multi-band imager MuSCAT2 \citep{Narita2019}, mounted on the 1.5 m Telescopio Carlos S\'{a}nchez (TCS) at Teide Observatory, Spain. MuSCAT2 is equipped with four CCDs, allowing for simultaneous imaging in the $g'$, $r'$, $i'$, and $z_s$ bands with minimal readout time. Each CCD consists of $1024 \times 1024$ pixels, covering a field of view of $7.4\arcmin \times 7.4\arcmin$.

The observations were conducted with the telescope in nominal focus. The exposure times were set to 50 seconds for the $g'$, $r'$, and $z_s$ bands, and 15 seconds for the $i'$ band. The raw data were processed using the MuSCAT2 pipeline \citep{Parviainen2019}, which performs dark and flat-field calibrations, aperture photometry, and transit model fitting while accounting for instrumental systematics.

\subsubsection{ExTrA}

The Exoplanets in Transits and their Atmospheres \citep[ExTrA,][]{Bonfils2015} is a low-resolution near-infrared (0.85 to 1.55~$\mu$m) multi-object spectrograph fed by three 60-cm telescopes located at La Silla Observatory in Chile. Eleven transits of TOI-5007.01 were observed using one, two, or three of the ExTrA telescopes. We used 8$\arcsec$ aperture fibers and the lowest-resolution mode ($R$$\sim$20) of the spectrograph, a combination that is optimal for the target's magnitude, with an exposure time of 60~seconds. Five fibers are positioned in the focal plane of each telescope to select light from the target and four comparison stars. The resulting ExTrA data were analyzed using custom data reduction software described in \cite{2021_Cointepas_Extra_reduction}.

\subsubsection{LCO-SSO-0m4}
We observed an ingress of TOI-5007.01 from the \citep[LCOGT;][]{Brown:2013} 0.4\,m network node at Siding Spring Observatory near Coonabarabran, Australia (SSO).
The telescope is equipped with  a $2048\times3072$ pixel SBIG STX6303 camera having an image scale of 0$\farcs$57 pixel$^{-1}$, and a FOV of $19\arcmin\times29\arcmin$.
The observation was carried out on UTC 2023\,Apr\,21 in the Sloan-$r'$ with an exposure time of 600s.
Science images were calibrated by the standard LCOGT {\sc BANZAI} pipeline \citep{McCully:2018}, and differential photometry was performed using {\sc AstroImageJ} ({\sc AIJ}) \citep{Collins:2017}. The light curve was not included in the global fit due to the low signal-to-noise ratio of the transit.

\subsection{Spectroscopy}
\label{sec:spec}

\begin{table*}
    \centering
    \begin{tabular}{cccccc}
    \midrule \midrule
         & \textbf{TOI-3288~A} & \textbf{TOI-4666} & \textbf{TOI-5007} & \textbf{TOI-5292~A} & \textbf{TOI-5916}  \\ \midrule
         \multicolumn{6}{c}{\textbf{Designation} } \\ \midrule 
    TIC$^1$  & 79920467 & 165202476 & 424196109 & 33397739 & 305506996 \\
    GAIA DR3$^2$ & 6685431748042347776 & 4855422771071903232 & 5833386583912265728 & 2610422023759160576 & 1741429438313012608 \\
    2MASS$^3$ & J19482648-4301044 & J03450956-3903470 & J15441536-6102284 & J22565337-0807108 & J21411189+0935570  \\ \midrule 
    \multicolumn{6}{c}{\textbf{Photometric magnitudes} } \\ \midrule 
    \textit{TESS} (mag)$^1$ & 13.296 $\pm 0.009 $ & 13.327 $\pm 0.008$& 13.637 $\pm 0.007$& 14.895 $\pm 0.008$& 14.525 $\pm 0.008$ \\
    B (mag)$^4$ & 16.485 $\pm 0.119 $ & 16.746 $\pm 0.084 $& 17.718 $\pm 0.169$& 18.329$\pm 0.162$& 18.286 $\pm 0.165 $ \\
    V (mag)$^4$ & 14.979 $\pm 0.195$& 15.333 $\pm 0.172$& 15.327$\pm 0.183$& 16.416 $\pm 0.206$& 16.602$\pm 0.057$ \\
    G (mag)$^2$ & 14.234 $\pm 0.001$& 14.391 $\pm 0.001$& 14.652$\pm 0.001 $&15.909$\pm 0.001$& 15.693$\pm 0.001$\\
    J (mag)$^3$ & 12.043 $\pm 0.027$& 11.978 $\pm 0.023$& 12.228$\pm 0.023$& 13.629 $\pm 0.026$& 13.118$\pm 0.026$\\
    H (mag)$^3$ & 11.421 $\pm 0.034$& 11.289 $\pm 0.022 $& 11.52 $\pm 0.022$&12.955$\pm 0.022$& 12.444$\pm 0.03$\\
    Ks (mag)$^3$ & 11.201 $\pm 0.026$& 11.085 $\pm 0.025$& 11.307$\pm 0.019$ &12.777$\pm 0.023$& 12.235 $\pm 0.029$\\
    W1 (mag)$^3$ & 11.035 $\pm 0.022$& 10.999 $\pm 0.023$ & 11.229 $\pm 0.023$& 12.693$\pm 0.024$& 12.091$\pm 0.023$\\
    W2 (mag)$^3$ & 11.067 $\pm 0.021$& 10.981 $\pm 0.021$& 11.268$\pm 0.022 $ & 12.681$\pm 0.027$& 12.016$\pm 0.022$\\
    W3 (mag)$^3$ & 10.924 $\pm 0.134$ & 10.844 $\pm 0.061$& 11.16$\pm 0.254$& $>$11.506 & 11.603$\pm 0.216$\\
    W4 (mag)$^3$ & $>$8.48   & $>$9.512 & $>$9.129&$>$8.537 & $>$8.851\\ \midrule
    \multicolumn{6}{c}{\textbf{Astrometric properties} } \\ \midrule 
    Distance (pc)$^5$ & 200.5 $\pm 1.55 $& 154.75 $\pm 0.54$& 203.35 $\pm 1.24$& 337.10 $\pm 8.16$& 199.724 $\pm 3.35$\\
    Parallax (mas)$^2$ & 4.969 $\pm 0.025$ & 6.488 $\pm 0.016$& 4.909 $\pm 0.022$& 2.790 $\pm 0.041$&5.097 $\pm 0.042$ \\
    RA (J2000)$^2$ & 19:48:26.48 & 03:45:09.57 & 15:44:15.33 & 22:56:53.37 & 21:41:11.88 \\
    DEC (J2000)$^2$ & -43:01:04.8 & -39:03:47.45 & -61:02:28.91 & -08:07:11.09 & +09:35:56.59 \\
    $\mu_\mathrm{RA}$$^2$ & -6.566 $\pm 0.022$& 4.122 $\pm 0.014$& -13.515 $\pm 0.026$& -8.733 $\pm 0.041$&-8.816 $\pm 0.039$ \\
    $\mu_\mathrm{DEC}$$^2$ & -20.080 $\pm 0.014$& -22.135$\pm 0.019$& -29.914 $\pm 0.025$& -6.862 $\pm 0.039$&-25.598 $\pm 0.040$\\
    RUWE$^2$ & 1.217 & 0.998 & 1.104 &1.041 & 0.984 \\ \midrule
    \multicolumn{6}{c}{\textbf{Adopted stellar parameters} } \\ \midrule 
    Opt.~SpT      & M0$\pm$0.5 & M2$\pm$0.5 & M1$\pm$0.5 & M0$\pm$0.5 & M2$\pm$0.5 \\
    NIR.~SpT      & M1.5$\pm$0.5 & K8$\pm$1 & M0$\pm$0.5 & M0.5$\pm$0.5 & M1$\pm$0.5 \\
    $\rm [Fe/H]$ (dex)  &$+0.65\pm0.12$ &$+0.79\pm0.14$ & $+0.43\pm0.12$&$+0.33\pm0.14$ & $+0.09\pm0.14$  \\
    $R_{\star}$ ($\mathrm{R}_\odot$) & $0.672\pm0.020$& $0.585\pm 0.018 $ &$0.656 \pm 0.019 $ & $0.611\pm0.019 $ &$0.480 \pm 0.015$ \\
    $M_{\star}$ ($\mathrm{M}_\odot$) & $0.643\pm0.014$ & $0.576\pm 0.012$& $0.632 \pm 0.013$&$0.596 \pm 0.013 $ &$0.477 \pm0.011 $ \\
    ${\rm T_{eff}}$ (K) & $4072\pm 80$& $3793\pm 80$& $3793\pm 80$&$3919\pm 80$ &$3613\pm 80 $ \\
    $\log g_\star$ (cgs)  & $4.592\pm0.028$& $4.664\pm 0.028 $&$4.604\pm 0.027$ & $ 4.642\pm 0.029 $& $4.755 \pm0.029 $\\
    $L_\star$ (L$_\odot$) & $0.1117 \pm 0.0110$ & $0.0638 \pm0.0066 $ & $0.0803\pm 0.0082 $ &$0.0792 \pm 0.0081 $ &$0.0353 \pm0.0038 $ \\ \vspace{0.1cm}
    Age (Gyr) & 
    $5\pm2$ & $3^{+3}_{-2} $ & $4\pm2 $ &$5\pm2 $ &$5\pm2 $ \\
    \vspace{0.1cm}
    $U$ ($\mathrm{km~s^{-1}}$) & $25.54\pm{0.02}$ & $10.27\pm{0.03}$ & 
    $-14.54^{+0.08}_{-0.07}$ & 
    $15.83\pm{0.25}$ & 
    $18.55^{+0.16}_{-0.14}$ \\\vspace{0.1cm}
    $V$ ($\mathrm{km~s^{-1}}$) & $-21.70^{+0.08}_{-0.10}$ & $-14.63^{+0.02}_{-0.04}$ & 
    $-24.16^{+0.09}_{-0.11}$ & 
    $-9.44^{+0.07}_{-0.10}$ & 
    $-15.85^{+0.11}_{-0.10}$ \\ \vspace{0.1cm}
    $W$ ($\mathrm{km~s^{-1}}$) & $-11.93^{+0.03}_{-0.02}$ & 
    $-3.45\pm{0.02}$ & 
    $-15.20^{+0.05}_{-0.08}$ & 
    $9.62^{+0.04}_{-0.06}$ & 
    $-6.63^{+0.05}_{-0.07}$ \\ \midrule
    \multicolumn{6}{c}{\textbf{SED fit} } \\ \midrule 
    $R_\star$ (R$_\odot$) & $0.664 \pm 0.020$ & $0.577 \pm 0.019$ & $0.647 \pm 0.019$ & $0.595 \pm 0.021$ & $0.475 \pm 0.018$ \\
    $M_\star$ (M$_\odot$) & $0.672 \pm 0.028$ & $0.598 \pm 0.024$ & $0.663 \pm 0.024$ & $0.615 \pm 0.026$ & $0.501 \pm 0.026$ \\
    $T_{\rm eff}$ (K) & $4054 \pm 62$ & $3786 \pm 59$ & $3947 \pm 73$ & $3896 \pm 61$ & $3609 \pm 66$ \\
    $\log g_\star$ (cgs) & $4.622 \pm 0.025$ & $4.692 \pm 0.025$ & $4.638 \pm 0.024$ & $4.678 \pm 0.025$ & $4.783 \pm 0.026$ \\
    $L_\star$ (L$_\odot$) & $0.1073 \pm 0.0046$ & $0.0616 \pm 0.0018$ & $0.0911 \pm 0.0051$ & $0.0735 \pm 0.0033$ & $0.0345 \pm 0.0015$ \\
    $\rm [Fe/H]$ (dex) & $+0.26 \pm 0.20$ & $+0.28 \pm 0.12$ & $+0.398 \pm 0.096$ & $+0.25 \pm 0.14$ & $+0.26 \pm 0.14$ \\
    $\rho_\star$ (g\,cm$^{-3}$) & $3.24 \pm 0.27$ & $4.38 \pm 0.38$ & $3.46 \pm 0.28$ & $4.12 \pm 0.37$ & $6.56 \pm 0.61$ \\
    $A_v$ (mag) & $0.137 \pm 0.083$ & $0.024 \pm 0.018$ & $0.23 \pm 0.15$ & $0.057 \pm 0.046$ & $0.088 \pm 0.071$ \\ \midrule
    \midrule
    \end{tabular}
    \caption{Stellar properties for all the systems considered in this work. Sources: 1: \protect\cite{Stassun2019}; 2: \protect\cite{gaiaDR3cat}; 3: \protect\cite{2masscat}; 4: \protect\cite{ucac4}; 5: \protect\cite{BJdist}}
    \label{tab:spectroscopic_params}
\end{table*}

\subsubsection{IRTF/SpeX}

We obtained near-infrared spectra of two targets and one companion star with the SpeX spectrograph \citep{Rayner2003} on the 3.2-m NASA Infrared Telescope Facility.
The short-wavelength cross-dispersed (SXD) mode was used with the $0\farcs{3} \times 15''$ slit aligned to the parallactic angle, yielding $R{\sim}2000$ spectra covering 0.80--2.42\,$\mu$m.
TOI-5292~A was observed on UTC 2023\,Oct\,31 under clear conditions with $0\farcs{6}$ seeing.
We collected 12 exposures of 300\,s at an average airmass of 1.1, nodding in an ABBA pattern.
We observed its faint, bound companion located $9\farcs42$ away, 2MASS J22565386-0807049 (TOI-5292~B), on UTC 2025\,Sep\,12, collecting $12 \times 300$\,s exposures under $1\farcs6$ seeing at an average airmass of 1.3.
Finally, TOI-5916 was observed on UTC 2024\,Jul\,11 under $1\farcs5$ seeing.
We again collected $12 \times 300$\,s exposures at an average airmass of 1.0.
After each science target, we collected a standard SXD calibration set and observed an A0\,V standard at a similar airmass.
The data were reduced with the Spextool v4.1 pipeline \citep{Cushing2004}, following the standard approach \citep[e.g.,][]{Barkaoui2024, Barkaoui2025, Ghachoui2024}.
The final spectra of TOI-5292~A, TOI-5292~B, and TOI-5916 have median SNRs per resolution element of 78, 9, and 76, respectively.

\subsubsection{SOAR/TripleSpec4.1}

We collected medium-resolution near-infrared spectra of TOI-3288~A, TOI-4666, and TOI-5007 with the TripleSpec4.1 spectrograph \citep{Schlawin2014} on the 4.1-m Southern Astrophysical Research (SOAR) telescope using the AEON queue.
The fixed 1\farcs{1} slit provides $R{\sim}3500$ spectra covering 0.80--2.47\,$\micron$, and all observations were carried out with the slit aligned to the parallactic angle and nodding in an ABBA pattern.
No significant contamination ($\lesssim5\%$ of the total flux) is expected from TOI-3288~B, the bound companion at $2\farcs16$, given this instrumental set up and the delta magnitude of 3.0 in the \textit{Gaia} band. 
TOI-4666 was observed on UTC 2023\,Oct\,6 under $1\farcs9$ seeing at an airmass of 1.2, with eight 30-s exposures yielding a median SNR per resolution element of 56.
TOI-3288~A and TOI-5007 were observed on UTC 2025\,Jul\,14 under $1\farcs3$ seeing at airmasses of 1.1 and 1.4, respectively, with four 120-s exposures for TOI-3288~A (median SNR per resolution element of 87) and four 300-s exposures for TOI-5007 (median SNR of 104). 

For each target, we also observed an A0\,V star at a similar airmass for telluric correction.
The data were reduced with Spextool v4.1 \citep{Cushing2004}, modified for TripleSpec4.1\footnote{\url{https://noirlab.edu/science/observing-noirlab/observing-ctio/observing-soar/data-reduction/triplespec-data}}, using the standard set of flat-field and arc-lamp exposures collected closest to the observations.

\subsubsection{Lick/Kast}
We obtained low-resolution red optical spectra of TOI-5916 and TOI-5292~A and B  with the Kast double spectrograph \citep{kastspectrograph} on the 3-m Shane telescope at Lick Observatory on UTC 2024\,Dec\,5 and UTC 2025\,Aug\,26, respectively. 
Observations in UTC 2024\,Dec were obtained in partly cloudy conditions with 0$\farcs$7 seeing; observations in UTC 2025\,Aug were obtained in scattered clouds with 0$\farcs$9 seeing. 
Both observations were conducted in a single longslit mask, with blue and red optical spectra split at 5700\,{\AA} by the d57 dichroic and dispersed by the 600/4310 grism and 600/7500 grating, respectively.
TOI-5916 was observed using the 1$\arcsec$ wide slit aligned to the parallactic angle to resulting in spectral resolutions of $\lambda/\Delta\lambda$ $\approx2200$ and $\approx2900$ for the blue and red spectra, respectively. We obtained a single 2400\,s exposure in the blue channel and two 1200\,s exposures in the red channel at an average airmass of 1.4. The G2\,V star HD 211476 ($V=7.0$) was observed at an airmass of 1.2 for telluric absorption calibration, and the sdO spectrophotometric calibrator Feige~110  \citep{Hamuy1992, Hamuy1994} was observed afterward for flux calibration. 
TOI-5292~A and B were observed using the 1$\farcs$5 wide slit aligned to the binary axis (position angle 232$^o$), resulting in spectral resolutions of $\lambda/\Delta\lambda$ $\approx1500$ and $\approx1950$ for the blue and red spectra, respectively. We obtained a single 3600\,s exposure in the blue channel and two 1800\,s exposures in the red channel at an average airmass of 1.45, and observed the G2\,V star HD 224383 ($V=8.8$) and the sdO spectrophotometric calibrator BD+28 4211 \citep{1990AJ.....99.1621O} for telluric and flux calibration, respectively.
HeHgCd and HeNeArHg arc lamp exposures were used to wavelength calibrate the blue and red data, and flat-field lamp exposures were used for pixel response calibration. Data were reduced using the kastredux code\footnote{\url{https://github.com/aburgasser/kastredux}.} using standard settings. 
The spectrum of TOI-5916 has median SNRs of
16 at 5425\,{\AA} and 73 at 7350\,{\AA}.
The spectrum of TOI-5292~A has median SNRs of
30 at 5425\,{\AA} and 95 at 7350\,{\AA}.
The spectrum of TOI-5292~B has median SNRs of
2 at 5425\,{\AA} and 26 at 7350\,{\AA}.

\subsubsection{Magellan/MagE}

We observed TOI-3288~A and TOI-4666 on UTC 2022\,Oct\,6 with the Magellan Echellette spectrograph \citep[MagE;][]{Marshall2008} on the 6.5-m Magellan Baade Telescope.
Conditions were clear, with seeing ranging from $0\farcs7$ to $1\farcs5$ over the night. 
We used the $0\farcs7 \times 10''$ slit, providing a resolving power of $\lambda / \Delta \lambda \approx 6000$ over 4000--9000\,\AA. 
For TOI-3288~A, we collected two 235-s exposures at an airmass of 1.4, and for TOI-4666, two 225-s exposures at an airmass of 1.0.
For flux calibration, we observed the spectrophotometric standard Feige 110 \citep{Hamuy1992, Hamuy1994}, gathering two 375-s exposures at an airmass of 1.1.
Wavelength calibration and pixel response corrections were performed using ThAr arcs, Xe flash lamps, bias, and incandescent lamp images.
Since we did not observe a telluric standard, prominent atmospheric absorption features from O$_2$ and H$_2$O remain in the final spectra.
Data reduction was carried out with PypeIt \citep{pypeit:zenodo, pypeit:joss_arXiv, pypeit:joss_pub} using standard settings.
The final spectra have median SNR of roughly 50 and 80 for TOI-3288~A and TOI-4666, respectively.

\subsubsection{SOAR/Goodman}

We observed TOI-5007 on UTC 2025\,Apr\,20 with the Goodman Spectrograph \citep{Clemens2004} on the SOAR telescope through the AEON queue. 
Conditions were clear with $1\farcs3$ seeing.
Using the red camera with the 400 lines mm$^{-1}$ grating, a $1''$ slit, and $2 \times 2$ binning, we obtained four 300-s exposures spanning 0.5--0.9\,$\mu$m at a resolving power of $R \approx 1000$. Standard sets of flat fields and arc lamps were collected immediately before and after the science sequence, and the spectrophotometric standard LTT 3218 was observed the same night for flux calibration. 
No telluric correction was applied. 
The data were reduced with PypeIt \citep{pypeit:zenodo, pypeit:joss_arXiv, pypeit:joss_pub}.
The final spectrum has a median signal-to-noise ratio of $S/N \approx 100$.

\subsubsection{VLT/ESPRESSO}
We obtained between 4 and 6 high-resolution ($\rm R \sim 140,000$) spectra of all targets using the ESPRESSO spectrograph, mounted at the VLT of the ESO Paranal observatory \citep{pepe21}. The observations were carried out between UTC 2023\,Dec and UTC 2025\,Sep. The brighter TOI-3288~A, TOI-4666, and TOI-5007 were observed with an exposure time of 600s, reaching on average a signal-to-noise ratio of 7.1, 4.6, and 4.1 at 550\,nm respectively. The fainter stars TOI-5292~A and TOI-5916 were observed with exposure times of 1200s and 900s, reaching an average signal-to-noise ratio of 5.3 and 2.3 respectively. All spectra were reduced by the ESO data reduction pipeline. Depending on the spectral type, we then used the M2 and K6 binary masks from the ESPRESSO pipeline\footnote{The ESPRESSO pipeline has been publicly released on \hyperlink{https://www.eso.org/sci/software/pipelines/espresso/espresso-pipe-recipes.html}{https://www.eso.org}.} to generate cross-correlation functions (CCFs). These are derived in wide borders from -200\,km\,s$^{-1}$ to 200\,km\,s$^{-1}$ at a resolution of 0.5\,km\,s$^{-1}$ for each spectral order and combined, weighted by the noise of each order. We used them to exclude any double-lined binaries and to derive a initial guess of the stellar radial velocity. We then derived precise radial velocities (RVs) of all targets, by fitting a double gaussian function introduced in \cite{sebastian24}. This function includes the presence of side-lobes in CCFs of M-dwarfs. Here we only use a 40\,km\,s$^{-1}$ wide part of the CCF, centered at the expected radial velocity of the star. We use these RVs in our global fit in Sec.~\ref{sec:global}, they are also reported in the Appendix, Table \ref{tab:RV_TOI-3288} -- \ref{tab:RV_TOI-5007}. Compared to the radial velocities, derived internally by the ESPRESSO pipeline, this additional step does not only allow us to measure or exclude any potential double-lined systems, it also does not require the knowledge of the stellar systemic velocity prior the derivation of the CCF. We compared RVs derived by both methods for our targets, but find no statistically significant difference.

\subsubsection{Magellan/PFS}
\label{sec:PFS}

We observed TOI-5007 with the Planet Finder Spectrograph \citep[PFS;][]{crane_carnegie_2006, crane_carnegie_2008, crane_carnegie_2010} on the 6.5 m Magellan II (Clay) telescope at Las Campanas Observatory. Between UTC 2023\,Apr\,8 and 2024\,Jul\,26 we obtained 7 visits, each consisting of three exposures of 1200 seconds, in $3\times3$ CCD binning mode with a $0\farcs3$ slit. This data was taken with the iodine gas absorption cell in the light path, which imprints a dense forest of molecular lines \citep{hatzes_iodine_2019} between 5000 to 6200 \AA. We also obtained one template spectrum without the iodine cell consisting of six exposures of 1200 seconds. The RVs were derived following the methodology of \cite{butler_attaining_1996}. As noted by \cite{Hartman2015} and \cite{bakos_hats-71b_2020}, due to the faintness of the target, and the optical region for the iodine region, the formal errors on the PFS RVs are likely underestimated. The final PFS RVs are included in \ref{tab:RV_TOI-5007} and as a machine readable table with the manuscript.





\section{Host Characterisation}
\label{sec:stellar_char}

In this section we combine spectroscopic, photometric and astrometric data with archival imagining to characterise the host stars. We begin with a search for line-of-sight or bound companions that could be blended in our photometry. We then determine physical parameters from spectroscopic analysis, scaling relations and spectral energy distribution (SED) analysis. Finally, we estimate kinematic ages for the stars in our sample. Adopted stellar parameters can be found in Table \ref{tab:spectroscopic_params}.

\subsection{Archival Imaging}
\label{sec:blends}
In order to check for possible blends between our target star and bound or background objects, we query the \textit{Gaia} DR3 Archive \citep{gaiaDR3cat} and investigate archival images for the systems discussed in this paper. All archival images can be found in Fig.~\ref{fig:archival_imaging}. 


For TOI-3288~A, we used data from the POSS-II/DSS \citep{poss} in the infrared filter taken in 1991 and in the red filter in 1995, as well as SSO in \textit{I+z} taken in 2022, thus spanning 31 years. The target has shifted \textcolor{black}{$0\farcs651$} from 1991 to 2022, due to its proper motion of 21\,mas\,yr$^{-1}$ \citep{gaiaDR3cat}. We find that there is a star very close to our target (Gaia DR3 6685431748040148992, TIC 1924436362, projected separation of $2\farcs16$) with matching parallaxes and proper motions and a delta magnitude of 3.0 in the $G$-band. We use \texttt{CoMover} \citep{2021_comover_software}, a Bayesian framework to determine the probability that stars are co-moving objects, and find the probability of these stars being a comoving pair is 97.25\%. We therefore determine TOI-3288 to be a wide binary.

Our ground based photometric instruments are unable to separate the two components, so our observations in redder bands are partially diluted. However, the $1\arcsec$ aperture fibre of ESPRESSO confirmed the planet host to be TOI-3288~A.

For TOI-4666, we used data taken in the blue filter of POSS-II/DSS in 1977, its red filter in 1991, and the \textit{I+z} filter of SSO in 2023. The target has shifted by \textcolor{black}{$1\farcs06$} in 46 years, due to its proper motion of 23\,mas\,yr$^{-1}$. No background objects are detected at the current position of the star. 

For TOI-5007, we used data taken in the infrared filter of POSS-II/DSS in 1979, its red filter in 1992, and the \textit{Sloan-i'} filter of SSO in 2023. The target has shifted by \textcolor{black}{$1\farcs52$} in 44 years, due to its proper motion of 33\,mas\,yr$^{-1}$. The nearby field of TOI-5007 is crowded, and we find that there is a line-of-sight blended star (TIC 1116679820) at a projected separation of $1\farcs85$. TIC 1116679820 is 5 \textit{TESS} magnitudes fainter than TOI-5007 and is therefore too faint to produce any significant contamination of our photometric or spectroscopic apertures.  

For TOI-5292~A, we used data from the POSS-I/DSS \citep{poss} taken in the red filter in 1953, the POSS-II/DSS infrared image taken in 1994 and the SPECULOOS-North image taken in the \textit{Sloan-z} filter in 2024. The target has shifted by \textcolor{black}{$0\farcs78$} in 71 years due to its proper motion of 11\,mas\,yr$^{-1}$.  We once again find a nearby star with a similar astrometric solution (Gaia DR3 2610422028054589824, TIC 33397741, projected separation of $9\farcs43$) which is 2.6 magnitudes fainter in the $G$-band. Using \texttt{CoMover} as above, we find the probability of these stars being a comoving pair is 99.99\%, and therefore determine TOI-5292 to be a wide binary. Given the larger projected separation of these stars, we are able to exclude the flux of the companion in our ground-based photometry.

For TOI-5916, we use images taken in 1953 (POSS-I/DSS, red filter), 1994 (POSS-II/DSS, infrared) and 2022 (SPECULOOS-North, \textit{Sloan-g'} filter). The target has shifted by \textcolor{black}{$1\farcs86$} in 69 years due to its proper motion of 27\,mas\,yr$^{-1}$, and no background objects are detected at the current position of the star.

\subsection{Spectroscopic Characterisation}

\begin{figure}
    \centering
    \includegraphics[width=\linewidth]{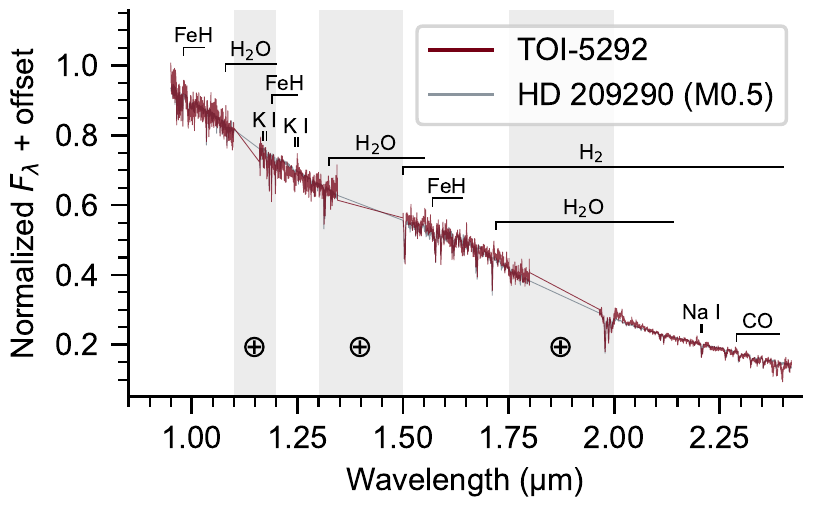}
    \includegraphics[width=\linewidth]{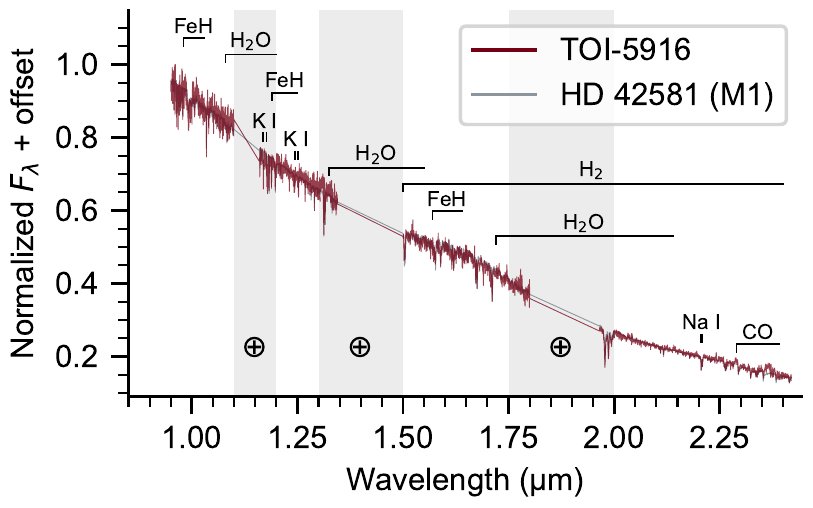}
    \caption{
    SpeX SXD spectra of TOI-5292~A (top) and TOI-5916 (bottom).
    The target spectra (red) are shown along with the closest spectral standard (grey).
    Spectral features of late-type stars are highlighted, and regions of strong tellurics are shaded.
    }
    \label{fig:spex}
\end{figure}

The SpeX spectra of TOI-5292~A and TOI-5916 are shown in Fig.\,\ref{fig:spex}.
Comparison with single-star standards from the IRTF Spectral Library \citep{Cushing2005, Rayner2009} using the SpeX Prism Library Analysis Toolkit \citep[SPLAT, ][]{splat} indicates the best matches to the M0.5\,V standard HD\,209290 and the M1\,V standard HD\,42581, and we adopt near-infrared spectral types of M0.5\,$\pm$0.5 and M1.0\,$\pm$0.5 for TOI-5292~A and TOI-5916, respectively.
From the $K$-band Na,\textsc{i} and Ca,\textsc{i} features and the H$_2$O--K2 index \citep{Rojas-Ayala2012}, we estimate iron enrichments of $\mathrm{[Fe/H]} = +0.33 \pm 0.14$ for TOI-5292~A and $\mathrm{[Fe/H]} = +0.09 \pm 0.14$ for TOI-5916, based on the \citet{Mann2013} relation.
The SpeX spectrum of TOI-5292~B, the companion to TOI-5292~A, allows us to assign a spectral type of M3.5$\pm$0.5, but it does not have sufficient SNR for a metallicity estimate.

\begin{figure}
    \centering
    \includegraphics[width=\linewidth]{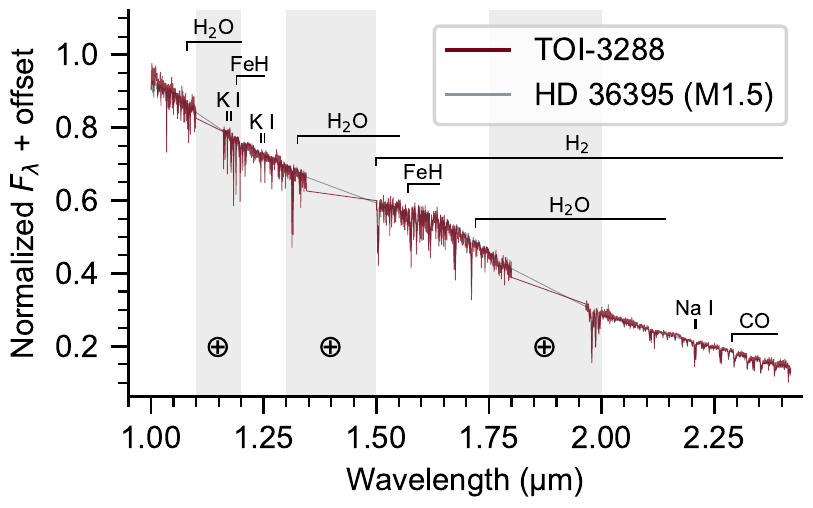}
    \includegraphics[width=\linewidth]{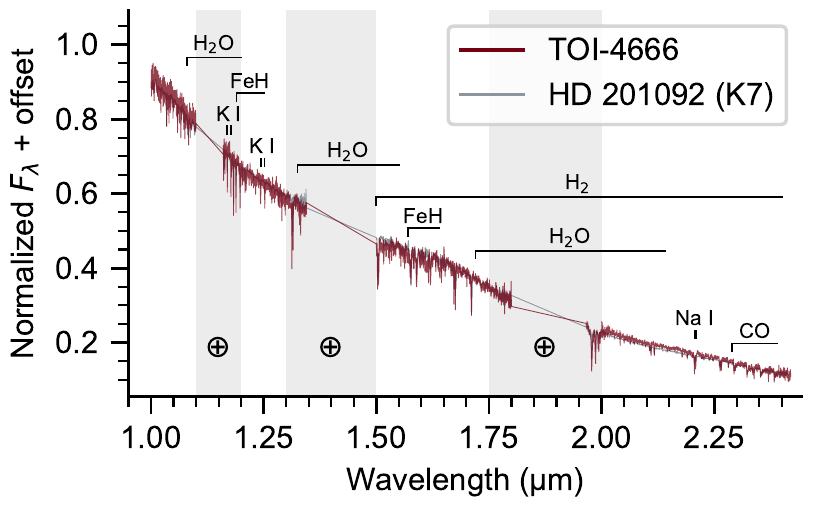}
    \includegraphics[width=\linewidth]{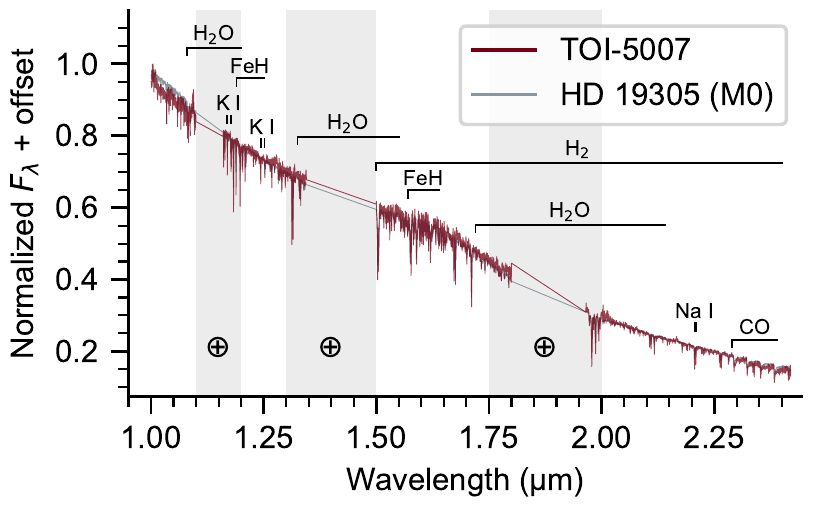}
    \caption{
    TripleSpec4.1 spectra of TOI-3288~A (top), TOI-4666 (middle), and TOI-5007 (bottom).
    The figure elements are the same as those of Fig.\,\ref{fig:spex}.
    Note that the TripleSpec4.1 spectra of the targets (red) have higher spectral resolving power than the SpeX spectra of the standards ($R{\sim}3500$ vs.\ $R{\sim}2000$).
    }
    \label{fig:tspec}
\end{figure}

The TripleSpec4.1 spectra of TOI-3288~A, TOI-4666, and TOI-5007 are shown in Fig.\,\ref{fig:tspec}.
For TOI-3288~A and TOI-5007, we find the best matches to the M1.5\,V standard HD\,36395 and the M0\,V standard HD\,19305, respectively, and adopt NIR spectral types of M1.5$\pm$0.5 and M0$\pm$0.5.
For TOI-4666, the best match is the K7\,V standard HD\,201092, but the next closest match is to the M0\,V standard because the IRTF Spectral Library does not include K8 or K9 standards.
Given this, we adopt an NIR spectral type of K8$\pm$1.
Using the \citet{Mann2013} relation, we estimate [Fe/H] values of $+0.65 \pm 0.12$, $+0.79 \pm 0.14$, and $+0.43 \pm 0.12$ for TOI-3288~A, TOI-4666, and TOI-5007, respectively.
Importantly, the metallicity estimates of TOI-3288~A and TOI-4666 lie outside the calibrated range of this relation ($-1.04 < \mathrm{[Fe/H]} <+0.56$) and should be treated with caution.

\begin{figure}
    \centering
    \includegraphics[width=\linewidth]{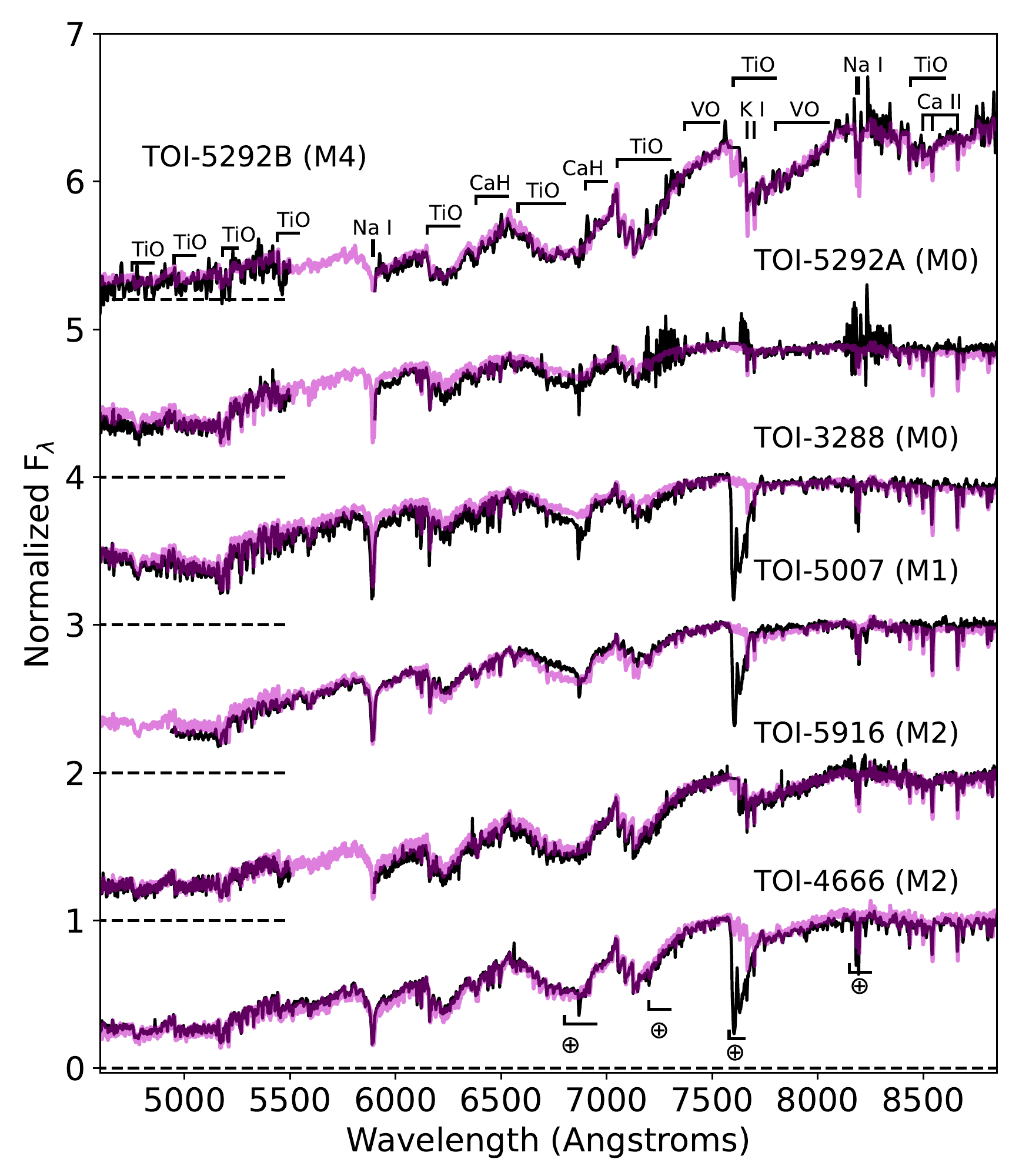}
    \caption{
    Optical spectra of (from bottom to top) 
    TOI-4666 (MagE), TOI-5916 (Kast), TOI-5007 (Goodman), TOI-3288~A (MagE), and TOI-5292~A and B (Kast; black lines)
    compared to their best-fit spectral standards from \citet[][magneta lines]{2007AJ....133..531B}. All spectra are normalized at 7400~{\AA} and offset by constants (dashed lines). Key spectral features are labeled, including uncorrected telluric absorption ($\oplus$) in the MagE and Goodman spectra.
    }
    \label{fig:optical}
\end{figure}

The optical spectra of TOI-3288~A, TOI-4666, TOI-5007, TOI-5292~A, TOI-5292~B, and TOI-5916 are shown in Figure~\ref{fig:optical}.
All of the spectra show spectral slopes and absorption features consistent with early mid-M dwarfs (M0--M2), with TOI-5292~B best matched to the M4 dwarf SDSS spectral template from 
\citet{2007AJ....133..531B}.
These classifications were confirmed using index-based methods described in 
\citet{1995AJ....110.1838R,1997AJ....113..806G}; and \citet{2003AJ....125.1598L}.
All of the spectra show weak absorption features of H~I, with the exception of TOI-4666 which exhibits emission in both H$\beta$ (equivalent width EW = $-$0.74$\pm$0.05~{\AA}) and H$\alpha$ (EW = $-$0.59$\pm$0.03~{\AA}), the latter indicating a relative H$\alpha$ luminosity of $\log{\left(L_{{\rm H}\alpha}/L_{\rm bol}\right)} = -4.51\pm0.05$ using the $\chi$ factor relation of \citet{2014ApJ...795..161D}. Magnetic emission in an M2 dwarf indicates an activity age $\lesssim$1.2~Gyr \citep{2008AJ....135..785W}, although there is no evidence of  Li\,\textsc{i} absorption at 6708~{\AA} ruling out an age less than $\sim30$\,Myr. Metallicity indices for all of the sources are measured to be $\zeta \gtrsim 1$ indicating solar to slightly supersolar metallicities (0 $\lesssim$ [Fe/H] $\lesssim$ +0.5). \\

We additionally used the empirical relations of \citet{Mann2015, Mann2019} to derive masses, radii, effective temperatures, surface gravities, and luminosities for TOI-3288~A, TOI-4666, TOI-5007, TOI-5292~A, and TOI-5916.
These relations use 2MASS $K_s$ magnitudes \citep{Skrutskie2006} and Gaia DR3 parallaxes and BR–RP colors \citep{GaiaCollaboration2023}.
[Fe/H] measurements may optimally be used, but given the need for additional scrutiny of some of the [Fe/H] estimates reported here, we opted not to include any in the empirical calibrations. 

We adopt the resulting stellar parameters in our analysis and present them in Table \ref{tab:spectroscopic_params}.


\subsection{Spectral Energy Distribution}

 
As an additional step to check our inferred stellar parameters, we used the publicly available exoplanet fitting suite EXOFASTv2 \citep{Eastman2017, Eastman2019}, which employs a differential evolution Markov chain Monte Carlo (DE-MCMC) method to jointly fit stellar and exoplanetary parameters. We simultaneously fit the stellar spectral energy distribution (SED) while incorporating the MESA Isochrones and Stellar Tracks (MIST; \citep{Paxton2015}. To sample the stellar SED, we used available broadband apparent magnitudes (see Fig.~\ref{fig:sed}), obtained using MKSED, a program distributed with EXOFASTv2 that automatically queries the most trusted photometric catalogs. Gaussian priors were imposed on the Gaia DR3 parallax \citep{Gaia2020}, which was corrected following the prescription of \cite{Lindegren2021}. An upper limit on the V-band extinction, $A_V$, was set based on reddening maps \citep{Schlegel1998, Schlafly2011}.

The results of the SED fitting are summarized in Table~\ref{tab:spectroscopic_params} for comparison.

\subsection{Age Estimation}

\begin{figure}
    \centering
    \includegraphics[width=\columnwidth]{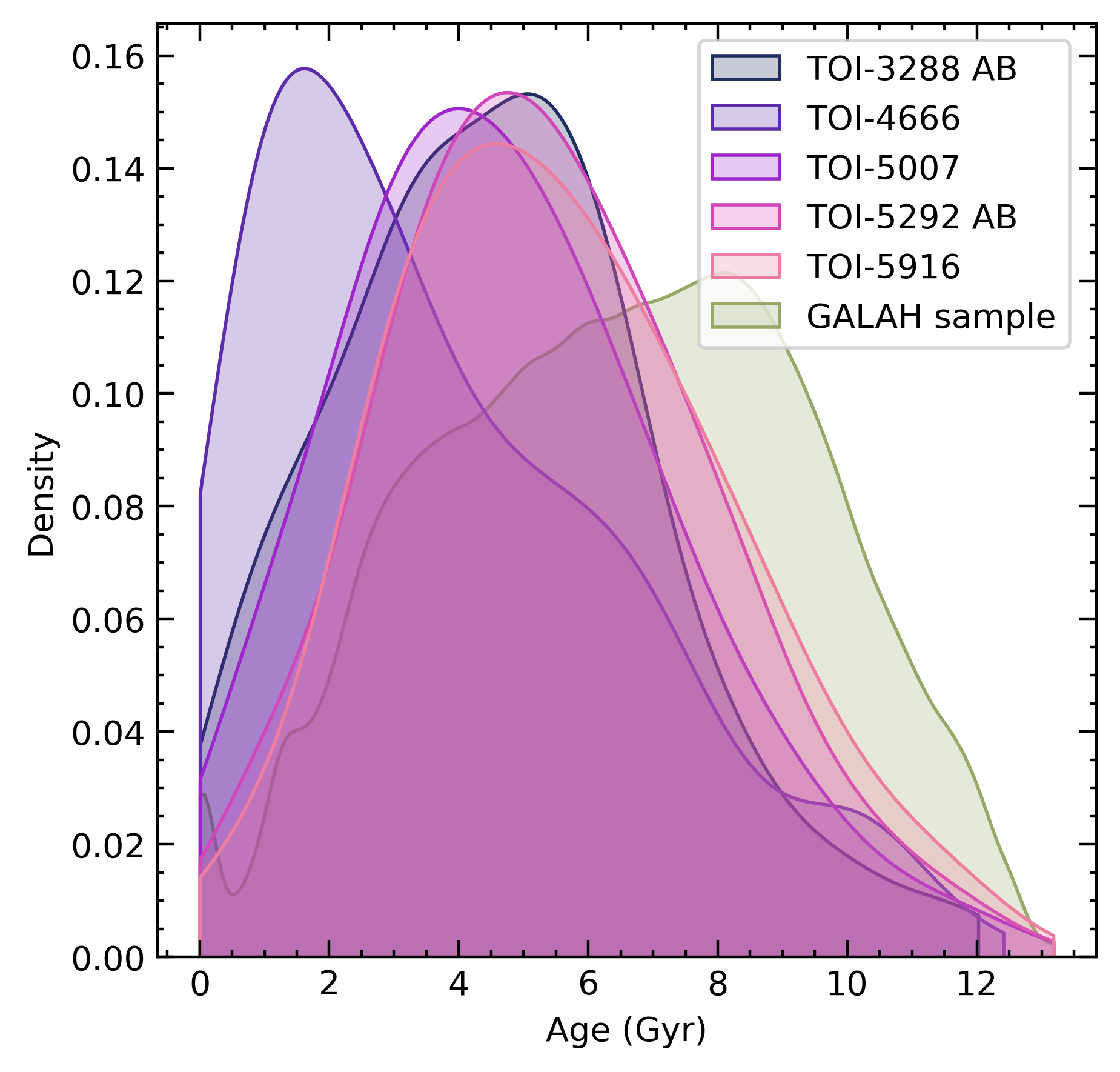}
    \caption{Distribution of ages for matched stars in the GALAH DR3 catalogue for each star in our sample in their respective colours. In the background in green we show the full GALAH sample for comparison. }
    \label{fig:ages}
\end{figure}

Following the methods outlined in \cite{Burgasser2017,Delrez2022,Dransfield2023}, we estimate the ages of the stars in our sample by comparing their $UVW$ velocities and metallicities to a selection of stars in the GALAH DR3 catalogue \citep{galah}. The comparison samples have ages estimated from isochrone fittings using Bayesian Stellar Parameter Estimation \citep[BSTEP,][]{bstep} and metallicities measured from high resolution spectroscopy. 

We compute the Galactic heliocentric orbital velocities $UVW$ for each star in our sample from their ESPRESSO radial velocities and their five-parameter \textit{Gaia} DR3 astrometric solutions, for which we correct the parallax for the noted zero-point offset following \citet{Lindegren2021}.
The adopted Galactic parameters are a solar position of $(X_{\odot}, Y_{\odot}, Z_{\odot}) = (8.2, 0, 0.0208)$~kpc, a Local Standard of Rest velocity of $240$~km/s, and a solar peculiar motion with respect to the LSR of $(U_{\odot}, V_{\odot}, W_{\odot}) = (11.1, 12.24, 7.25)$~km/s \citep{schonrich2010, bovy2015, bennett2019, gravity2019}.
The resulting $UVW$ are reported in Table~\ref{tab:spectroscopic_params}. The five stars show kinematics typical of the Galactic thin disk, which hosts relatively young, near-solar-metallicity stars like the Sun. Their supersolar metallicities are consistent with this classification and suggest ages younger than $\sim4$ Gyr, in line with Galactic chemical enrichment models \citep{spitoni2019}.


We select a matching sample from GALAH DR3 for each of our stars, with distances $\leq 250$ pc, $UVW$ velocities agreeing to within 15~km\,s$^{-1}$, and metallicities agreeing to within 0.3~dex. 

For TOI-3288~A we find 369 matched stars with a median age of $4.53^{+1.60}_{-1.69}\rm ~Gyr$. For TOI-4666 we widened the constraint on matching velocity to 20~km\,s$^{-1}$ as only 129 matching stars were found initially. With the constrained thus relaxed, we found 277 matching stars with a median age of $3.10^{+2.61}_{-1.66}\rm ~Gyr$. For TOI-5007 we find 6121 matching stars with a median age of $4.48^{+1.88}_{-1.60}\rm ~Gyr$. For TOI-5292~A we find 2124 matching stars with a median age of $5.18^{+1.84}_{-1.60}\rm ~Gyr$. Finally for TOI-5916 we find 11674 matching stars with a median age of $5.31^{+1.97}_{-1.70}\rm ~Gyr$.

We present the estimated ages in Table \ref{tab:spectroscopic_params} and the age distributions for matched stars as well as the full GALAH sample in Fig.~\ref{fig:ages}. Given the challenges associated with estimated precise stellar ages, we report slightly larger uncertainties of 2~Gyr for all our stars, except TOI-4666 where we increase this to 3~Gyr.


\section{Global analysis}
\label{sec:global}
We use {\sc Allesfitter} \citep[][and references therein]{AllesfitterSoft,AllesfitterPaper} to jointly model all available photometry and radial velocity datasets, following the procedures detailed in \cite{TOI282,Dransfield2023} and described here in brief. 

We adopt the signal parameters reported by {\sc SPOC} \citep{SPOC, Caldwell2020} as uniform priors and the stellar parameters reported in Table \ref{tab:spectroscopic_params} as normal priors. Additionally, we use {\sc PyLDTK} \citep{pyldtk} and the Phoenix stellar atmosphere models \citep{phoenix} to calculate quadratic limb darkening parameters which we adopt as normal priors after reparametrising them in the \citep{kippingldcs} parametrisation.

We fit for all transit parameters ($R_{\rm p}/R_\star$, $(R_\star+R_{\rm p})/a$, $\cos\,i$, $T_0$, and $P$) and allow the eccentricity to vary as a free parameter, parametrised as $\sqrt{e_{\rm b}} \cos{\omega_{\rm b}}$ and $\sqrt{e_{\rm b}} \sin{\omega_{\rm b}}$ {\citep[following][]{Triaud2011}}. We expect all orbits to be circular given the short orbital periods \citep[e.g.][]{Guillot1996}; therefore, any robustly detected eccentricity could indicate an additional perturbing body in the system.

Baseline trends in the photometry are modeled using {\sc Allesfitter}'s `hybrid spline' functionality, and the `hybrid offset' is used for the RVs. Finally, we fit a jitter term which is added in quadrature to the RV errors and an `error scaling' term to account for white noise in the photometric data. We use the nested sampling algorithm {\sc Dynesty} \citep{dynesty} in all our fits. The results of our global fits can all be found in Table \ref{tab:fit_results}.

\begin{table*}
    \centering
    \begin{tabular}{cccccc}
    \toprule \toprule
    \multicolumn{6}{c}{\textbf{Fitted Parameters} } \\ \midrule \midrule
         & \textbf{TOI-3288~Ab} & \textbf{TOI-4666~b} & \textbf{TOI-5007} & \textbf{TOI-5292~Ab} & \textbf{TOI-5916~b} \\ \midrule  \vspace{0.1cm}
       $R_b / R_\star$  & $0.16701_{-0.00080}^{+0.00090}$ & $0.19653\pm0.00042$ & $0.1551_{-0.0014}^{+0.0015}$& $0.1897_{-0.0023}^{+0.0024}$ & $0.2169\pm0.0016$\\ \vspace{0.1cm}
       $(R_\star + R_b) / a_b$  & $0.1592_{-0.0030}^{+0.0047}$ & $0.09610\pm0.0010$ & $0.1172_{-0.0038}^{+0.0043}$ & $0.1211_{-0.0039}^{+0.0046}$ & $0.1032_{-0.0059}^{+0.013}$\\ \vspace{0.1cm}
       $\cos{i_b}$  & $0.0175\pm{-0.011}$ & $0.0036_{-0.0024}^{+0.0034}$ &$0.0933_{-0.0064}^{+0.0076}$ & $0.020\pm0.012$ & $0.0140_{-0.0081}^{+0.010}$\\ 
       $T_{0;b}~(\mathrm{BJD})$  & $2460343.9069\pm0.0001$ & $2460265.13619\pm0.00007$  & $2460068.8110\pm{-0.0001}$  & $2460551.1467\pm0.0002$ & $2460255.6055\pm0.0002$\\ 
       $P_b~(\mathrm{d})$  & $1.4338647\pm0.0000003$ & $2.9089165\pm0.0000007$ & $2.543372\pm0.000001$ & $2.021910\pm0.000003$ & $2.367128\pm0.000001$\\ 
       $K_b~(\mathrm{km\,s^{-1}})$  & $0.433_{-0.017}^{+0.016}$ & $0.103\pm0.013$ & $0.1327_{-0.0057}^{+0.0058}$ & $0.3115\pm0.010$ & $0.173_{-0.019}^{+0.037}$\\  
       $\sqrt{e_b} \cos{\omega_b}$  & $-0.03_{-0.10}^{+0.11}$ & $-0.01_{-0.11}^{+0.15}$  & $-0.019_{-0.072}^{+0.070}$ & $-0.107_{-0.060}^{+0.080}$ & $-0.06_{-0.20}^{+0.23}$\\ 
       $\sqrt{e_b} \sin{\omega_b}$  & $0.01_{-0.12}^{+0.14}$ & $-0.008\pm0.065$ & $0.308_{-0.074}^{+0.057}$ & $-0.115_{-0.091}^{+0.13}$  & $0.17_{-0.24}^{+0.20}$ \\\midrule
    \multicolumn{6}{c}{\textbf{Derived Parameters} } \\ \midrule
        $R_\mathrm{b}$ ($\mathrm{R_{jup}}$) & $1.092\pm0.033$ & $1.118\pm0.035$ & $0.991\pm0.030$ & $1.128_{-0.037}^{+0.039}$ & $1.013_{-0.033}^{+0.032}$\\
        $M_\mathrm{b}$ ($\mathrm{M_{jup}}$) & $1.687_{-0.109}^{+0.114}$ & $0.489_{-0.066}^{+0.067}$ & $0.684_{-0.047}^{+0.051}$ & $1.299_{-0.074}^{+0.081}$ & $0.713_{-0.114}^{+0.241}$ \\
        $a_\mathrm{b}$ (AU) & $0.02285_{-0.00092}^{+0.00089}$ & $0.0340\pm0.0011$ & $0.0.0300{-0.0013}^{+0.0014}$ & $0.0279\pm{-0.0013}$ & $0.0262_{-0.0029}^{+0.0021}$\\
        $a_\mathrm{b}/R_\star$ & $7.33_{-0.21}^{+0.14}$ & $12.45\pm0.13$ & $9.85_{-0.35}^{+0.33}$ & $9.82_{-0.35}^{+0.33}$ & $11.79_{-1.30}^{+0.71}$ \\
        $i_\mathrm{b}$ (deg) & $89.00\pm{-0.61}$ & $89.79_{-0.19}^{+0.14}$ & $84.65_{-0.44}^{+0.37}$ & $88.87_{-0.72}^{+0.69}$ & $89.20_{-0.58}^{+0.47}$ \\
        $e_\mathrm{b}$ & $0.021_{-0.015}^{+0.026}$ & $0.023_{-0.016}^{+0.033}$ & $0.100_{-0.038}^{+0.037}$ & $0.034_{-0.020}^{+0.024}$ & $0.090_{-0.062}^{+0.114}$\\
        $w_\mathrm{b}$ (deg) &$174_{-108}^{+91}$& $183_{-49}^{+50}$  &$93_{-13}^{+14}$ & $227_{-53}^{+29}$ & $114_{-54}^{+89}$\\
        $b_\mathrm{tra;b}$ & $0.128_{-0.077}^{+0.074}$ & $0.045_{-0.030}^{+0.042}$ & $0.831\pm{0.007}$ & $0.199_{-0.119}^{+0.116}$ & $0.158_{-0.091}^{+0.088}$\\
        $T_\mathrm{tot;b}$ (h) & $1.737_{-0.007}^{+0.008}$  & $2.137\pm0.004$ & $1.446_{-0.015}^{+0.016}$ & $1.884_{-0.019}^{+0.023}$ & $1.774_{-0.013}^{+0.015}$\\
        $T_\mathrm{full;b}$ (h) &$1.229_{-0.009}^{+0.007}$   & $1.432\pm{-0.004}$ & $0.280_{-0.082}^{+0.062}$ & $1.257_{-0.029}^{+0.017}$ & $1.125_{-0.016}^{+0.015}$\\
        $\rho_\mathrm{b}$ ($\rm g\,cm^{-3}$) & $1.60_{-0.20}^{+0.23}$ & $0.43_{-0.07}^{+0.08}$ & $0.87_{-0.11}^{+0.13}$ & $1.12_{-0.14}^{+0.16}$ & $0.86_{-0.18}^{+0.31}$\\
        $T_\mathrm{eq;b}$ (K) & $973_{-23}^{+24}$ & $695\pm15$ & $782\pm21$ & $809_{-21}^{+22}$ & $683_{-29}^{+43}$\\
        $S$ (S$_{\oplus}$)  & $213.8\pm27.2$ & $55.6\pm6.8$ & $89.5\pm12.2$ & $102.0\pm14.2$ & $51.7\pm12.6$ \\
         \midrule
    \end{tabular}
    \caption{Fitted and derived parameters from our global fits.}
    \label{tab:fit_results}
\end{table*}

\section{Results}
\label{sec:results}

In this section we summarise the results for each system, highlighting key parameters and standout features. 

\subsection{TOI-3288~Ab}

\begin{figure}
    \centering
    \includegraphics[width=0.8\columnwidth]{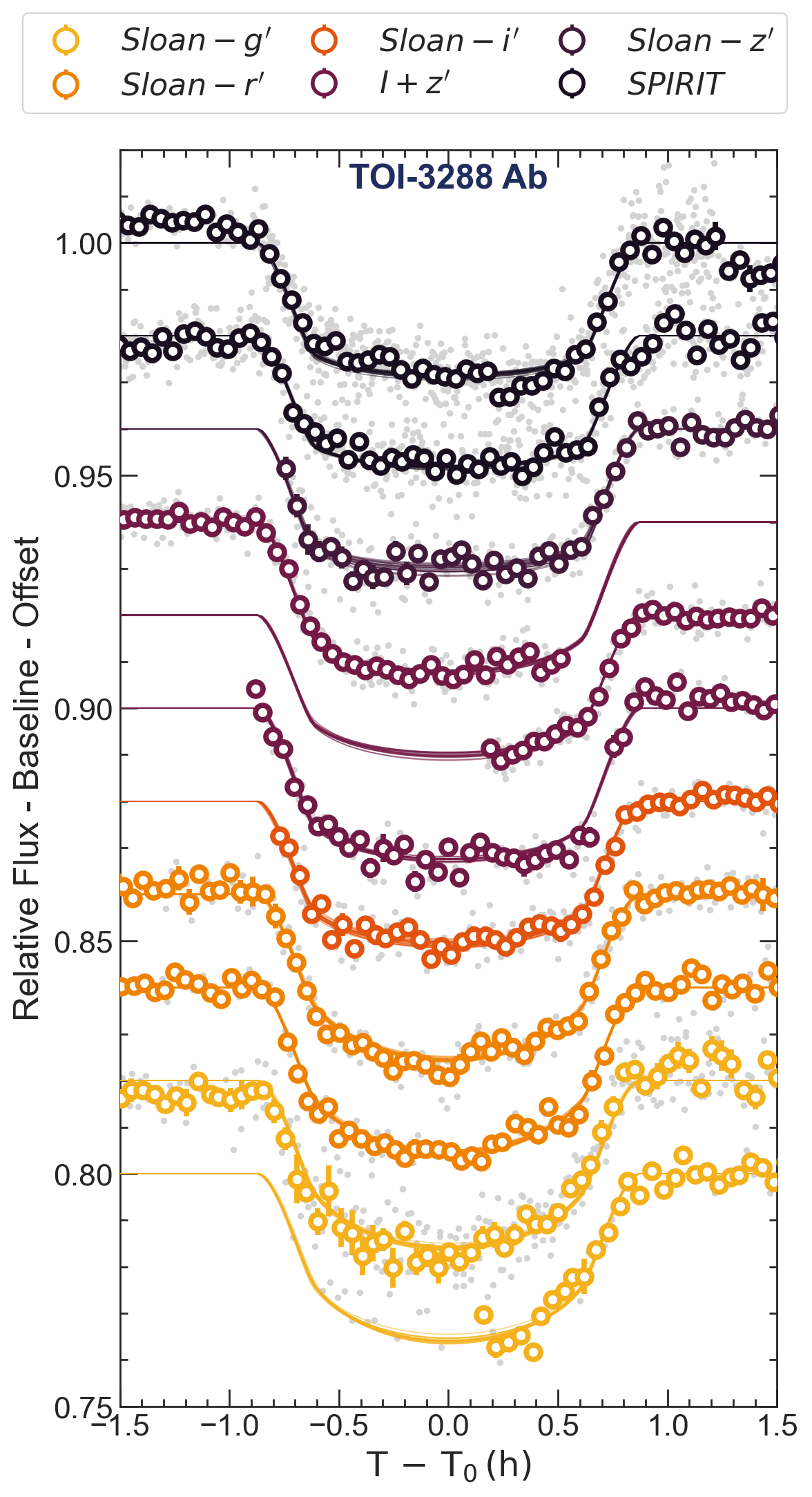}
    \includegraphics[width=0.8\columnwidth]{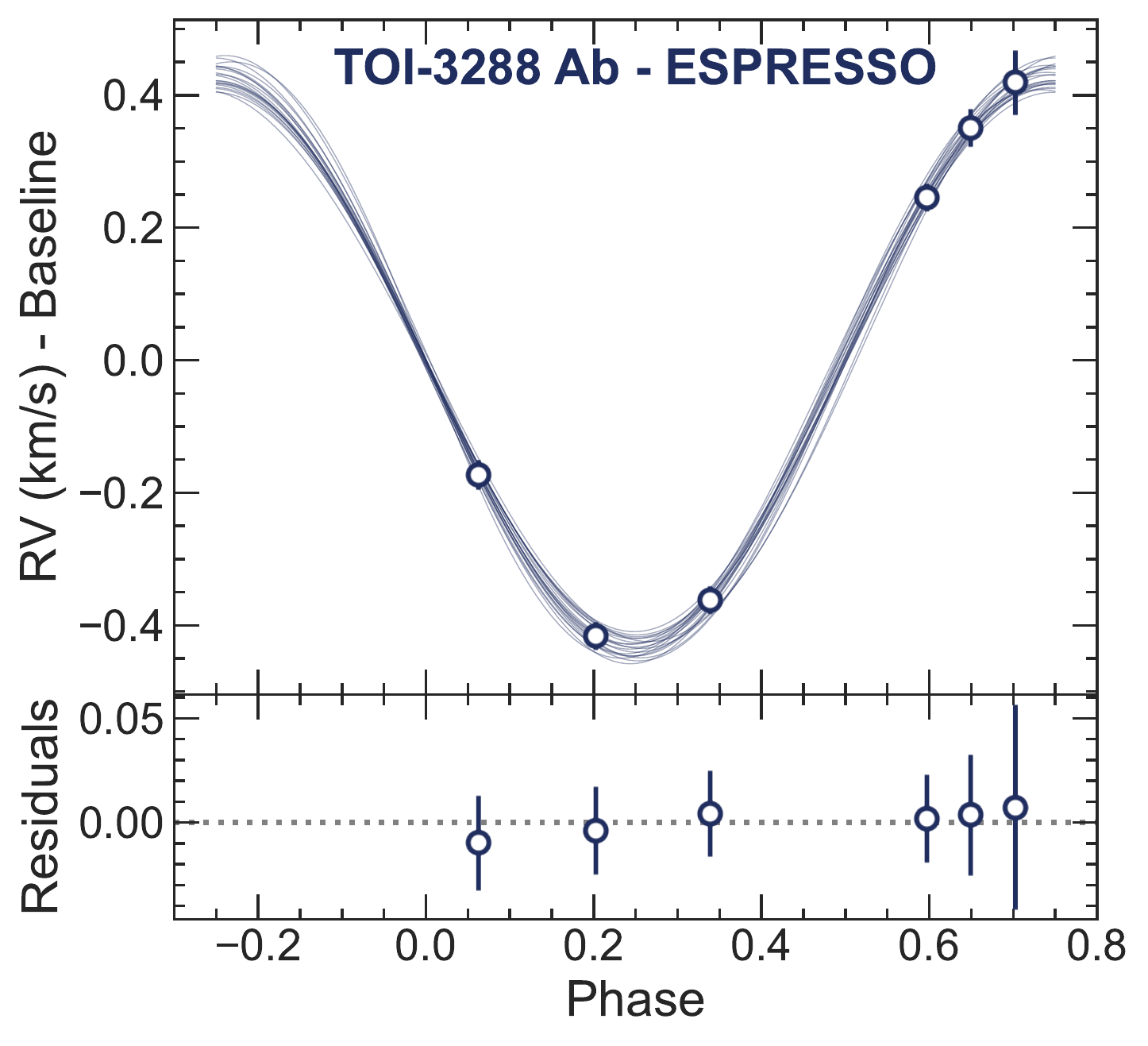}
    \caption{Models for TOI-3288~Ab obtained from a joint photometry+RV fit.
    \textit{Top:} Photometry of TOI-3288~Ab along with best fitting models. Raw flux is plotted in grey and 15-minute binned flux in white circles outlined in colours to indicate the wavelength band the observation was taken in. Transit models are corrected for by subtracting the baseline. All transits have a relative offset applied for plotting. The model lines are 20 random draws from the posterior transit model. Where multiple transits were observed by an instrument in the same filter, these are phase-folded.  
    \textit{Bottom:} Phase-folded radial velocities of TOI-3288~A from ESPRESSO. Model lines are 20 random draws from posterior radial velocity model. RV models is corrected by subtracting the baseline model.}
    \label{fig:3288_phot}
\end{figure}

TOI-3288~Ab is a $1.092\pm0.033~\mathrm{R_{jup}}$ planet with a mass of $1.687_{-0.109}^{+0.114}~\mathrm{M_{jup}}$. As such, it is the densest planet in our sample. Our model finds an eccentricity consistent with zero at the $2\sigma$ level ($0.021_{-0.015}^{+0.026}$). We calculate the tidal circularisation timescale for TOI-3288~Ab following \cite{Guillot1996}, taking Jupiter's values for the tidal dissipation factor, $Q\sim 10^5$, and the planet's primordial rotation rate, $\omega_{\rm p} \sim 1.7\times 10^{-4} \rm s^{-1}$. We also use the value of tidal dissipation factor, $Q\sim 10^{6.5}$, reported in \cite{Jackson2008} to account for the difficulty in accurately deriving this quantity. We find that the planet would have circularised in $0.05-1.5~\rm Gyr$, which is consistent with our finding of a circular orbit given our system age estimation. In Fig \ref{fig:3288_phot} we present the transits used in our fits along with best fitting models, as well as the radial velocity points from ESPRESSO.

As described above, TOI-3288~A is in a close binary system and most of our ground-based photometry is contaminated by the secondary. Inspection of the radial PSF plots showed that the companion is not detected in our bluest band, $Sloan-g'$. We therefore fixed the dilution parameter in our fits to zero for this band, and set a uniform prior of $\mathcal{U}(0,0.1)$ in remaining bands. This prior was informed by calculating the maximum dilution expected in our reddest bands based on the fluxes of both components measured by \textit{Gaia} in the $G_{rp}$ band. The fitted dilutions for each band are presented in Table \ref{tab:ldcs_3288Ab}, and the result of adding these parameters is that the planet radius increased by $\sim10\%$ compared with the initial model without dilution.

\begin{figure}
    \centering
    \includegraphics[width=0.8\columnwidth]{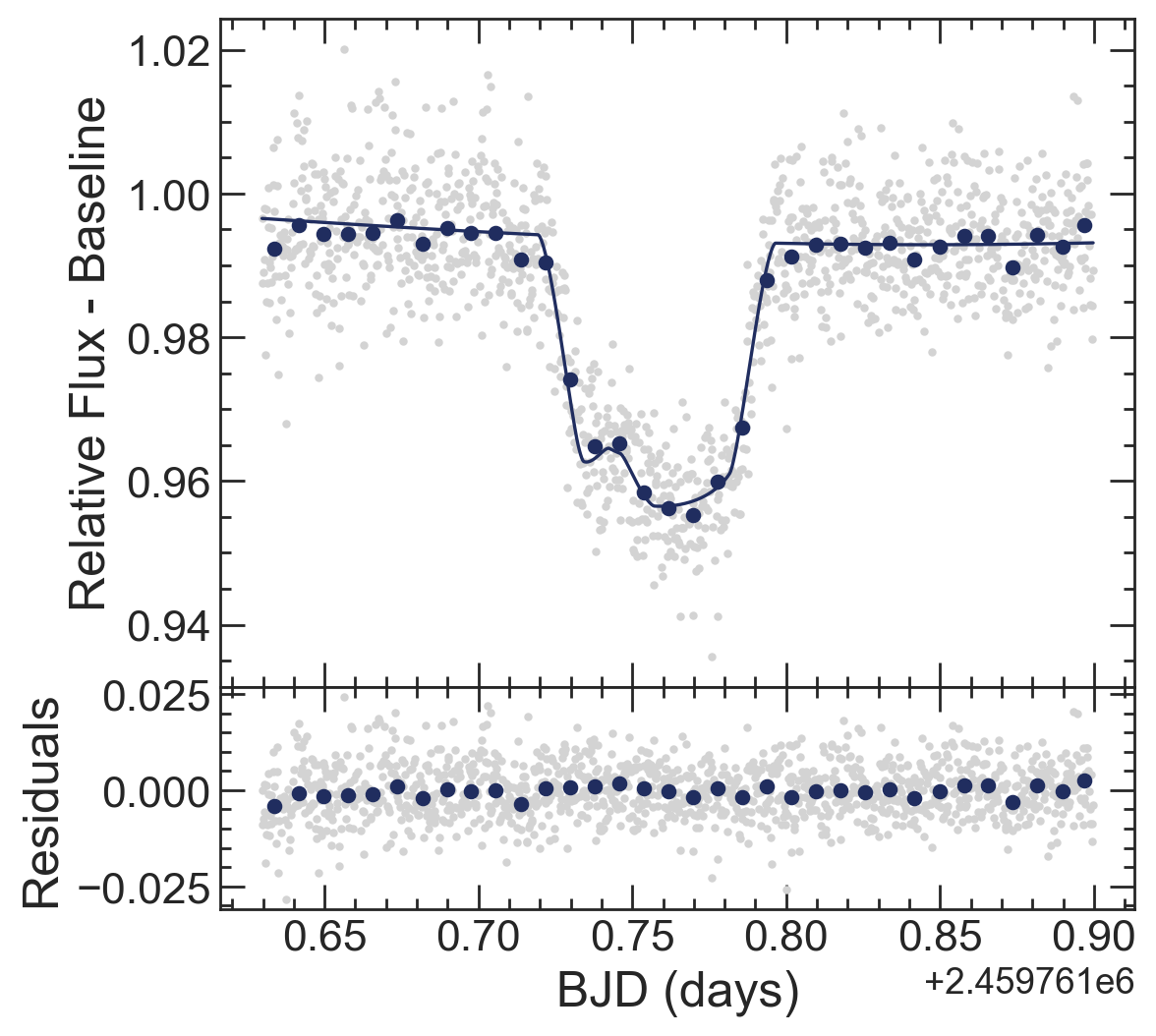}
    \includegraphics[width=0.8\columnwidth]{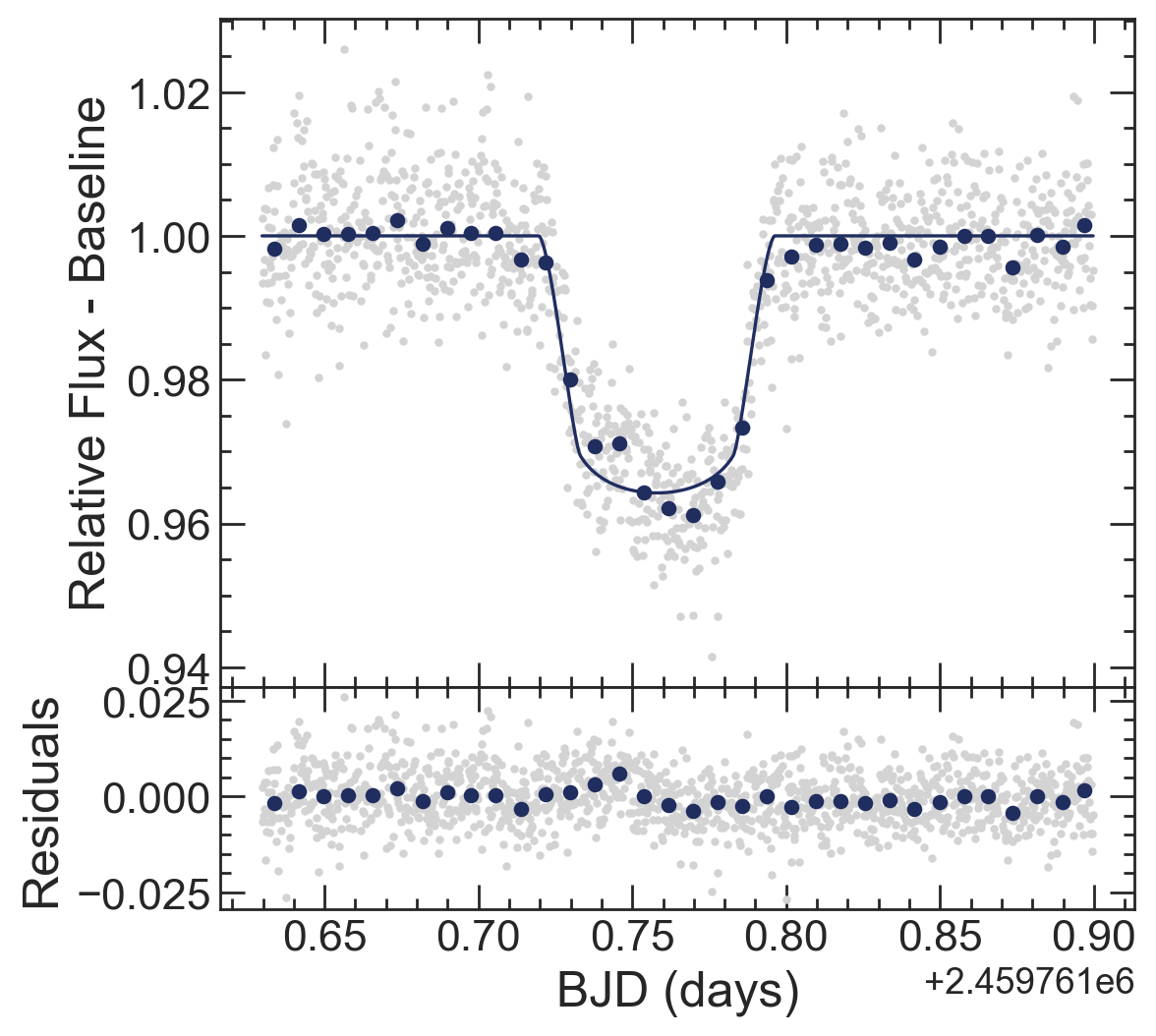}
    \caption{Comparison of fits with and without a physical spot model for one transit epoch of TOI-3288~Ab. The upper figure is the alternative hypothesis which includes a single starspot, while the bottom figure is the null hypothesis. }
    \label{fig:3288_spot}
\end{figure}

The planetary radius measurement could still be under or overestimated, as several of the transits of TOI-3288~Ab show spot crossings. In fact, of the systems presented in this work, TOI-3288~A is the most consistently spotty host. Comprehensive analysis of the spot parameters (latitudes, longitudes, and contrasts, and sizes) is beyond the scope of this work, and will be presented in a future paper by Davoudi, F. (in prep.). However, to illustrate the potential effects of spot crossings on the inferred planetary radius, we select one transit epoch observed with SSO/Callisto in $zYJ$ that has a clear spot crossing event. We fit a model including {\sc Allesfitter}'s physical spot model, and compare it to the null hypothesis, which consists of a transit model with no starspot. We compare the Bayesian evidence for the models by calculating the Bayes Factor, $\Delta \ln Z$. In both cases, we fit circular models with fixed $T_0$, $P$, and limb darkening coefficients. We present both fitted models in Fig~\ref{fig:3288_spot}.

We find that the model including one spot is preferred with a Bayes Factor $\Delta \ln Z = 23.9$. In the interpretation of \cite{kastspectrograph}, $\Delta \ln Z > 5$ is considered decisive evidence. The radius derived from the null is $0.03~\rm R_\oplus$ smaller (2.5\%) than the alternative model radius. This is less than a $1\sigma$ difference, which indicates that the scatter introduced by spot crossings in our transits is well accounted for in our uncertainties. The effect of spot crossings is also mitigated by phase folding repeated observations taken with the same instrument using the same filter. 

We note that as we have not used any of the \textit{TESS} data in our analysis, our results are completely independent of those presented in Frensch et al. (submitted). 

\subsection{TOI-4666 b}

\begin{figure}
    \centering
    \includegraphics[width=0.8\columnwidth]{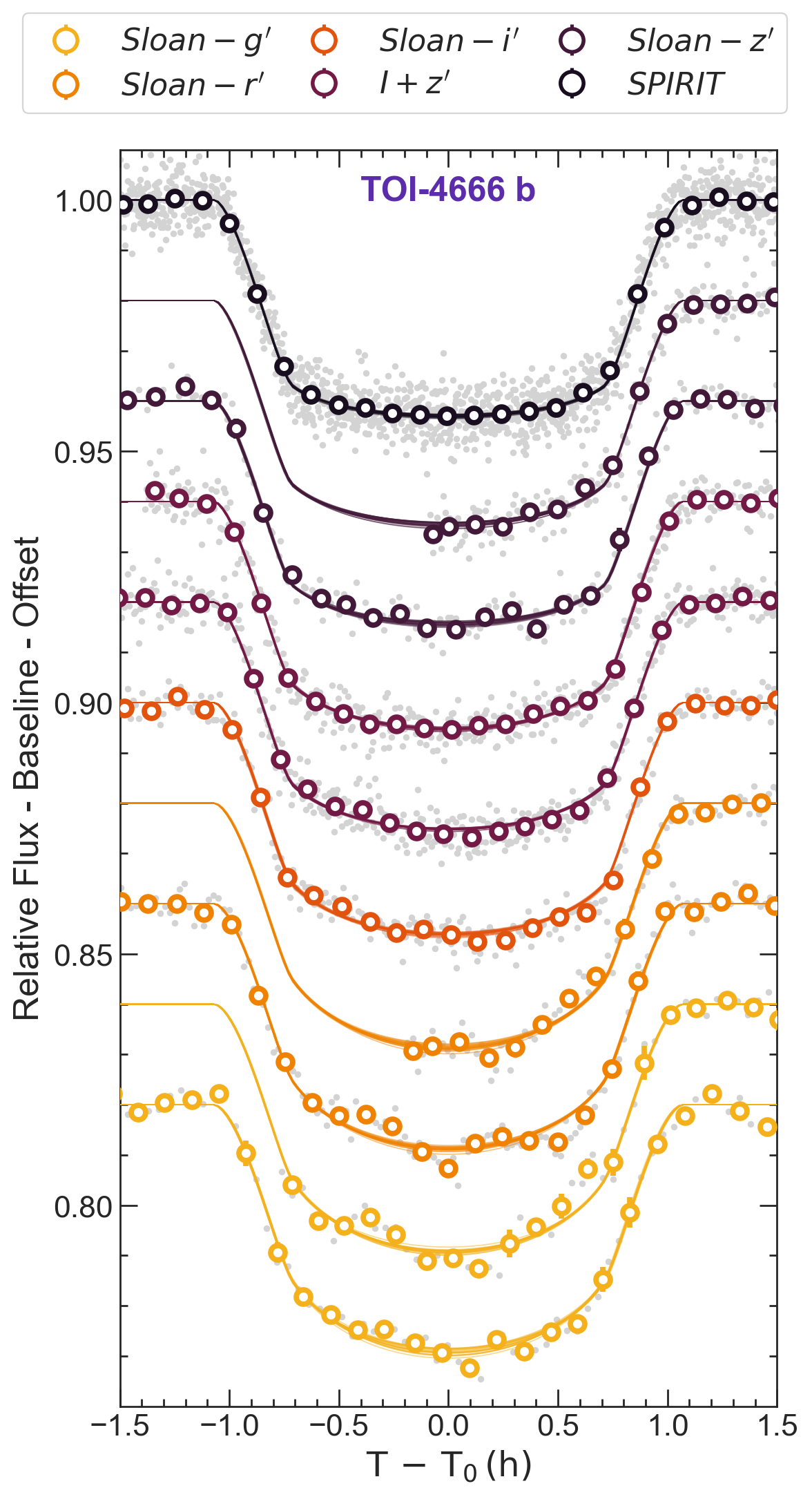}
    \includegraphics[width=0.8\columnwidth]{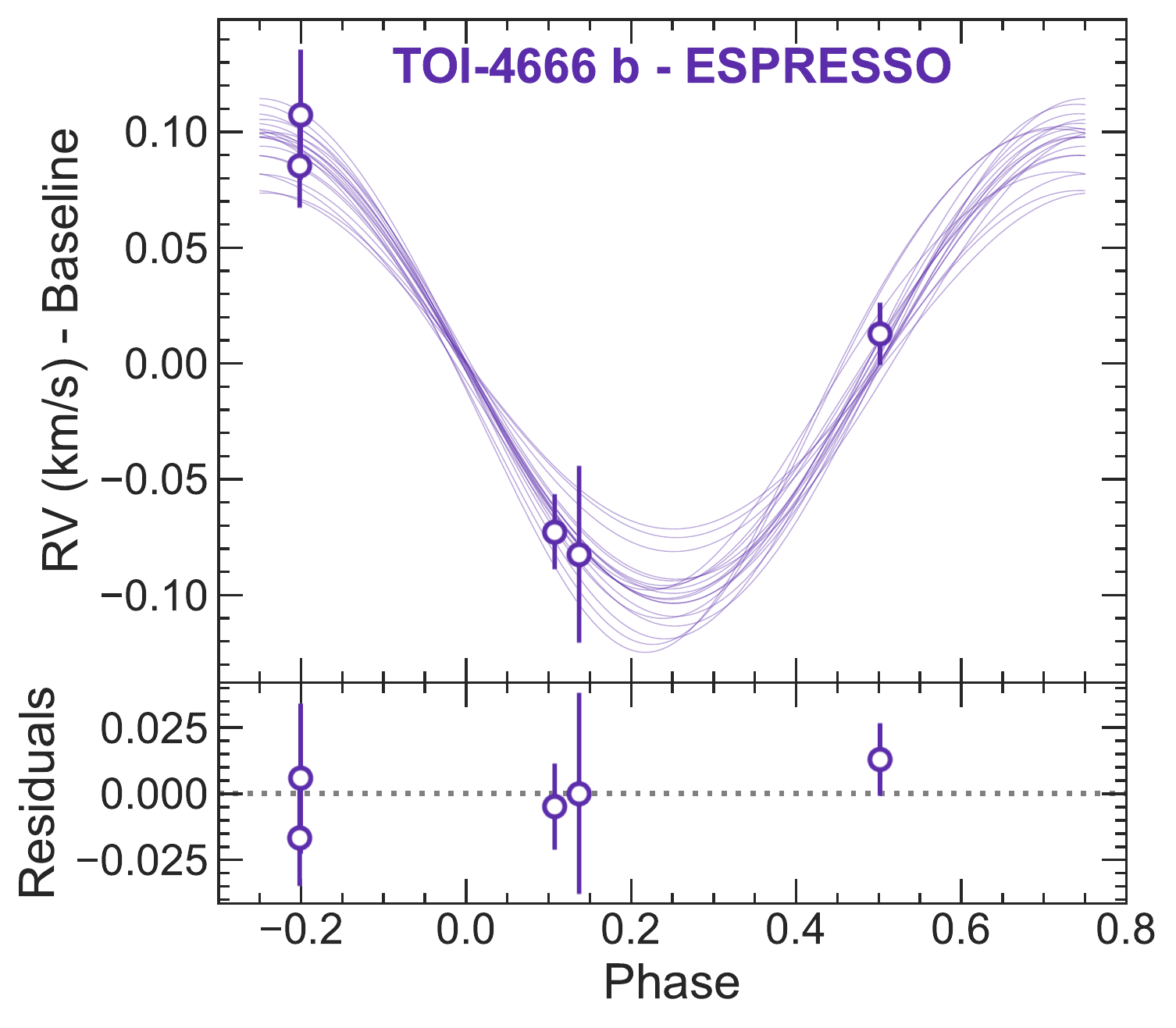}
    \caption{Same as Fig.~\ref{fig:3288_phot} but for TOI-4666.
    }
    \label{fig:4666_phot}
\end{figure}

TOI-4666~b is a $1.118\pm0.035~\mathrm{R_{jup}}$ planet with a mass of $0.489_{-0.066}^{+0.067}~\mathrm{M_{jup}}$. It the lowest density planet in our sample ($\rho_\mathrm{b}$ = $0.43_{-0.07}^{+0.08}~\mathrm{g\,cm^{-3}}$), with a density $\sim 60\%$ that of Saturn ($0.69~\mathrm{g\,cm^{-3}}$). Like TOI-3288~Ab, there are spot crossings evident on many of the individual transits, leading to some additional scatter in the data. All transits and radial velocity points used in this can be found in Fig. \ref{fig:4666_phot}.

The eccentricity derived from our fit is once again consistent with zero; however, as can be seen in the bottom panel of Fig. \ref{fig:4666_phot}, the sampling of the orbit is uneven. In the case of TOI-4666~b, further RV points would be beneficial to confirm the low eccentricity in light of the low system age revealed in Section \ref{sec:spec}. Calculating the tidal circularisation timescale as above, using the same values of $Q$, we find a circularisation timescale for TOI-4666~b of $0.2-6.5$ Gyr. Therefore, without more data we cannot be certain whether the orbit is circular or not. 

As with TOI-3288~Ab, the results presented here are completely independent of those presented in Frensch et al. (submitted). 

\subsection{TOI-5007 b}

\begin{figure}
    \centering
    \includegraphics[width=0.8\columnwidth]{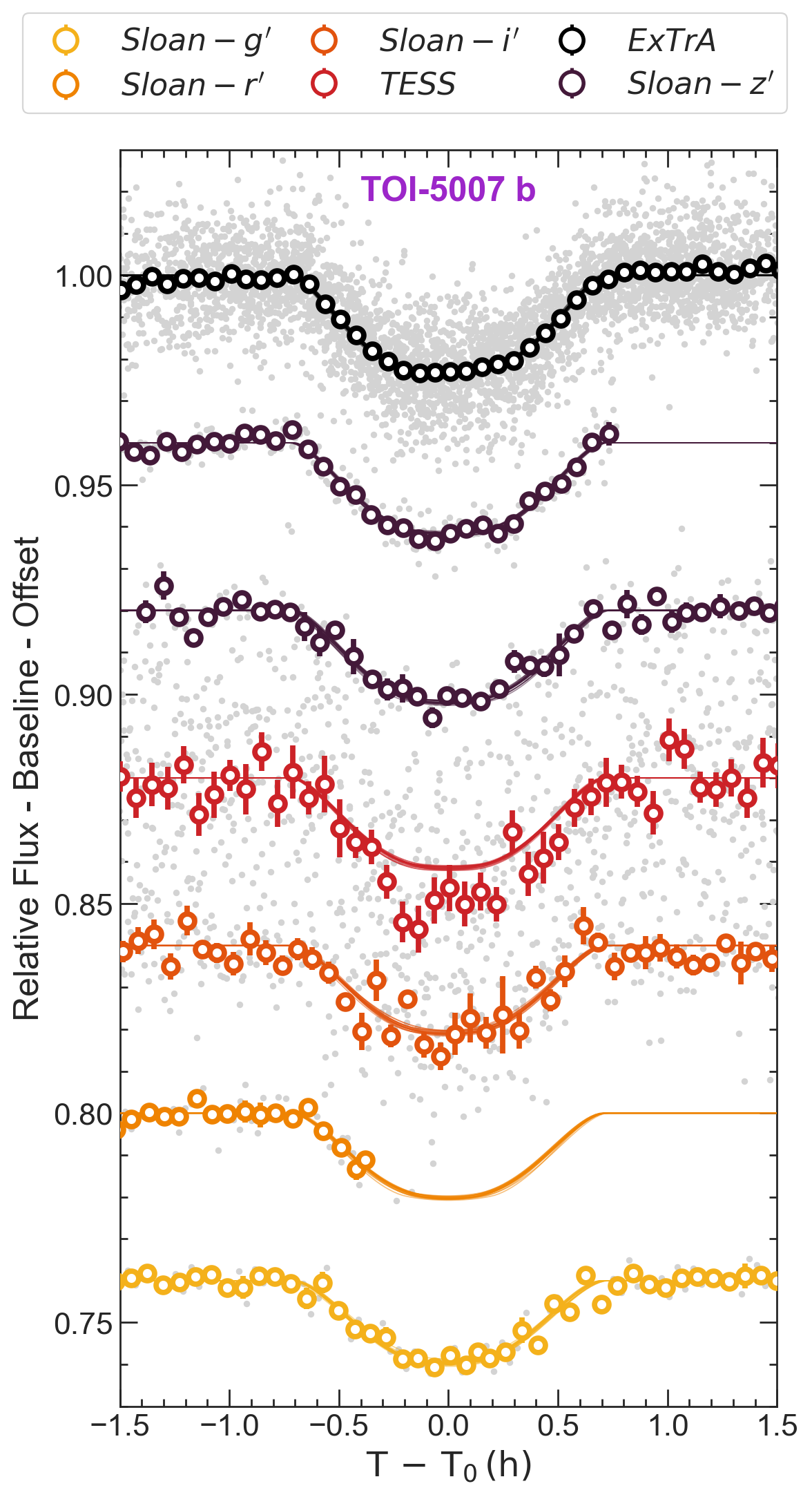}
    \includegraphics[width=0.8\columnwidth]{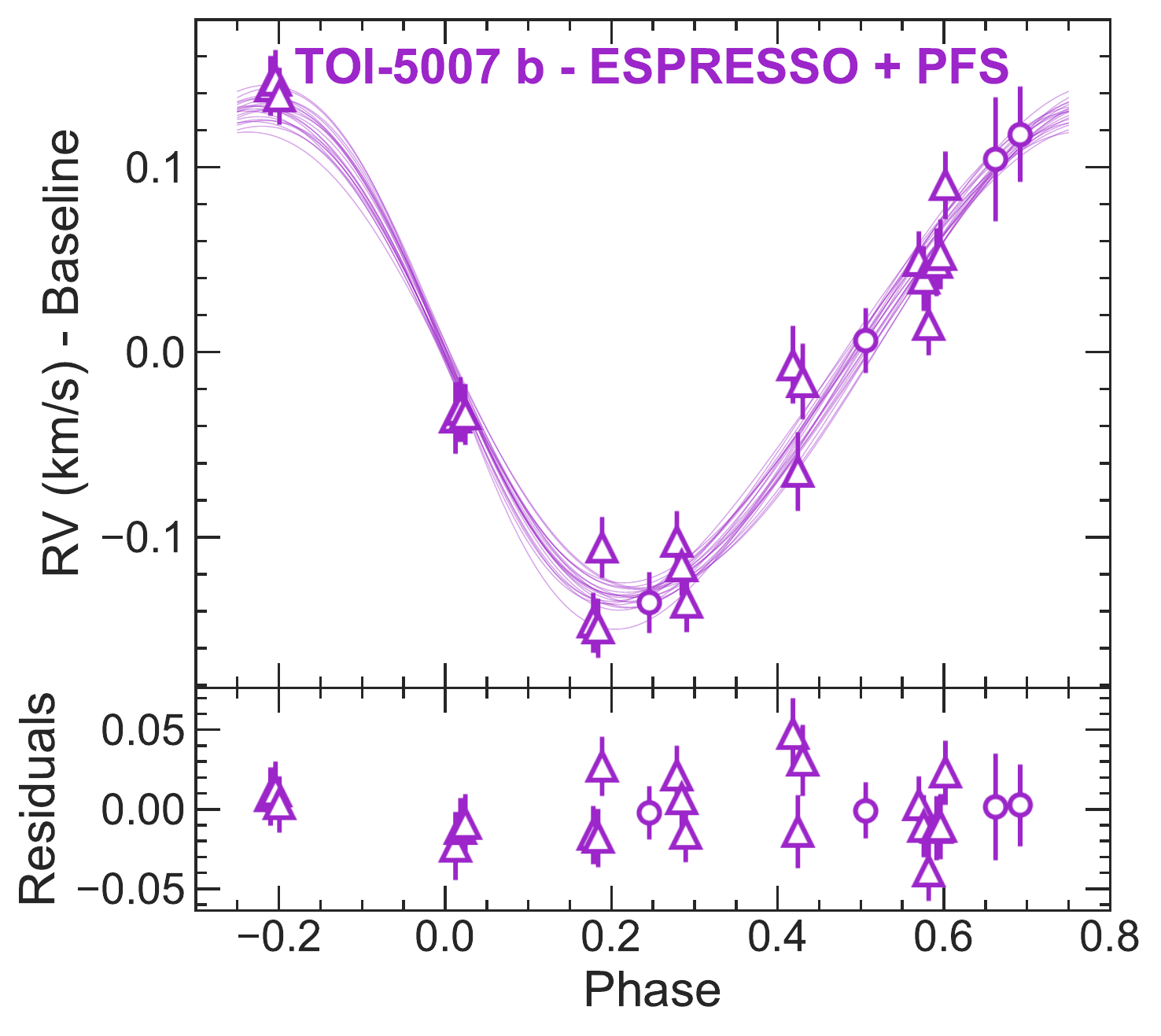}
    \caption{Same as Fig.~\ref{fig:3288_phot} but for TOI-5007. In the bottom panel we plot the PFS RV points as triangles and the ESPRESSO points as circles.}
    \label{fig:5007_phot}
\end{figure}

TOI-5007~b is a $0.991\pm0.030~\mathrm{R_{jup}}$ planet with a mass of $0.684_{-0.047}^{+0.051}~\mathrm{M_{jup}}$. In Fig \ref{fig:5007_phot} we present the transits used in our fits along with best fitting models, as well the radial velocity points from ESPRESSO and PFS. 

For this planet, our fit finds a small but non-zero eccentricity at the $3\sigma$ level of $e_\mathrm{b}=0.100_{-0.038}^{+0.037}$. As above we calculate the tidal circularisation timescale and find it to be $\tau \approx \rm 1.7 - 5.4 Gyr$. Given our system age estimation, it is possible that this planet could still be undergoing tidal circularisation. 

However, an alternative explanation for the detected eccentricity could be an additional body in the system which continues to force the eccentricity. We searched for additional signals in the radial velocity residuals using a Generalised Lomb-Scargle periodogram \citep{2009_Generalized_periodogram} and do not find any additional periodicities. We also searched the residuals of the \textit{TESS} data using {\sc TLS} \citep{tls} and find no evidence for additional transiting planets. However, given that only one sector of \textit{TESS} data with 2-minute cadence is available, this is by no means conclusive. 

Further scrutiny of the RV data and models in Fig \ref{fig:5007_phot} reveals that the orbit is not evenly sampled. In particular, there is limited sampling between phases $-0.2$ and $0.2$, and the scatter around phase 0 is smaller than the rest of the PFS data. All 20 models drawn from the posterior in this figure pass through these data at phase 0.0, so this is likely driving the fit. As described in Section \ref{sec:PFS} there are 21 PFS spectra collected across seven nights, effectively producing seven distinct measurements. 

To test the impact of intra-night variation on the derived eccentricity, we average the three nightly spectra. While the loss of time resolution is unlikely to have an impact, the risk of averaging the measurements is artificially shrinking the uncertainties. To be as conservative as possible, we calculate an unweighted mean of the three nightly RVs and estimated the uncertainty as the root-mean-square (RMS) of the individual uncertainties. We rerun our global fit of the data with the averaged PFS data and find that the eccentricity reduces to $0.080_{-0.041}^{+0.034}$, which is consistent with zero at the $2\sigma$ level.

Given the uncertainty associated with this planet's eccentricity, we report in Table \ref{tab:fit_results} the value derived from modelling all the available data, and emphasise that more RV points sampling the full orbital phase will be crucial to conclusively rule in or out an eccentric orbit for this planet. 

TOI-5007~b has the highest impact parameter in our sample, with a configuration that is almost grazing. In fact, the large number of transits collected by ExTrA with their near-infrared capabilities was critical in constraining the shape of the transit and therefore the planetary radius. 

No dilution was included in our model as the blended non-bound companion is $>5$ magnitudes fainter and is not detected in any of our photometry. The \textit{TESS} PDCSAP photometry used in our fit is already corrected for the contamination of nearby stars in the aperture, however there is additional scatter in the \textit{TESS} photometry most likely attributable to the very crowded field. This can be seen clearly in Fig.~\ref{fig:5007_phot}.

\subsection{TOI-5292~Ab}

\begin{figure}
    \centering
    \includegraphics[width=0.8\columnwidth]{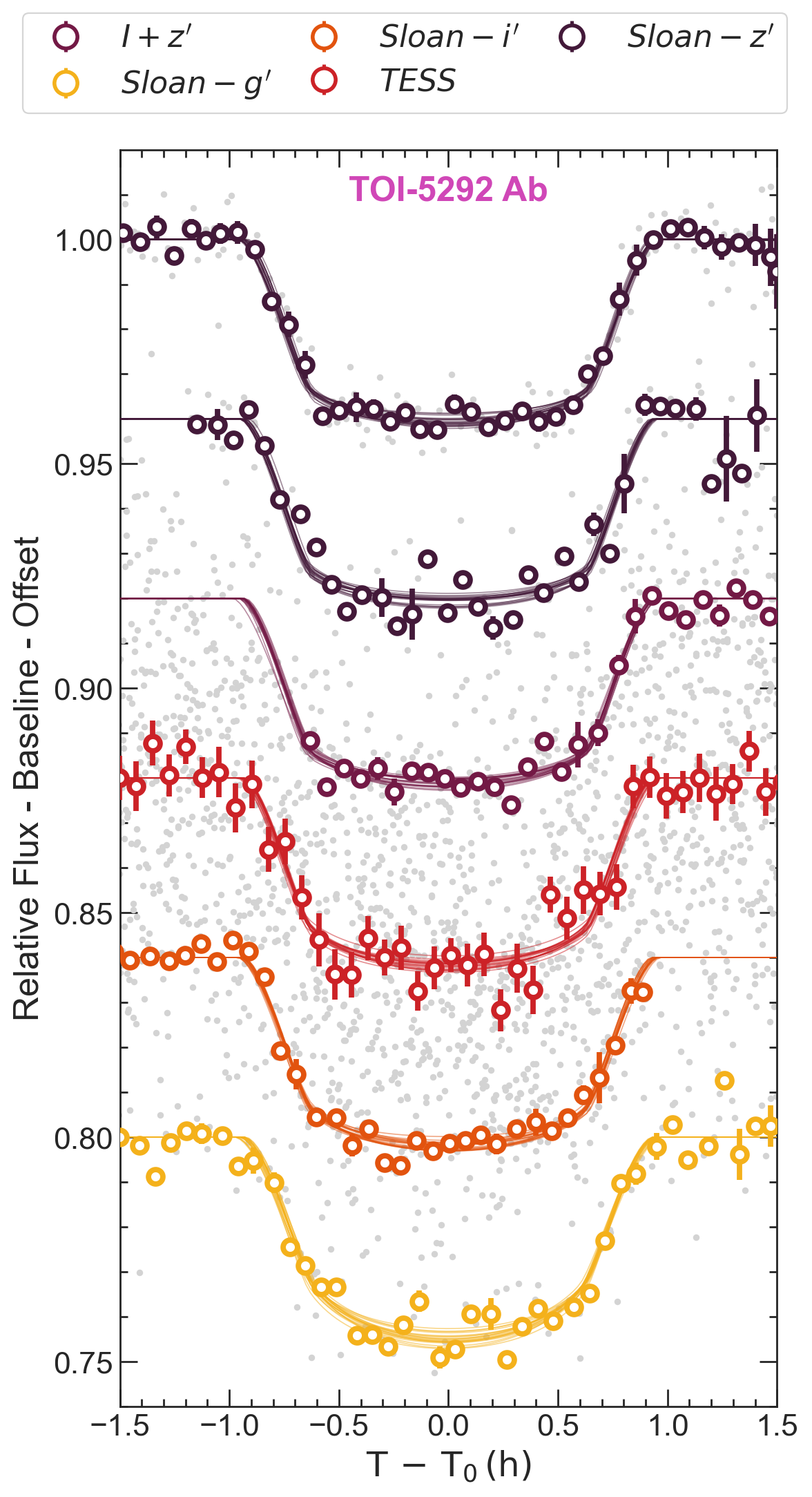}
    \includegraphics[width=0.8\columnwidth]{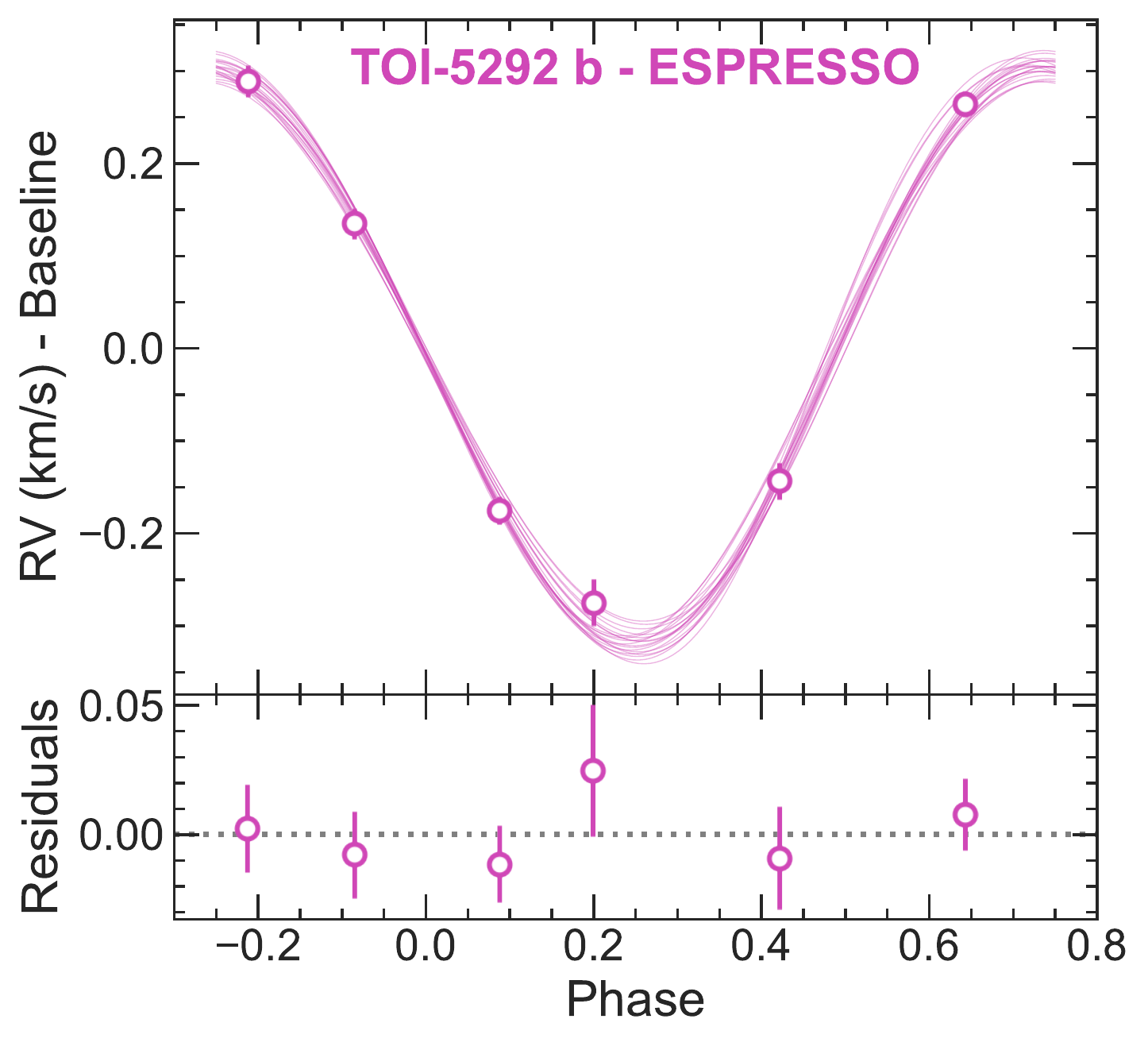}
    \caption{Same as Fig.~\ref{fig:3288_phot} but for TOI-5292~A.}    \label{fig:5292_phot}
\end{figure}

TOI-5292~Ab is a $1.128_{-0.037}^{+0.039}~\mathrm{R_{jup}}$ planet with a mass of $1.299_{-0.074}^{+0.081}~\mathrm{M_{jup}}$, making it the second highest density planet in our sample ($\rho_{\rm b}=1.12_{-0.14}^{+0.16}~\rm g\,cm^{-3}$) and the closest to Jupiter ($1.33~ \rm g\,cm^{-3}$). In Fig \ref{fig:5292_phot} we present the transits used in our fits along with best fitting models, as well the radial velocity points from ESPRESSO. 

The eccentricity of the planet is found to be consistent with zero at the $2\sigma$ level, and as can be seen in the lower panel of Fig. \ref{fig:5292_phot} the orbit is very evenly sampled. As before, we calculate $\tau$ which gives a range $0.14-4.6\rm\,Gyr$ suggesting that tidal circularisation would indeed be complete given the system age of $5.23^{+1.77}_{-1.65}~\rm Gyr$. 

As the bound companion is well separated in the photometry, we do not include any dilution in our model. 

\subsection{TOI-5916 b}

\begin{figure}
    \centering
    \includegraphics[width=0.8\columnwidth]{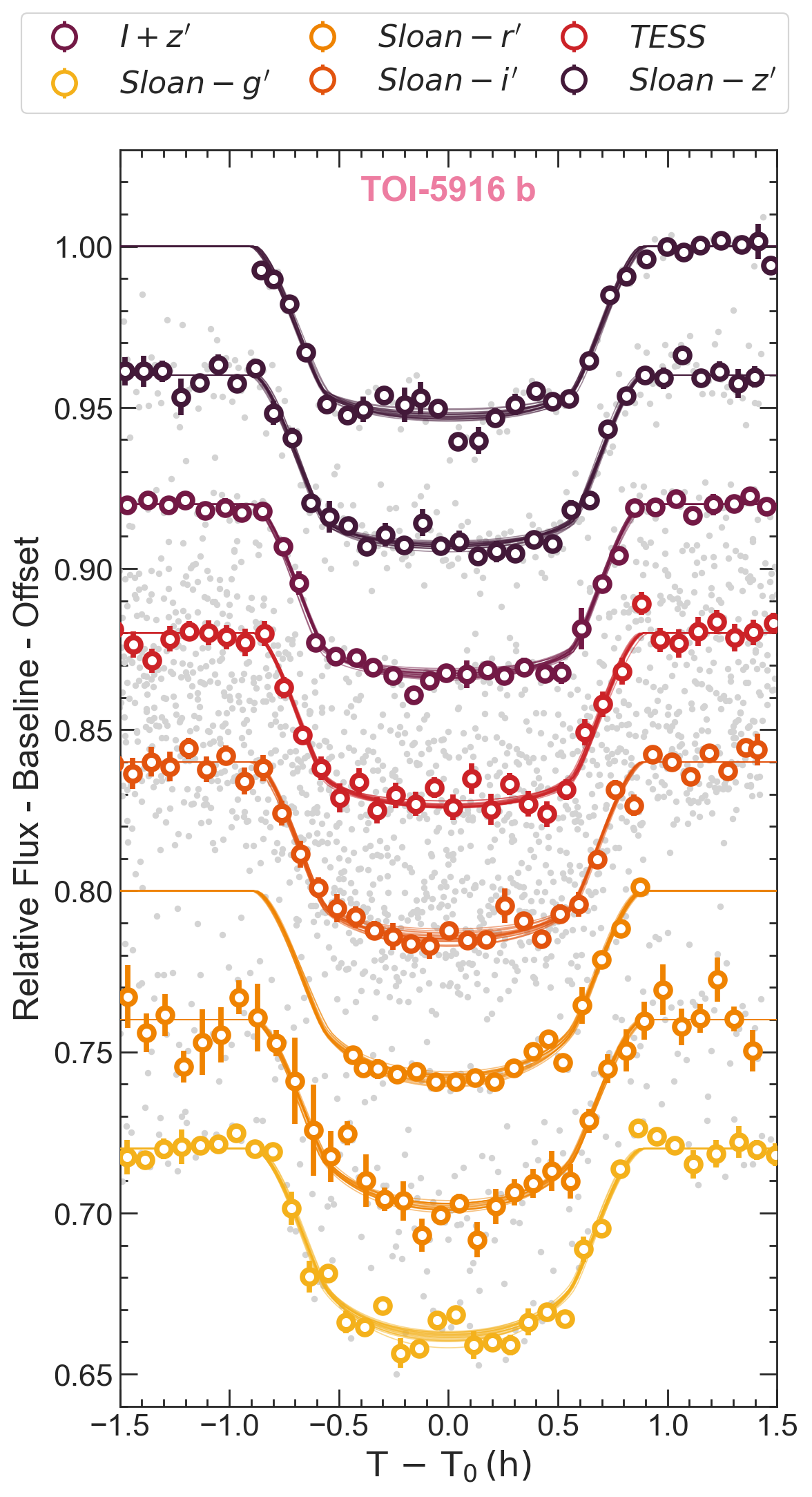}
    \includegraphics[width=0.8\columnwidth]{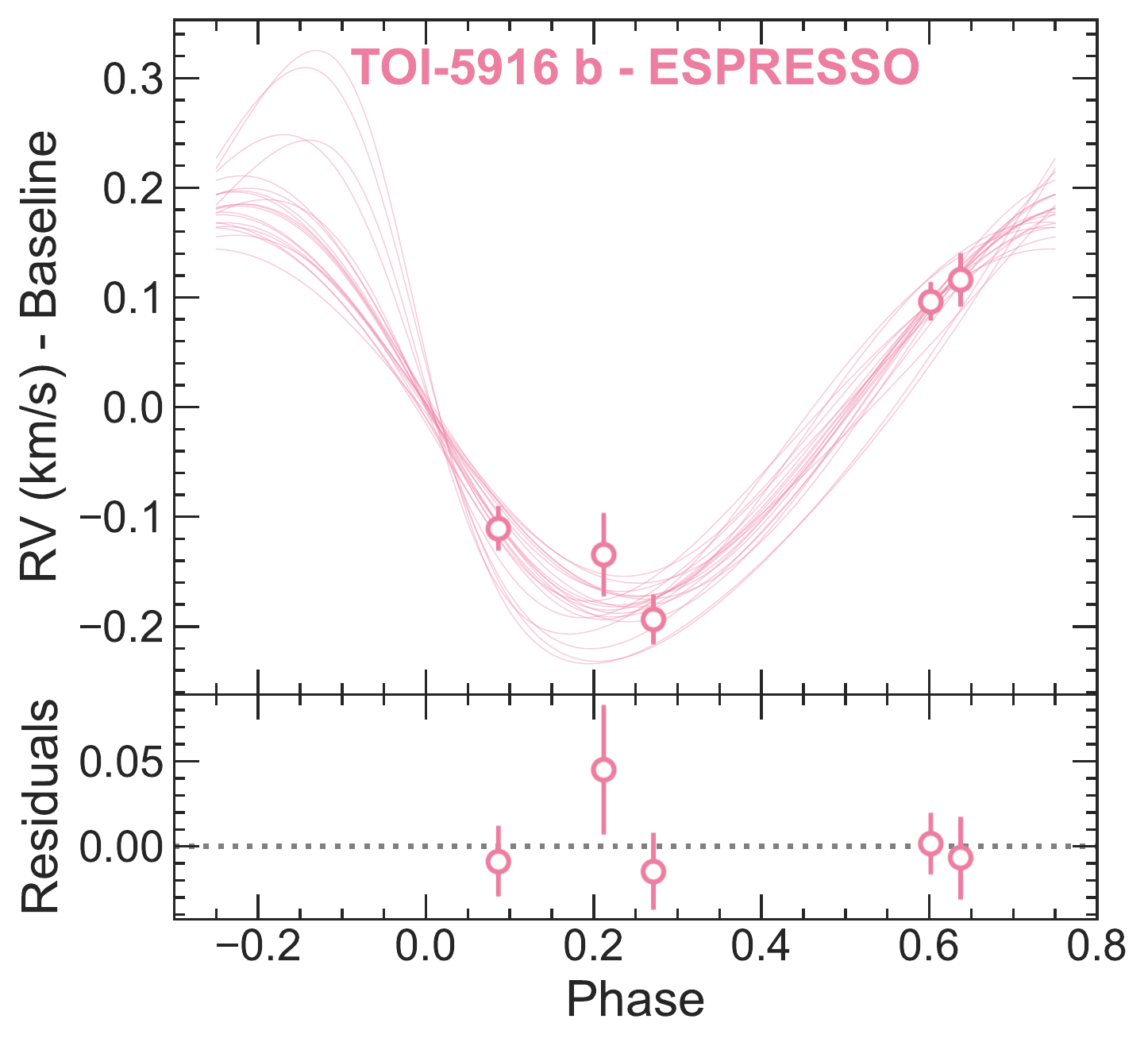}
    \caption{Same as Fig.~\ref{fig:3288_phot} but for TOI-5916.}
    \label{fig:5916_phot}
\end{figure}

TOI-5916~b is a $1.013_{-0.033}^{+0.032}~\mathrm{R_{jup}}$ planet with a mass of $0.713_{-0.114}^{+0.241}~\mathrm{M_{jup}}$. In Fig \ref{fig:5916_phot} we present the transits used in our fits along with best fitting models, as well the radial velocity points from ESPRESSO.

Our fit finds an eccentricity of $e_{\rm}=0.090_{-0.062}^{+0.114}$, which is the second highest we measure, but is still consistent with zero at the $2\sigma$ level. The eccentricity cannot be well constrained by our fit as the radial velocity sampling is not even enough in the orbit. However, with a calculated tidal circularisation timescale of $<3.9~\rm Gyr$, we would not expect this planet to be eccentric. 

\section{Discussion}
\label{sec:discuss}

\begin{figure}
    \centering
    \includegraphics[width=\columnwidth]{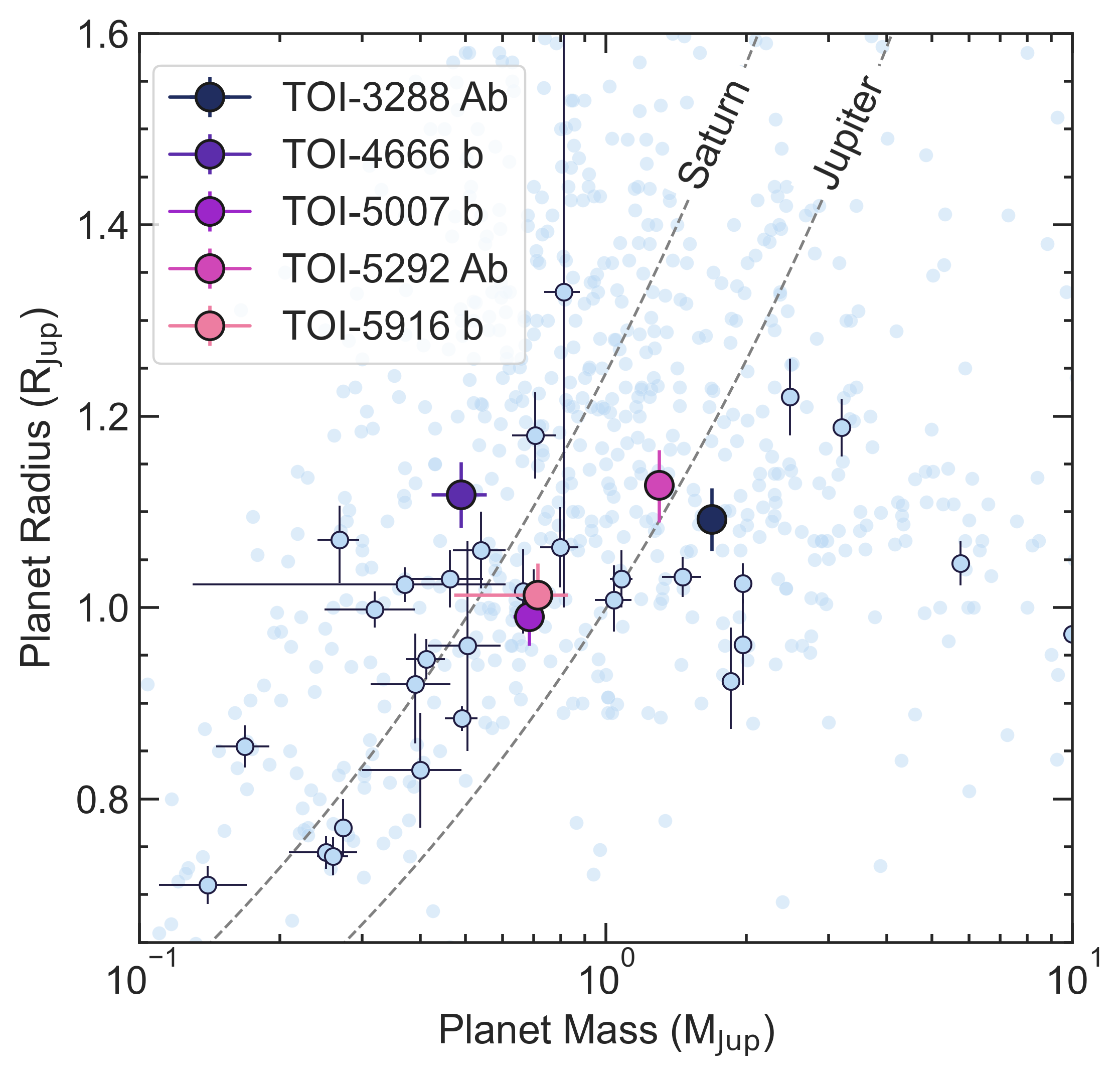}
    \caption{Mass-radius of transiting giant exoplanets with measured masses, as queried from the NASA Exoplanet Archive. In the background in light blue are giant planets orbiting FGK stars. Light blue planets with black edges and error bars are MANGOS-type planets: giants orbiting M dwarfs with periods $<7$ days. The five planets presented in this work are shown in their respective colours. The two dashed lines show the density contours at Saturn and Jupiter's densities.}
    \label{fig:mr}
\end{figure}

In this section we contextualise the five new MANGOS systems presented in this work. 
In Fig.~\ref{fig:mr} we present the sample of giant transiting planets as queried from the NASA Exoplanet Archive\footnote{Accessed on UTC 2025 Sep 25} \citep{Christiansen_2025}. In the background we show giants orbiting FGK stars, and in the foreground we present the planets discovered so far that meet the MANGOS program criteria ($P<7$ days, $R_{\rm p}>7~\rm R_{\oplus}$, \teff$<4000$~K). The five planets presented in this work are shown in their respective colors as well as contours indicating the densities of Jupiter and Saturn. 

TOI-4666~b is the only planet in our sample with a density lower than Saturn, sharing the parameter space with 11 other MANGOS-type planets, the most similar being TOI-519~b \citep{519a,519b} and TOI-5293~Ab \citep{5293}. Conversely, two of the planets in our sample (TOI-3288~Ab and TOI-5292~Ab) are slightly denser than Jupiter, similar to seven other giants orbiting low-mass stars. The remaining two have densities intermediate between the solar system gas giants. The planets are very similar to each other, and also to the recently discovered TOI-7149~b \citep{2025_TOI-7149b}. However we note that the mass of TOI-5916~b is the least constrained in our sample, as described above.

\subsection{Could these MANGOS have formed by core accretion?}

One of the motivations for searching for giant planets orbiting M dwarfs is the challenge they pose to our understanding of planet formation \citep[e.g.][]{Bryant2023,Kanodia2024}. In the current paradigm, giant planets on short orbits form via core accretion, where small solids (pebbles or planetesimals) will gradually come together to form a planetary embryo \citep{1996_Pollack_core_accretion,2020_Liu_Ji_core_accretion}. Once these embryos become massive enough, runaway gas accretion can kick in, allowing large envelopes of H/He gas to be acquired \citep{Ida2005}. 

The challenge to this model when it comes to MANGOS is that on average the mass of the protoplanetary disc scales with the mass of the host star \citep[e.g.][]{Andrews2013,Ansdell2017}, which means that formation of close-in giant planets becomes increasingly difficult with decreasing stellar mass due to lower availability of solids. A further challenge is that the runaway gas accretion phase has to complete before the gas in the disc dissipates, but the timescale for these cores to form is significantly longer around M stars than around FGK stars \citep{Laughlin2004}.

Two host stars in our sample have masses $M_\star > 0.6 \rm\,
M_\odot$: TOI-3288~A and TOI-5007. The closest host mass considered by \cite{Burn2021} in their population synthesis is $0.7\,\rm M_\odot$, and at this mass the formation of giant planets can easily be reconciled with core accretion. In fact, their simulations indicate that 9\% of stars in this mass range would host giant planets. TOI-4666 and TOI-5292~A have masses $<0.6~\rm\,M_\odot$ making them more comparable to the $0.5~\rm M_\odot$ hosts in the simulations. Here \cite{Burn2021} find that only 2\% of systems host giant planets due to the decreased efficiency of core accretion. 

TOI-5916 has a mass of $0.47\pm0.01~\rm M_\odot$ - the lowest in our sample. For masses $<0.5~\rm M_\odot$ no giant planets were formed in the population synthesis of \cite{Burn2021}. However, TOI-5916~b is the 11$^{\rm th}$ giant planet discovered orbiting a star in this mass range. Assuming a maximum disc-to-star mass ratio of 0.6\% \citep{Andrews2013} and a dust-to-gas mass ratio of 1\%, we find that the maximum amount of dust available in TOI-5916's protoplanetary disc is $\sim9.6~\rm M_\oplus$. Given that a core of $\sim 10~\rm M_\oplus$ is needed to trigger runaway gas accretion \citep{Mordasini2008}, this implies that 100\% of available solids in the disc would be used in the formation of this core.

Therefore, in order for TOI-5916~b to have formed via core accretion, we would require a higher star-to-disc mass ratio, a very dust rich disc and highly efficient core formation.

\subsection{The impact of host metallicity on planet bulk density}

\begin{figure}
    \centering
    \includegraphics[width=\columnwidth]{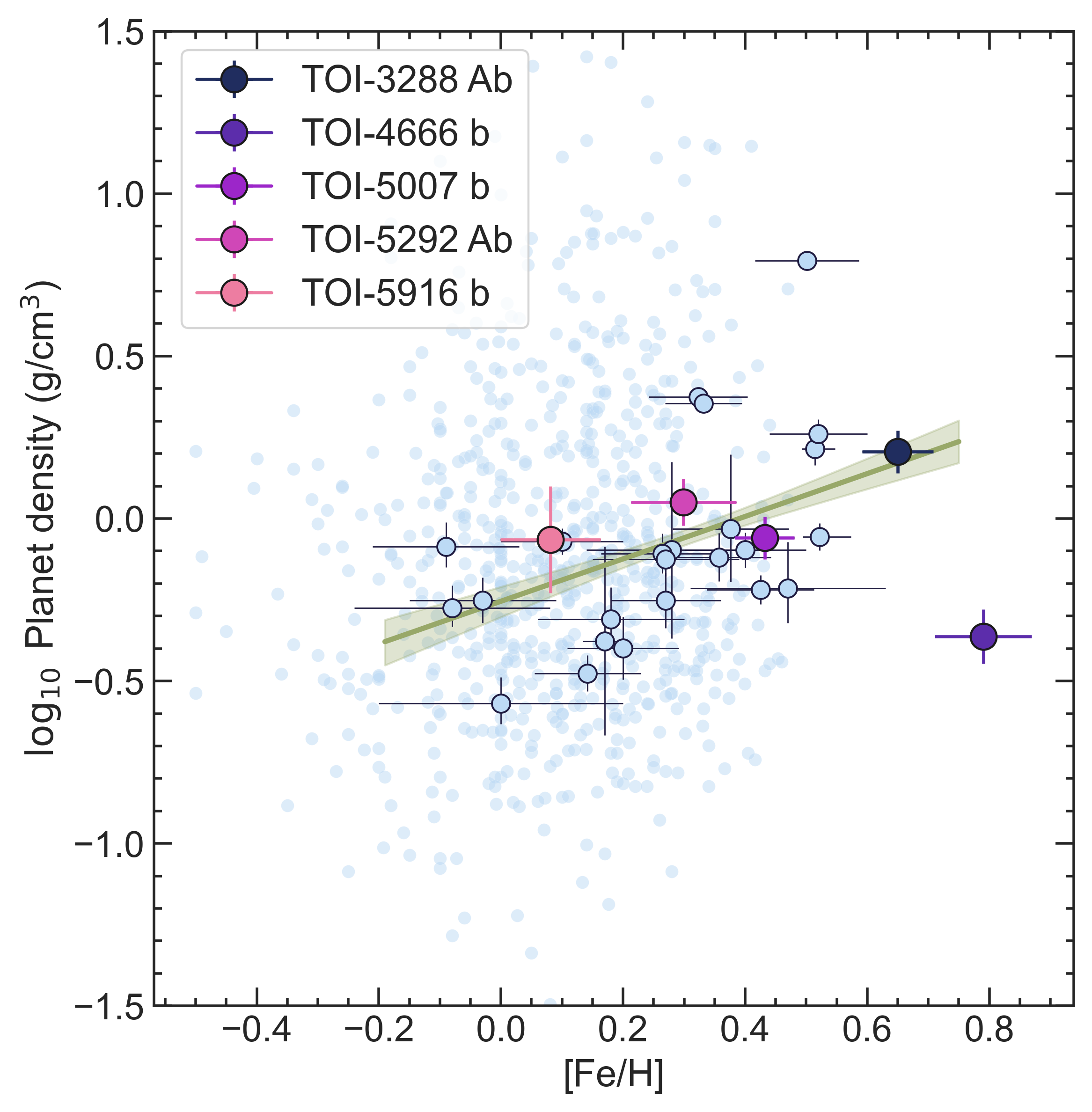}
    \caption{Planet density vs. host star metallicity. All planets shown in this figure are transiting planets with measured masses as queried from the NASA Exoplanet Archive. In the background in light blue are giant planets orbiting FGK stars. Light blue planets with black edges and error bars are MANGOS-type planets: giants orbiting M dwarfs with periods $<7$ days. The five planets presented in this work are shown in their respective colours. We show the best fit regression line from our analysis in green, as well as the 68\% confidence interval.}
    \label{fig:dens_met}
\end{figure}

In Fig. \ref{fig:dens_met} we show $\log_{10}$ planet density vs. host star metallicity for all planets in the NASA Exoplanet Archive that meet MANGOS programme criteria. In the background we plot all other giant transiting planets, and we overplot the five new MANGOS presented in this work. 

For this discussion on stellar metallicity, we note that the difficulty of accurately measuring host metallicities of M dwarfs is well documented in the literature \citep[e.g.][]{Lindgren2017}. The challenges lie in the vast amount of molecular lines present in M-dwarf spectra, as well as the wide variety of approaches used to extract stellar metallicity. \cite{Passegger2022} found that different methods can lead to discrepancies of up to 0.3~dex. The values for our host stars used in this analysis are those presented in Table \ref{tab:spectroscopic_params} which were extracted from near-infrared spectra (see Section \ref{sec:observations}). The values for other MANGOS-type planets are those presented in the NASA Exoplanet Archive Composite Data table, and they were extracted using various methods. We therefore emphasise that caution is needed when looking at relationships involving the metallicity of low-mass host stars. 

With the exception of TOI-4666~b\footnote{We exclude TOI-4666 from this analysis given the very high metallicity we derived, and the fact it may be young.}, we find that the planets in this work are consistent with the existing population of MANGOS-type planets on Fig. \ref{fig:dens_met}. The planet-metallicity relation is well studied \citep[e.g.][]{2010_Johnson,Thorngren2016}, and recent works have investigated whether this extends to low-mass stars \citep{2025_Gan_metallicity}. 

We investigate a potential correlation between host metallicity and planet bulk density for MANGOS planets. First we calculate the Pearson correlation coefficient to check for a statistically significant correlation, finding values of $r = 0.558$, $p = 0.00202$, where a $p$ value <0.05 is considered statistically significant. 

We then use a Monte Carlo linear regression to account for the measurement uncertainties in both host metallicity and planet bulk density. As the density uncertainties are asymmetric, we select the larger of the two values in each case. For each system, we draw 10,000 random values of $\log_{10}$~density and metallicity from Gaussian distributions centered on their measured values, fitting a straight line to each simulated dataset. In this way we build up posterior distributions for the slope and intercept allowing us to extract best-fitting values. We measure a slope of $0.654\pm 0.139$ and an intercept of $-0.253\pm 0.048$, which gives the following relationship between host metallicity and planet bulk density:
\begin{equation}
    \rho = 0.558 \times 10^{0.654 \times \feh} .
\end{equation}
We overplot the median and 68\% confidence intervals for this relationship on Fig. \ref{fig:dens_met}. 

\cite{Thorngren2016} searched for a link between planetary heavy element mass and host star metallicity; despite previous studies finding such a link with smaller sample sizes \citep[e.g.][]{Miller2011}, this result was not replicated. More recently \cite{Teske2019} carried out a similar study with homogenously determined metallicities and also found no link. 

Planet bulk density is a good proxy for planet heavy element mass, as can be seen from the data presented by \cite{Thorngren2016}. Given our result, it is possible that such a correlation between heavy element mass and stellar metallicity exists for giant planets in short orbits around low-mass stars. Given that MANGOS planets are cooler on average, we might expect them to track the initial metallicity of the protoplanetary disc more securely. Such an investigation is beyond the scope of this paper, but it presents an exciting avenue of investigation for future studies.


\subsection{The possible eccentricity of TOI-5007~b}

Our global analysis of TOI-5007~b revealed a small but non-zero eccentricity. From our calculation of tidal circularisation timescales and our measured kinematic age of the star, we were unable to confirm whether this planet is expected to have circularised yet. However, two other MANGOS-type planets with $3\sigma$ detected eccentricities exist. 

TOI-2379~b \citep{2024_TOI-2379} is a super-Jupiter with a robustly detected eccentricity of $0.3420\pm0.0039$, and the authors calculate a maximum tidal circularisation timescale of $\tau = 16.6~\rm Gyr$. This is significantly longer than the system age they determine from isochrone fitting ($13.8\pm 4.1~\rm Gyr$). TOI-6330~b is another super-Jupiter \citep{6330} with a measured eccentricity of $0.34\pm0.01$; once again the authors derive a value for $\tau$ that is longer than the system age of $7.6\pm 4.9~\rm Gyr$. Both of these planets have masses significantly larger than Jupiter ($5.76\pm0.2~\mathrm{M_{jup}}$ and $10.0\pm0.32~\mathrm{M_{jup}}$ respectively), while TOI-5007~b has a more modest mass of $0.684_{-0.047}^{+0.051}~\mathrm{M_{jup}}$. Their higher masses could be indicative of different formation pathways such as gravitational instability, given the need for significantly more solids in the protoplanetary disc \citep{2020_Mercer_Stamatellos}. Therefore the eccentricity of TOI-5007~b might have a different origin. 

It would be beneficial to collect further radial velocity points for this system in order to confirm the eccentricity and place better constraints on additional planets in the system. Thus far, no MANGOS-type planets have been found to reside in multi-planet systems, so small eccentricities provide an exciting avenue for exploration. Given that warm Jupiters around FGK stars very often have planetary siblings \citep{Huang2016}, understanding whether this is true or not of MANGOS will help reveal similarities between these different flavours of giant planets.

\subsection{How might we further characterise these MANGOS?}

Given the exceptionally deep transits produced by MANGOS planets and the ease with which masses can be constrained, they appear at first glance to be an ideal population for atmospheric investigations via transmission spectroscopy. The transmission spectroscopy metric  \citep[TSM,][]{kempton2018} is a convenient way to quantify the suitability of an exoplanet for investigations of this kind. We calculate the TSM for all the planets in our sample, and find modest values of between 21--53 for all planets except TOI-4666~b, which has a TSM of 116 due to its low density. These values are largely consistent with the rest of the MANGOS-type planets, for which we calculate TSMs of between 3-350. For context, \cite{kempton2018} suggest a cut-off value of 96 to pursue observations of this kind for planets in the radius regime of MANGOS.

Nevertheless, temperate giant planets are expected to have interesting molecular features such as C, N and O bearing species \citep{Fortney2020} which would allow determination of C/O and N/O ratios if detected. Both of these quantities contain critical information about the formation environment and location in the disc of a planet \citep[e.g.][]{Oberg2011, Madhusudhan2017,Turrini2021, Molliere2022,Walsh2025}. Given the enduring mysteries surrounding the origin of giant planets in tight orbits around M dwarfs, atmospheric investigations are a crucial piece of the puzzle, and well worth pursuing. 

The difficulty comes in dealing with the stellar contamination likely to be present in the observed spectra \citep[e.g.,][]{Rackham2018, Rackham2019}. Indeed, first results from the GEMS (Giant Exoplanets around M dwarf Stars) programme on \textit{JWST} \citep{canas2025} revealed significant stellar contamination, leading to degeneracies in the inferred abundances of $\rm H_2O$. Despite these challenges, they made robust detections of CH$_4$ and H$_2$S. 

Given that we detect persistent spot crossings on at least two planets presented in this work, including TOI-4666~b which has the largest TSM, more work needs to be done to develop mitigation techniques for stellar contamination \citep[e.g.,][]{Iyer2020, Rackham2023, Rackham2024}.

Spots are not always a hindrance, though. Another crucial piece of the formation history puzzle can come from measuring the sky-projected angle between a planet's orbit and the stellar spin, known as the stellar obliquity.  While stellar obliquity measurements are often obtained by observing a radial velocity anomaly during transit \citep{Triaud2018, Albrecht2022}, they can also be constrained by mapping out stellar rotation in systems with persistent star spots \citep[e.g.][]{Sanchis2013,Valio2022,Biagiotti2024, Sagynbayeva2025}. Several MANGOS targets are amenable to these kinds of investigations, providing another path to revealing the formation history of these planets.

Whether a system is aligned or misaligned can provide insights into its migration history -- which, together with constraints on the orbital eccentricity, may offer an avenue to distinguish between the hot Jupiter formation scenarios of in situ formation, disk migration, or high-eccentricity migration \citep{Dawson2018, Rice2022}. Such M-dwarf-hosted planets may be particularly useful for distinguishing between models for the tidal interactions that sculpt the stellar obliquity distribution for hot Jupiters, as M dwarfs lack radiative zones that are required for certain proposed tidal damping mechanisms to operate \citep{Zanazzi2024}.

\section{Conclusions}
\label{sec:we_conclude}

In this work we have presented our project called MANGOS: \textbf{M} dwarfs \textbf{A}ccompanied by close-i\textbf{N} \textbf{G}iant \textbf{O}rbiters with \textbf{S}PECULOOS, a programme designed to discover giant planets, BDs and M dwarfs orbiting low-mass stars, with the goal of better understanding the various formation environments around M dwarfs and calibrating the mass-radius relation at the bottom of the main sequence. 

We have presented the discovery and confirmation of five new MANGOS planets: TOI-3288~Ab ($R_{\rm b}=1.092\pm0.033~\mathrm{R_{jup}}$, $M_{\rm b}=1.687_{-0.109}^{+0.114}~\mathrm{M_{jup}}$), TOI-4666~b ($R_{\rm b}=1.118\pm0.035~\mathrm{R_{jup}}$, $M_{\rm b}=0.489_{-0.066}^{+0.067}~\mathrm{M_{jup}}$), TOI-5007~b ($R_{\rm b}=0.991\pm0.030~\mathrm{R_{jup}}$), $M_{\rm b}=0.684{-0.047}^{+0.051}~\mathrm{M_{jup}}$, TOI-5292~Ab ($R_{\rm b}=1.128_{-0.037}^{+0.039}~\mathrm{R_{jup}}$, $M_{\rm b}=1.299_{-0.074}^{+0.081}~\mathrm{M_{jup}}$), and TOI-5916~b ($R_{\rm b}=1.013_{-0.033}^{+0.032}~\mathrm{R_{jup}}$, $M_{\rm b}=0.713_{-0.114}^{+0.241}~\mathrm{M_{jup}}$). TOI-3288~Ab and TOI-4666~b show persistent spot crossings on their transits, making them amenable to characterisation of their stellar obliquities. Our modelling of TOI-5007~b revealed a $3\sigma$ detection of an eccentric orbit ($e_\mathrm{b}=0.100_{-0.038}^{+0.037}$), but further radial velocity points are needed to confirm this detection and investigate if it is caused by an additional body in the system. 

We find that all but one of the systems (TOI-5916) are consistent with formation via core accretion. TOI-5916~b could have formed via core accretion by invoking higher disc-to-star and gas-to-dust mass ratios, and very high core formation efficiencies.

We detect bound stellar companions for two stars in our sample (TOI-3288~A and TOI-5292~A), and in the case of TOI-5292~A we are able to characterise both components of the binary. Our spectroscopic characterisation revealed high metallicities for several of our targets, which we interpret with caution. These measurements emphasise the challenge of acquiring robust metallicities for M dwarfs. 

We also show that for MANGOS planets, there is a strong correlation between host star metallicity and planet bulk density. Again, this should be interpreted with caution, but it could be indicative of a relation between stellar metallicity and heavy element mass in planets. 

Our work brings the total number of MANGOS-type planets to 35. Continued detection of new systems, and their follow-up characterisation, will enable more robust measurements of occurrence rates and improve our understanding of how giant planets form in diverse stellar environments.

\section*{Acknowledgments}

GD acknowledges funding from Magdalen College, Oxford.
Funding for KB was provided by the European Union (ERC AdG SUBSTELLAR, GA 101054354).
This material is based upon work supported by the National Aeronautics and Space Administration under Agreement No.\ 80NSSC21K0593 for the program ``Alien Earths''.
The results reported herein benefited from collaborations and/or information exchange within NASA’s Nexus for Exoplanet System Science (NExSS) research coordination network sponsored by NASA’s Science Mission Directorate.
We acknowledge financial support from the Agencia Estatal de Investigaci\'on of the Ministerio de Ciencia e Innovaci\'on MCIN/AEI/10.13039/501100011033 and the ERDF “A way of making Europe” through project PID2021-125627OB-C32, and from the Centre of Excellence “Severo Ochoa” award to the Instituto de Astrofisica de Canarias.
MG is F.R.S-FNRS Research Director.
YGMC is partially supported by UNAM PAPIIT-IG101224
JDH and GAB acknowledge funding from NASA grant No. 80NSSC22K0315
EJ is F.R.S-FNRS Research Director.
AS acknowledges support from the European Research Council Consolidator Grant funding scheme (project ASTEROCHRONOMETRY, G.A. n. 772293, http://www.asterochronometry.eu).
Visiting Astronomer at the Infrared Telescope Facility, which is operated by the University of Hawaii under contract 80HQTR24DA010 with the National Aeronautics and Space Administration.
Based in part on observations obtained at the Southern Astrophysical Research (SOAR) telescope, which is a joint project of the Minist\'{e}rio da Ci\^{e}ncia, Tecnologia e Inova\c{c}\~{o}es (MCTI/LNA) do Brasil, the US National Science Foundation’s NOIRLab, the University of North Carolina at Chapel Hill (UNC), and Michigan State University (MSU).
The paper is based on observations made with the Kast spectrograph on the Shane 3m telescope at Lick Observatory. A major
upgrade of the Kast spectrograph was made possible through generous gifts from the Heising-Simons Foundation and William and
Marina Kast. We acknowledge that Lick Observatory sits on the unceded ancestral homelands of the Chochenyo and Tamyen Ohlone
peoples, including the Alson and Socostac tribes, who were the
original inhabitants of the area that includes Mt. Hamilton.
J.d.W. and MIT gratefully acknowledge financial support from the Heising-Simons Foundation, Dr. and Mrs. Colin Masson and Dr. Peter A. Gilman for Artemis, the first telescope of the SPECULOOS network situated in Tenerife, Spain.
This work is partly supported by JSPS KAKENHI Grant Number JP24H00017,
JP24K00689, and JSPS Bilateral Program Number JPJSBP120249910.
This article is based on observations made with the MuSCAT2
instrument, developed by ABC, at Telescopio Carlos Sánchez operated on
the island of Tenerife by the IAC in the Spanish Observatorio del
Teide.
The ULiege's contribution to SPECULOOS has received funding from the European Research Council under the European Union's Seventh Framework Programme (FP/2007-2013) (grant Agreement n$^\circ$ 336480/SPECULOOS), from the Balzan Prize and Francqui Foundations, from the Belgian Scientific Research Foundation (F.R.S.-FNRS; grant n$^\circ$ T.0109.20), from the University of Liege, and from the ARC grant for Concerted Research Actions financed by the Wallonia-Brussels Federation. 
The Cambridge contribution is supported by a grant from the Simons Foundation (PI Queloz, grant number 327127).
J.d.W. and MIT gratefully acknowledge financial support from the Heising-Simons Foundation, Dr. and Mrs. Colin Masson and Dr. Peter A. Gilman for Artemis, the first telescope of the SPECULOOS network situated in Tenerife, Spain. 
The Bern contribution is supported by the Swiss National Science Foundation (PP00P2-163967, PP00P2-190080 and the National Centre for Competence in Research PlanetS). The Birmingham contribution to SPECULOOS has received fund from the European Research Council (ERC)
 under the European Union's Horizon 2020 research and innovation programme (grant agreement n$^\circ$ 803193/BEBOP), from the MERAC foundation, and from the Science and Technology Facilities Council (STFC; grant n$^\circ$ ST/S00193X/1) and from the ERC/UKRI Frontier Research Guarantee programme (EP/Z000327/1/CandY).
TRAPPIST is funded by the Belgian Fund for Scientific Research (Fond National de la Recherche Scientifique, FNRS) under the grant PDR T.0120.21, with the participation of the Swiss National Science Fundation (SNF). TRAPPIST-North is funded by the University of Liège in collaboration with the Cadi Ayyad University of Marrakech.
We acknowledge funding from the European Research Council under the ERC Grant Agreement n. 337591-ExTrA.

We acknowledge the use of public TESS data from pipelines at the TESS Science Office and at the TESS Science Processing Operations Center.
Resources supporting this work were provided by the NASA High-End Computing (HEC) Program through the NASA Advanced Supercomputing (NAS) Division at Ames Research Center for the production of the SPOC data products.

\section*{List of Affiliations}
$^{1}$ \oxfordastro \\
$^{2}$ \oxfordmag \\
$^{3}$ \birmingham \\
$^{4}$ \liege \\
$^{5}$ \tautenburg \\
$^{6}$ \miteaps \\
$^{7}$ \mitkavli \\
$^{8}$ \sandiego \\
$^{9}$ \iac \\
$^{10}$ \grenoble \\
$^{11}$ \geneva \\
$^{12}$ \leicester \\
$^{13}$ \cfa \\
$^{14}$ \komaba \\
$^{15}$ \astrojp \\
$^{16}$ \cambridge \\
$^{17}$ \carnegie \\
$^{18}$ \ull \\
$^{19}$ \princeton \\
$^{20}$ \zoe \\
$^{21}$ \sharjah \\
$^{22}$ \cadi \\
$^{23}$ \seti \\
$^{24}$ \carnegieobs \\
$^{25}$ \bern \\
$^{26}$ \mariecurie \\
$^{27}$ \aim \\
$^{28}$ \mex \\
$^{29}$ \liegestar \\
$^{30}$ \colorado \\
$^{31}$ \ethch \\
$^{32}$ \iaa \\
$^{33}$ \yale \\
$^{34}$ \ohio \\

\section*{Data Availability}

\textit{TESS} data products are available via the MAST portal at \url{https://mast.stsci.edu/portal/Mashup/Clients/Mast/Portal.html}. Follow-up photometry and high resolution imaging data for TOIs are available on ExoFOP at \url{https://exofop.ipac.caltech.edu/tess/}. These data are freely accessible to ExoFOP members immediately and are publicly available following a one-year proprietary period. Radial velocities are provided in tables in the appendices of this manuscript.



\bibliographystyle{mnras}
\bibliography{example} 




\appendix
\section{TESS target pixel files}
\begin{figure}
\includegraphics[width=0.9\columnwidth]{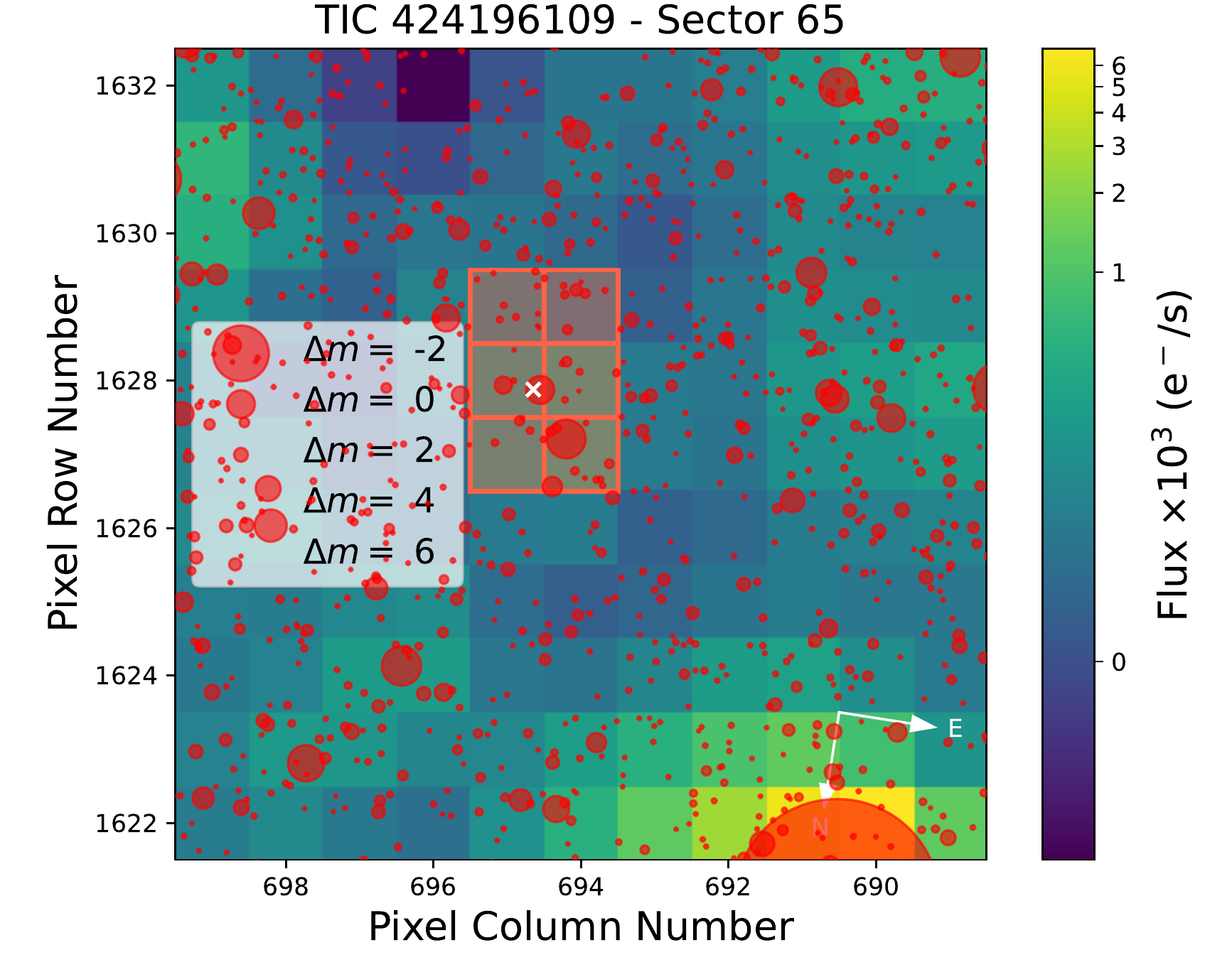}
\includegraphics[width=0.9\columnwidth]{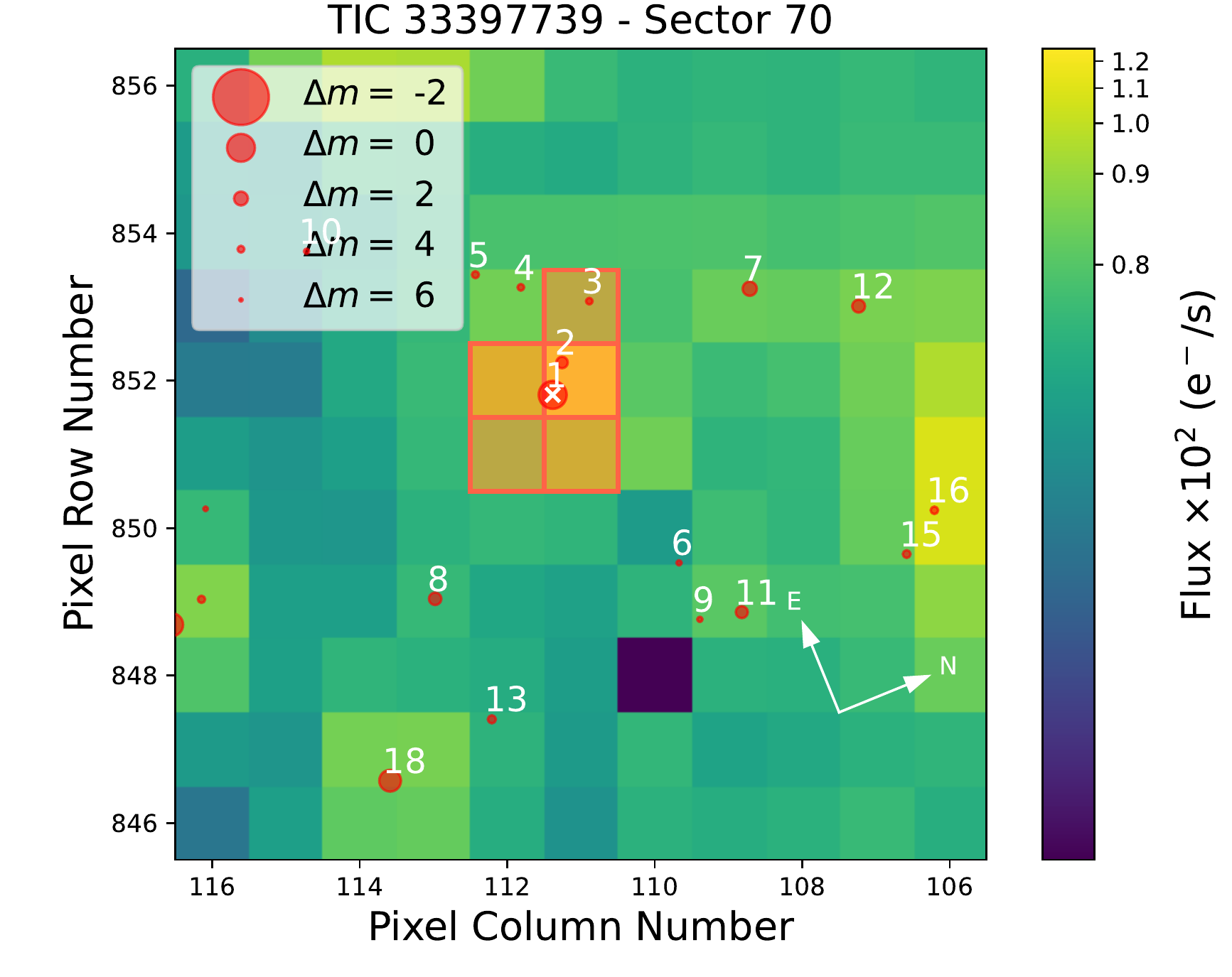}
\includegraphics[width=0.9\columnwidth]{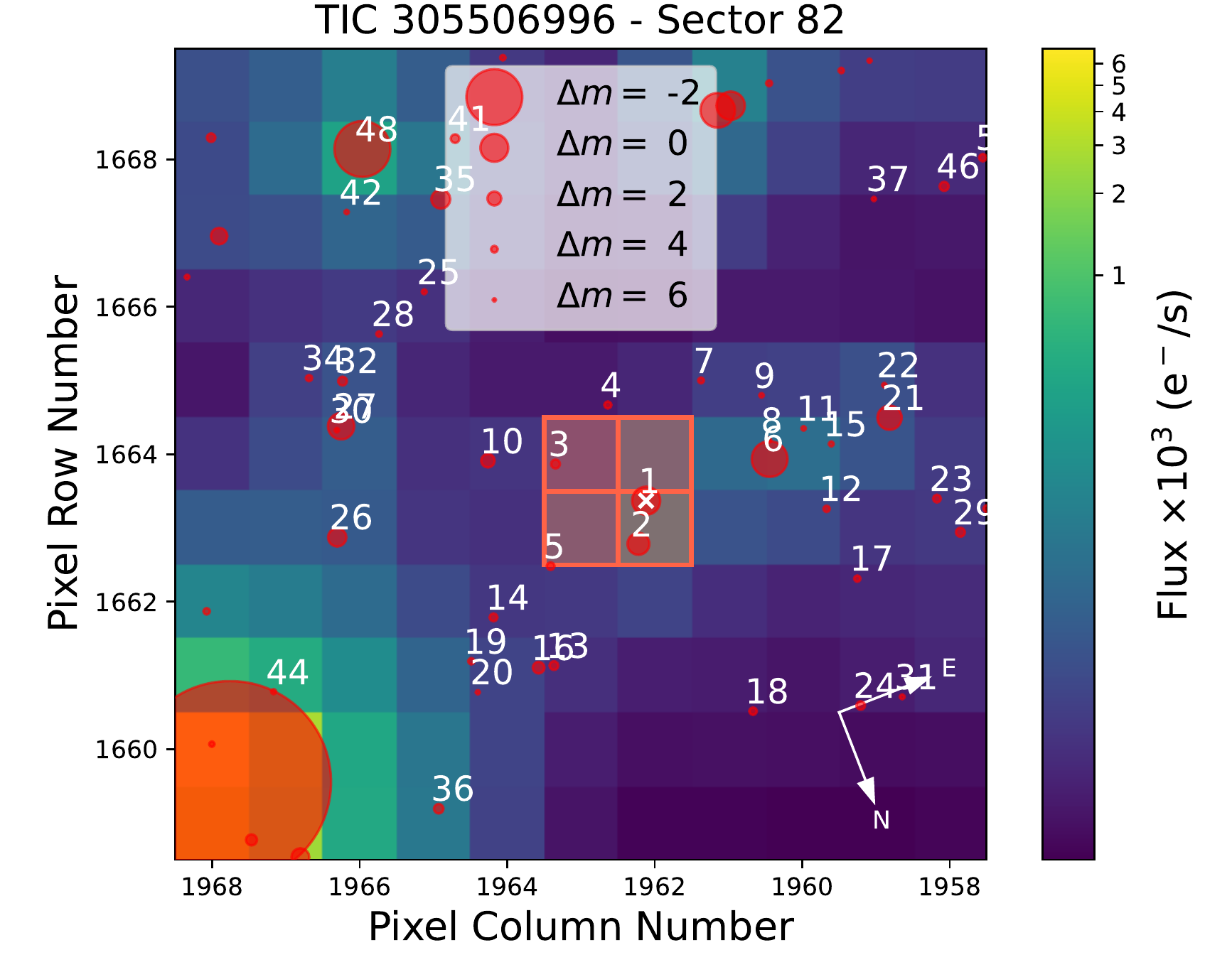}
\caption{TESS target pixel files for TOI-5007 (left), TOI-5292~A (middle) and TOI-5916 (right). Red grids show default \textit{TESS} apertures corresponding to the \textit{TESS} photometry used in this work. For TOI-5292~A, only one sector's TPF is shown. Red circles depict other GAIA sources within the field, with sizes corresponded to their magnitude difference compared to the target star. For TOI-5007, we remove the labeling of these sources for visual clarity, as a result of the field being very crowded.}
\label{fig:tpfs}
\end{figure}
In Fig.~\ref{fig:tpfs} we present a target pixel file for each of the stars in our sample where we used \textit{TESS} photometry (TOI 5007, TOI 5292~A and TOI-5916). Each has the SPOC pipeline aperture overplotted for reference.

\section{Summary of ground-based observations}
In Tables \ref{tab:photometry_summary} and \ref{tab:photometry_summary2} we present a journal of our ground-based photometric observations collected for this work. 
\begin{table*} 
\centering
\caption{Summary of ground-based follow-up observations obtained for TOI-3288~Ab, TOI-4666 b, and TOI-5007 b.}
\begin{tabular}{@{}cccccccccc@{}}
\midrule \midrule
 \textbf{Observatory} & \textbf{Filter} & \textbf{Date}     & \textbf{Coverage} & \textbf{Exposure (s)} & \textbf{FWHM (\arcsec) } & \textbf{Aperture (\arcsec)} & \textbf{Measurements} \\ \midrule
\multicolumn{8}{c}{\textbf{TOI-3288~Ab} } \\ \midrule
 SSO/Callisto & {\it zYJ} & 2022 Jun 30 & Full & 18 & 1.72 & 2.80 & 1290 \\ 
 SSO/Io & {\it Sloan-$g'$} & 2022 Jul 10 & Full & 120 & 3.71 & 2.38 & 181 \\
 SSO/Europa & {\it Sloan-$r'$} & 2022 Jul 20 & Full & 74 & 1.69 & 1.44 & 218 \\
 SSO/Callisto & {\it zYJ} & 2022 Aug 05 & Egress & 44 & 1.84 & 4.00 & 339 \\
 SSO/Europa & {\it Sloan-$g'$} & 2022 Aug 05 & Egress & 120 & 1.47 & 2.50 & 97 \\
 SSO/Io & {\it I+z} & 2022 Aug 05 & Egress & 17 & 1.78 & 2.64 & 431 \\
 SSO/Io & {\it I+z} & 2022 Aug 07 & Secondary & 17 & 1.47 & 2.23 & 1278 \\
 SSO/Callisto & {\it zYJ} & 2022 Aug 12 & Full & 40 & 2.21 & 2.80 & 441 \\ 
 SSO/Europa & {\it I+z} & 2022 Aug 12 & Egress & 17& 1.28 & 2.16  & 961 \\
 SSO/Ganymede & {\it Sloan-$r'$} & 2022 Aug 12 & Full & 74 & 2.47 & 3.33 & 212 \\
 SSO/Io & {\it Sloan-$g'$} & 2022 Aug 12 & Full & 120 & 2.23 & 2.16 & 120 \\
 SSO/Callisto & {\it zYJ} & 2022 Nov 12 & Full & 18 & 2.26 & 2.80 & 439\\ 
 SSO/Io & {\it Sloan-$g'$} & 2022 Nov 12 & Full & 112 & 2.15 & 0.61 & 68 \\
 SSO/Callisto & {\it Sloan-$z'$} & 2023 Apr 07 & Partial & 27 & 2.08 & 2.66 & 368 \\
 SSO/Europa & {\it Sloan-$r'$} & 2023 Apr 07 & Partial & 74 & 1.24 & 2.83 & 161 \\
 SSO/Ganymede & {\it Sloan-$i'$} & 2023 Apr 07 & Partial & 37 & 1.36 & 3.45 & 290 \\
 SSO/Io & {\it Sloan-$g'$} & 2023 Apr 07 & Partial & 120 & 1.84 & 2.13 & 104\\
 TRAPPIST-South & {\it I+z} & 2023 Sep 12 & Full & 60 & 2.87 & 4.14 & 200 \\
SSO/Callisto & {\it zYJ} & 2025 Sep 15 & Full & 18 & 1.56 & 2.80 & 931 \\ 
\midrule
\multicolumn{8}{c}{\textbf{TOI-4666 b} } \\ \midrule
 SSO/Europa & {\it I+z} & 2022 Oct 10 & Full & 16 & 1.45 & 2.32 & 822 \\
 SSO/Callisto & {\it zYJ} & 2022 Dec 16 & Full & 16 & 2.38 & 2.80 & 1412 \\
 SSO/Europa & {\it Sloan-$r'$} & 2022 Dec 16 & Egress & 70 & 1.53 & 2.83 & 217 \\
 SSO/Ganymede & {\it Sloan-$z'$} & 2022 Dec 16 & Egress & 25 & 1.35 & 3.17 & 488 \\
 SSO/Io & {\it Sloan-$g'$} & 2022 Dec 16 & Egress & 160 & 1.87 & 2.05 & 115 \\
 SSO/Callisto & {\it zYJ} & 2023 Jan 17 & Full & 16 & 2.31 & 2.80 & 1212 \\
 SSO/Europa & {\it Sloan-$g'$} & 2023 Jan 17 & Full & 160 & 1.36 & 2.52 & 111 \\
 SSO/Ganymede & {\it Sloan-$r'$} & 2023 Jan 17 & Full & 70 & 1.44 & 1.94 & 246 \\
 TRAPPIST-South & {\it Sloan-$z'$} & 2023 Jan 17 & Full & 130 & 2.52 & 3.40 & 118 \\
 SSO/Callisto & {\it I+z} & 2023 Sep 08 & Secondary & 16 & 2.55 & 4.71 & 822 \\
 SSO/Europa & {\it Sloan-$i'$} & 2023 Sep 08 & Secondary & 35 & 1.73 & 3.21 & 476 \\
 SSO/Europa & {\it Sloan-$i'$} & 2023 Sep 24 & Full & 35 & 1.27 & 2.04 & 493 \\
 SSO/Io & {\it I+z} & 2023 Sep 24 & Full & 16 & 1.93 & 3.09 & 927 \\
 SSO/Europa & {\it I+z} & 2023 Oct 13 & Secondary & 16 & 1.26 & 2.33 & 521 \\
 TRAPPIST-South & {\it Sloan-$z'$} & 2024 Dec 18 & Full & 120 & 2.32 & 6.03 & 144 \\
\midrule
\multicolumn{8}{c}{\textbf{TOI-5007 b} } \\ \midrule
 TRAPPIST-South & {\it I+z} & 2022 May 02 & Full & 60 & 1.93 & 3.52 & 209 \\
 TRAPPIST-South & {\it I+z} & 2022 Jul 02 & Full & 110 & 2.58 & 3.26 & 140 \\
 SSO/Io & {\it Sloan-$g'$} & 2022 Aug 27 & Full & 120 & 1.79 & 1.94 & 114 \\
 TRAPPIST-South & {\it Sloan-$z'$} & 2023 Feb 03 & Full & 110 & 2.61 & 3.3 & 89 \\
 SSO/Europa & {\it Sloan-$i'$} & 2023 May 26 & Full & 44 & 1.18 & 1.27 & 317 \\
 SSO/Europa & {\it Sloan-$r'$} & 2025 Mar 12 & Ingress & 88 & 1.01 & 1.27 & 78 \\
 SSO/Ganymede & {\it Sloan-$z'$} & 2025 Mar 12 & Full & 32 & 2.15 & 2.61 & 263 \\
 SSO/Callisto & {\it zYJ} & 2025 Sep 14 & Full & 40 & 2.25 & 2.80 & 176 \\
 LCO-SSO-0m4 & {\it Sloan-$r'$} & 2023 Apr 21 & Ingress & 600 & 5.19 & 3.40 & 12 \\
 ExTrA/T3 & {\it 1.41 $\mu$m} & 2022 May 22 & Full & 60 & 1.38 & 8.00 & 163 \\ 
ExTrA/T2 & {\it 1.41 $\mu$m} & 2022 July 02 & Full & 60 & 1.48 & 8.00 & 330 \\ 
ExTrA/T2 & {\it 1.41 $\mu$m} & 2022 Aug 27 & Full & 60 & 1.23 & 8.00 & 218 \\ 
ExTrA/T1+T2+T3 & {\it 1.41 $\mu$m} & 2023 Apr 28 & Full & 60 & 1.40 & 8.00 & 218 \\ 
ExTrA/T1+T2+T3 & {\it 1.41 $\mu$m} & 2023 May 03 & Full & 60 & 1.69 & 8.00 & 143 \\ 
ExTrA/T1+T2+T3 & {\it 1.41 $\mu$m} & 2023 May 21 & Partial & 60 & 1.51 & 8.00 & 119 \\ 
ExTrA/T1+T2+T3 & {\it 1.41 $\mu$m} & 2023 May 26 & Full & 60 & 1.33 & 8.00 & 207 \\ 
ExTrA/T1+T2+T3 & {\it 1.41 $\mu$m} & 2023 May 31 & Full & 60 & 1.41 & 8.00 & 218 \\ 
ExTrA/T1+T2+T3 & {\it 1.41 $\mu$m} & 2023 Jul 16 & Full & 60 & 1.67 & 8.00 & 237 \\
ExTrA/T1+T2+T3 & {\it 1.41 $\mu$m} & 2023 Aug 13 & Full & 60 & 1.22 & 8.00 & 191 \\ 
ExTrA/T1+T2 & {\it 1.41 $\mu$m} & 2023 Aug 26 & Full & 60 & 1.46 & 8.00 & 257 \\
\end{tabular}
\label{tab:photometry_summary}
\end{table*}

\begin{table*} 
\centering
\caption{Summary of ground-based follow-up observations obtained for TOI-5292~Ab and TOI-5916 b.}
\begin{tabular}{@{}ccccccccc@{}}
\midrule \midrule
 \textbf{Observatory} & \textbf{Filter} & \textbf{Date}     & \textbf{Coverage} & \textbf{Exposure (s)} & \textbf{FWHM (\arcsec) } & \textbf{Aperture (\arcsec)} & \textbf{Measurements} \\ \midrule

 \multicolumn{8}{c}{\textbf{TOI-5292~Ab} } \\ \midrule
 TRAPPIST-North & {\it I+z} & 2023 Dec 09 & Egress & 120 & 2.13 & 2.77 & 93 \\
 TRAPPIST-North & {\it Sloan-$z'$} & 2023 Dec 11 & Full & 150 & 2.43 & 2.63 & 69 \\
 SNO/Artemis & {\it Sloan-$g'$} & 2024 Sep 27 & Full & 160 & 1.37 & 1.94 & 87 \\
 SNO/Artemis & {\it Sloan-$z'$} & 2024 Oct 03 & Full & 60 & 1.54 & 1.83 & 193 \\
 SNO/Artemis & {\it Sloan-$z'$} & 2024 Dec 18 & Full secondary & 60 & 2.26 & 1.39 & 75  \\
 SNO/Artemis & {\it Sloan-$i'$} & 2024 Dec 19 & Full & 120& 1.51 & 2.33 & 77 \\ \midrule
\multicolumn{8}{c}{\textbf{TOI-5916 b} } \\ \midrule
 SNO/Artemis & {\it Sloan-$g'$} & 2022 Nov 28 & Full & 140 & 1.21 & 1.12 & 85 \\
 TRAPPIST-North & {\it I+z} & 2022 Nov 28 & Full & 120 & 2.27 & 3.58 & 114 \\
 TCS-MuSCAT2 & {\it $g'$} & 2024 Aug 16 & Full & 50 & - & 6.96 & 227 \\
 TCS-MuSCAT2 & {\it $r'$} & 2024 Aug 16 & Full & 50 & - & 6.96 & 231 \\
 TCS-MuSCAT2 & {\it $i'$} & 2024 Aug 16 & Full & 15 & - & 6.96 & 732 \\
 TCS-MuSCAT2 & {\it $zs'$} & 2024 Aug 16 & Full & 50 & - & 6.96 & 231 \\
\end{tabular}
\label{tab:photometry_summary2}
\end{table*}

\section{Archival Imaging}

\begin{figure*} 
\includegraphics[width=0.6\textwidth]{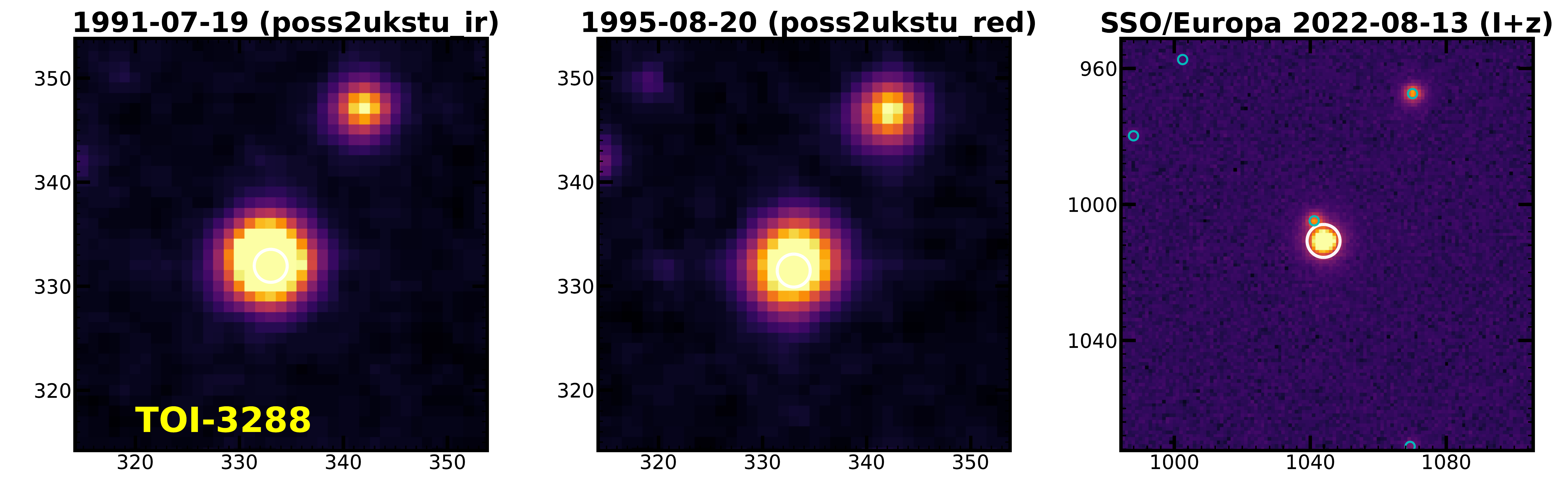} \vspace{0.5cm}
\includegraphics[width=0.6\textwidth]{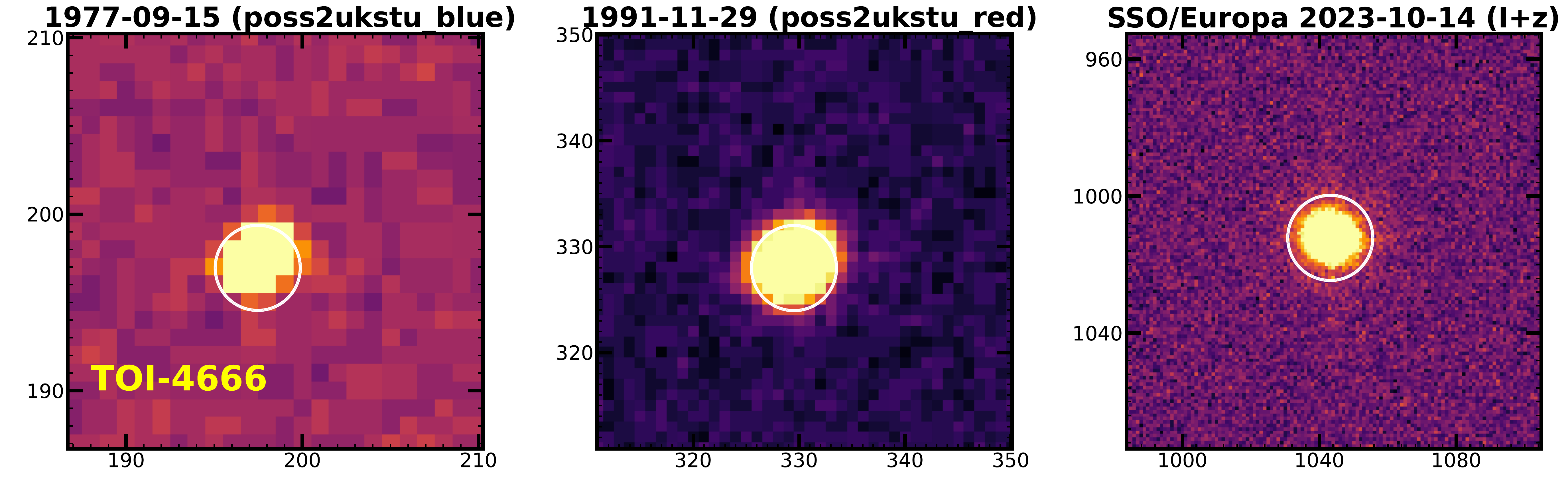} \vspace{0.5cm}
\includegraphics[width=0.6\textwidth]{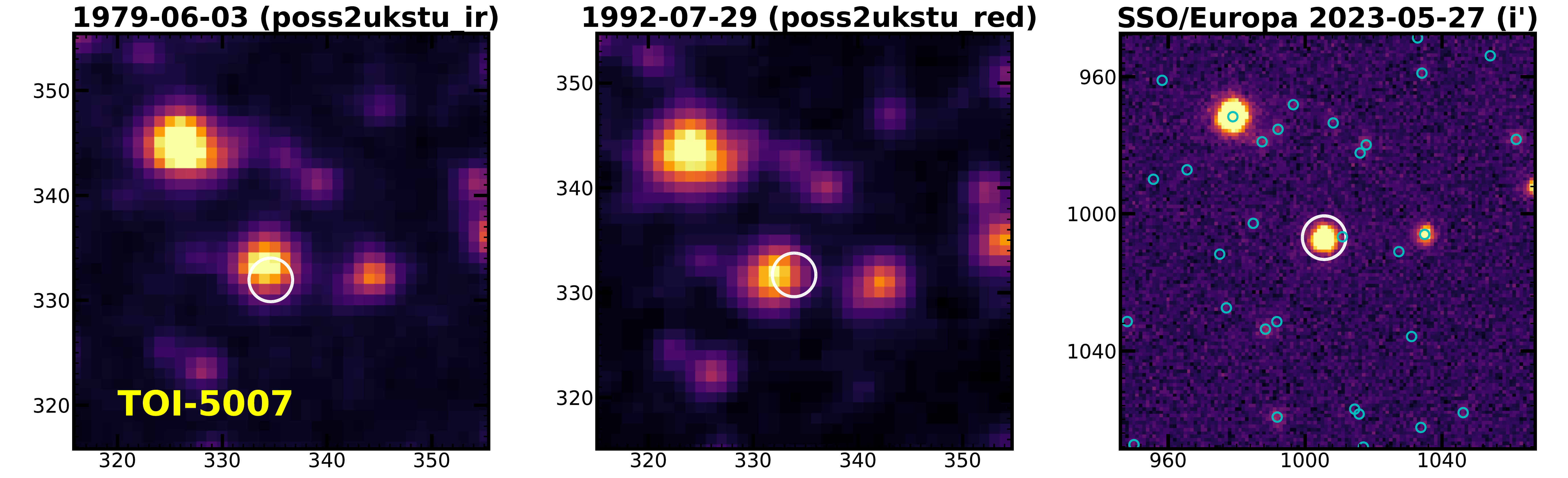} \vspace{0.5cm}
\includegraphics[width=0.6\textwidth]{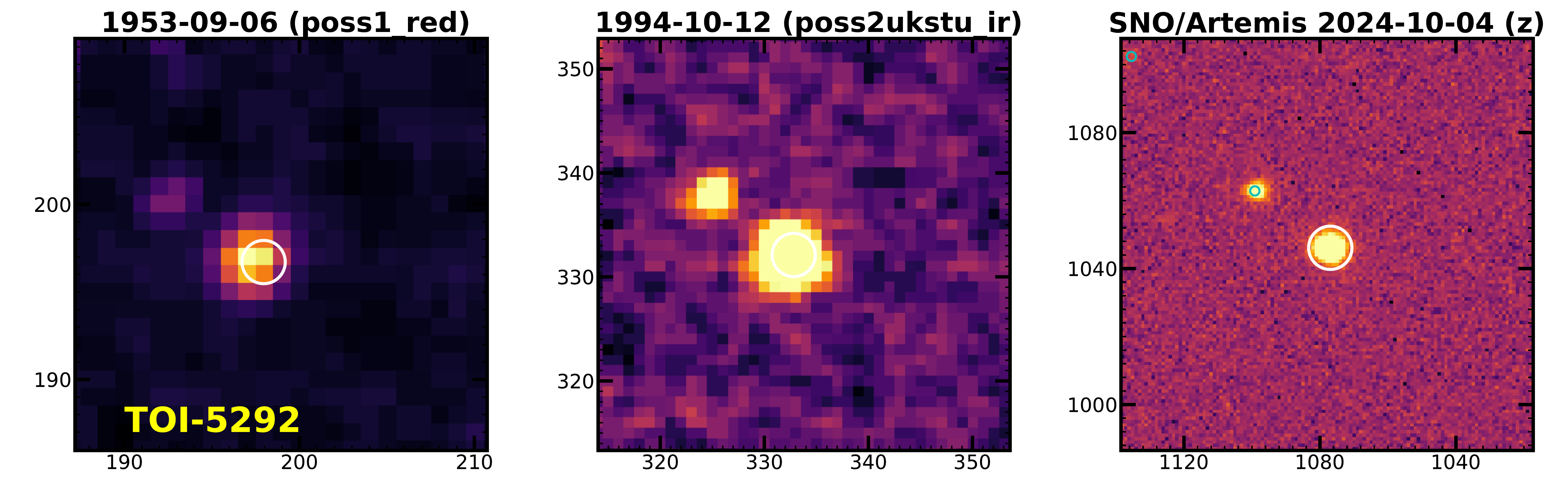} \vspace{0.5cm}
\includegraphics[width=0.6\textwidth]{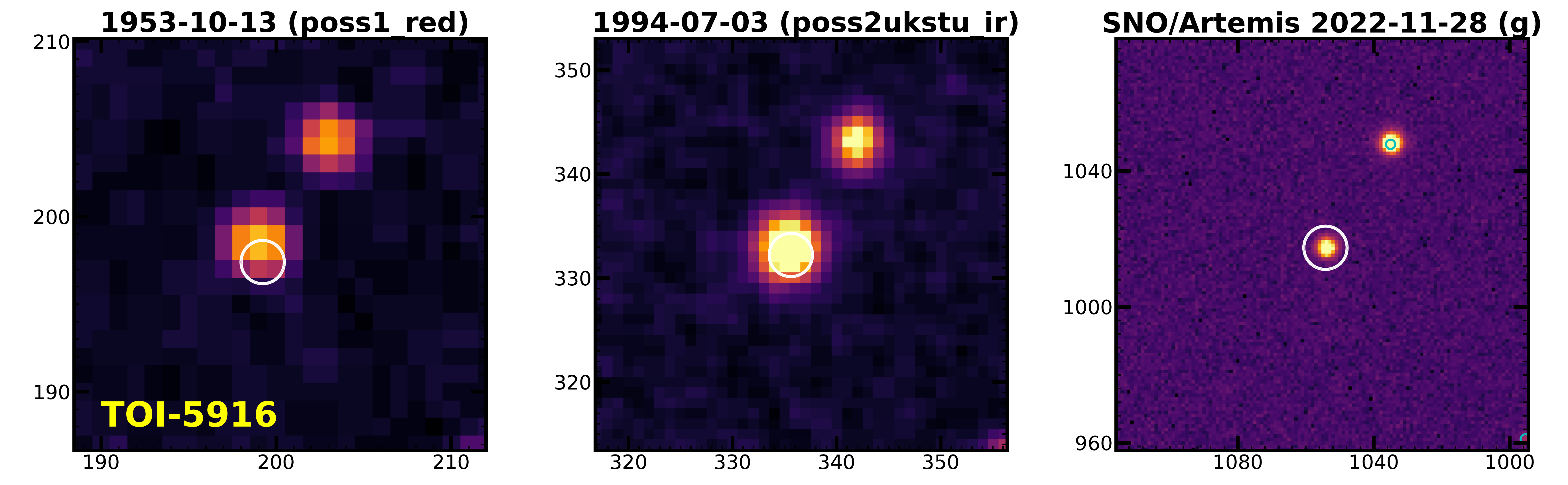}
\caption{Archival images for the systems discussed in the paper. From top to bottom, the row of three images show: TOI-3288~A, TOI-4666, TOI-5007, TOI-5292~A and TOI-5916. Each plot is titled with the date and instrument used to take the image. White circles show the position of each respective star at its most recent image position (i.e., in the SPECULOOS images), thus showing how its position has evolved.}
\label{fig:archival_imaging}
\end{figure*}

In Fig.~\ref{fig:archival_imaging} we present the archival images described in Section \ref{sec:blends}. 

\section{SED fits}
\begin{figure*}
    \centering
    \includegraphics[width=\columnwidth]{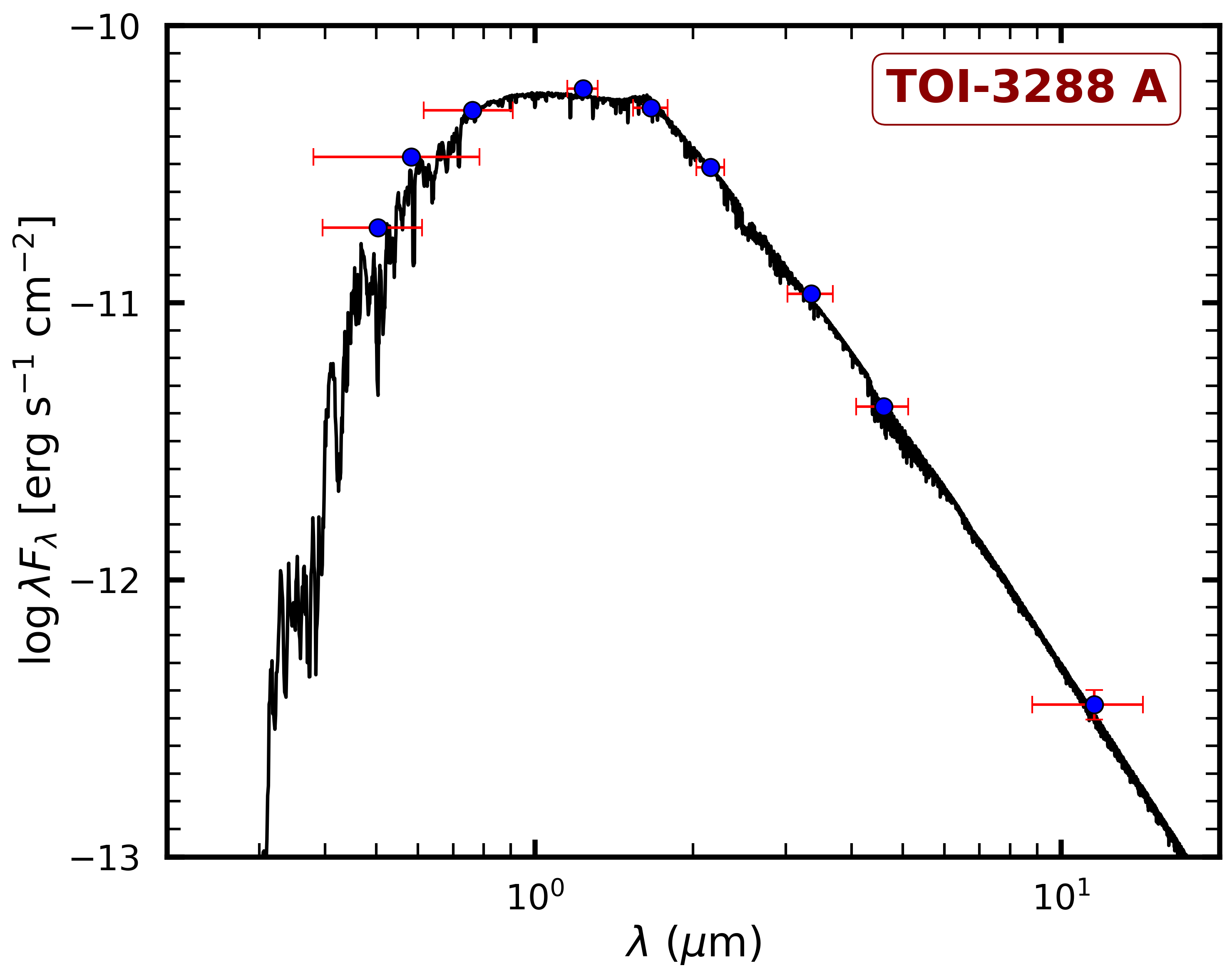}
    \includegraphics[width=\columnwidth]{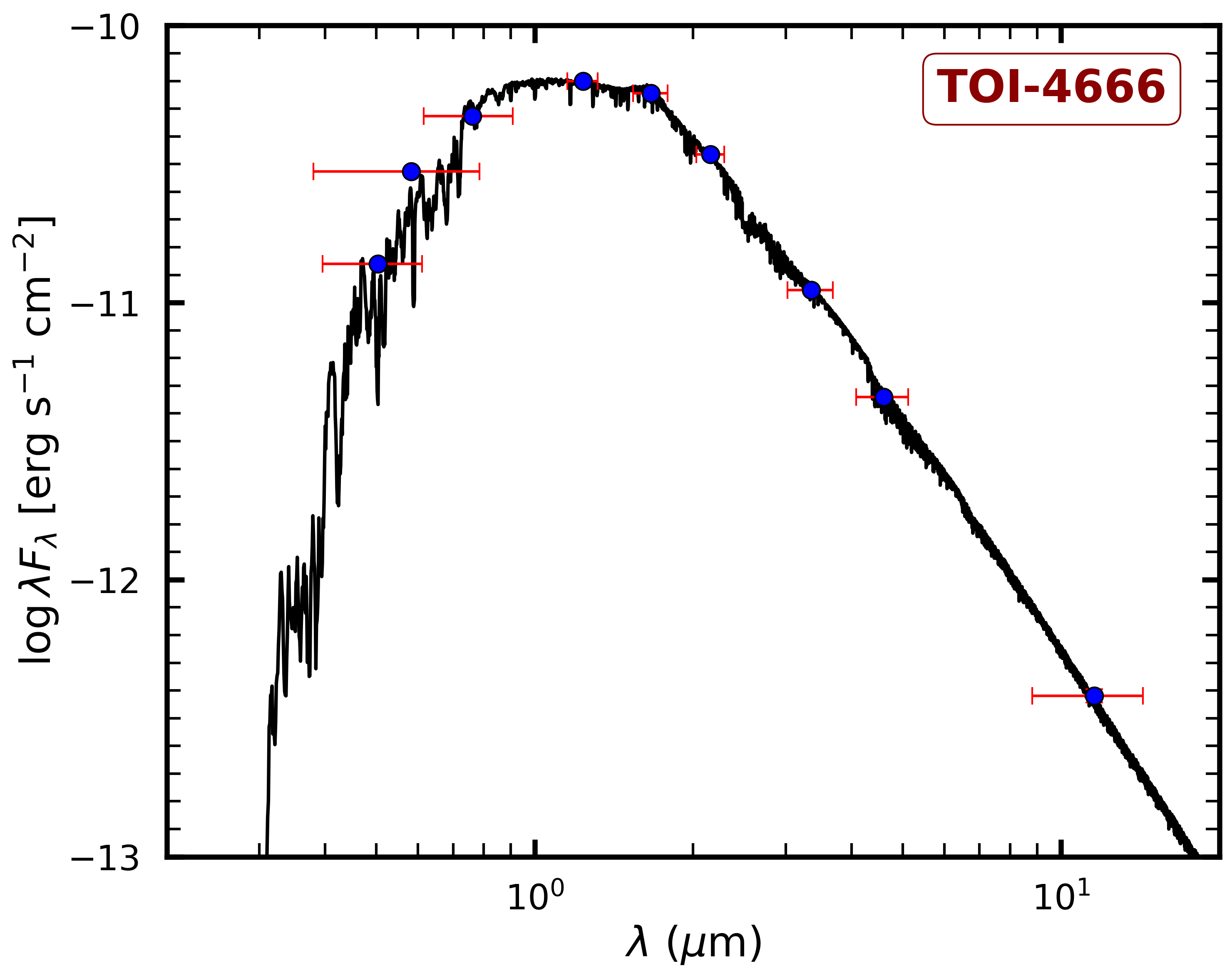}
    \includegraphics[width=\columnwidth]{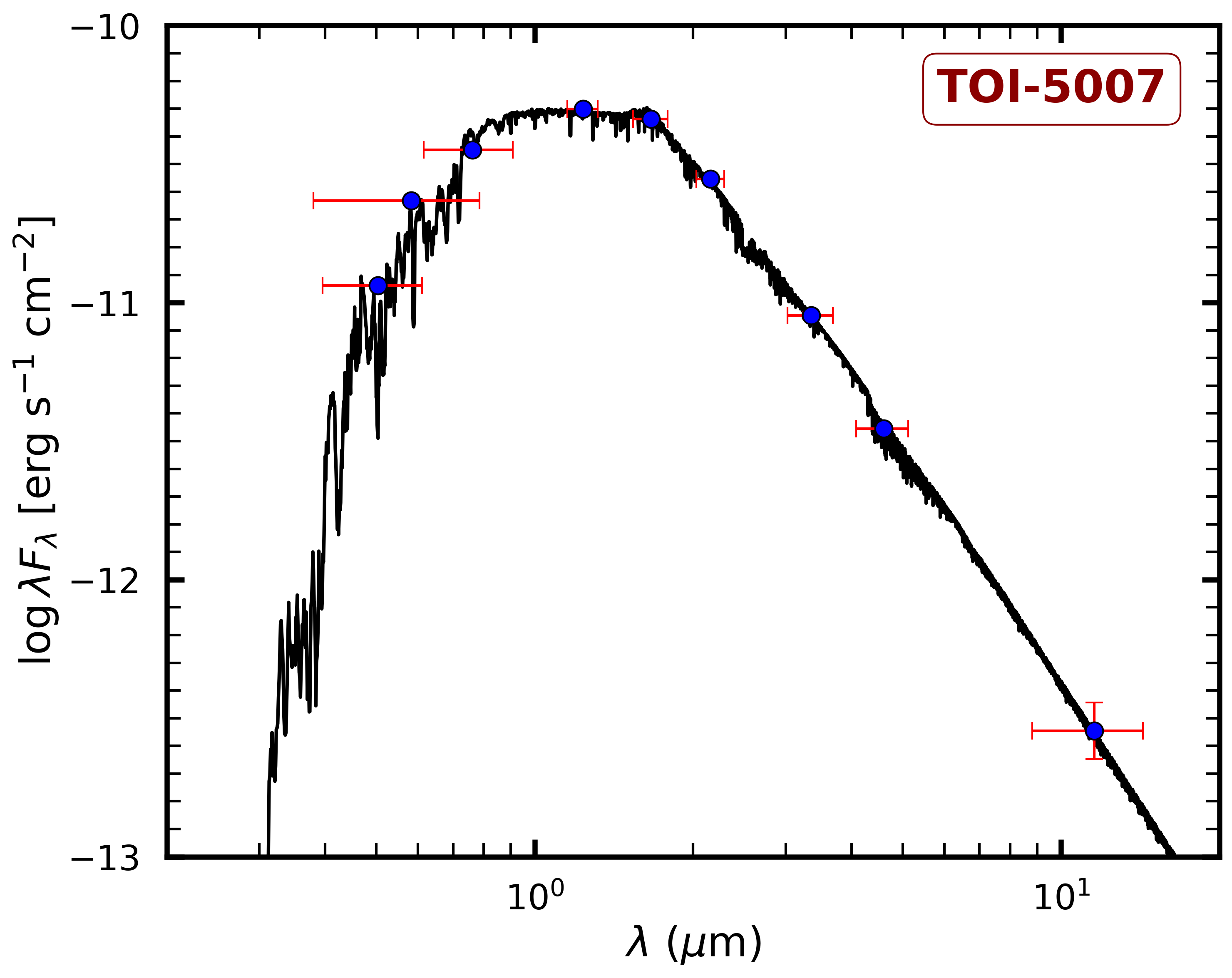}
    \includegraphics[width=\columnwidth]{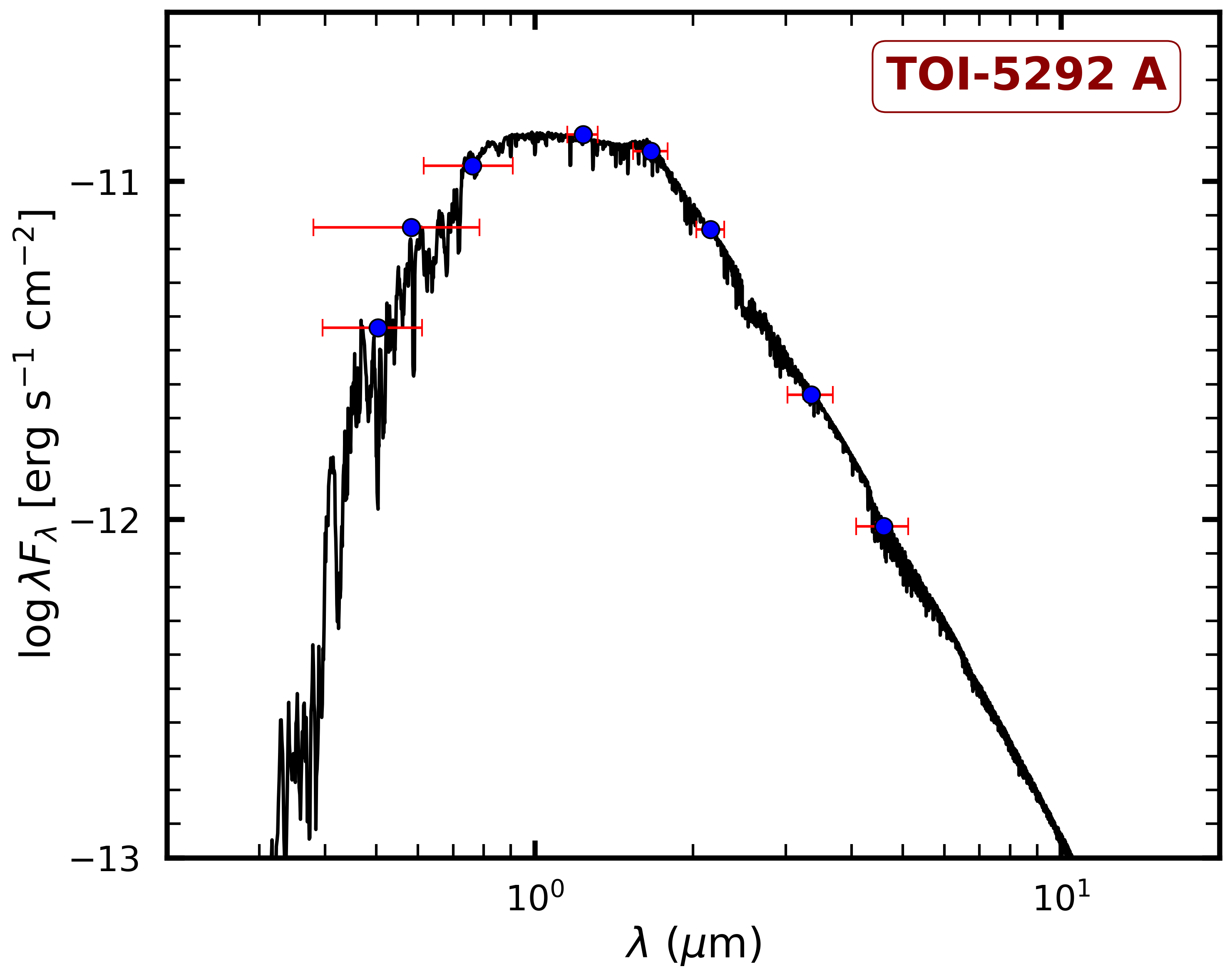}
    
    \includegraphics[width=\columnwidth]{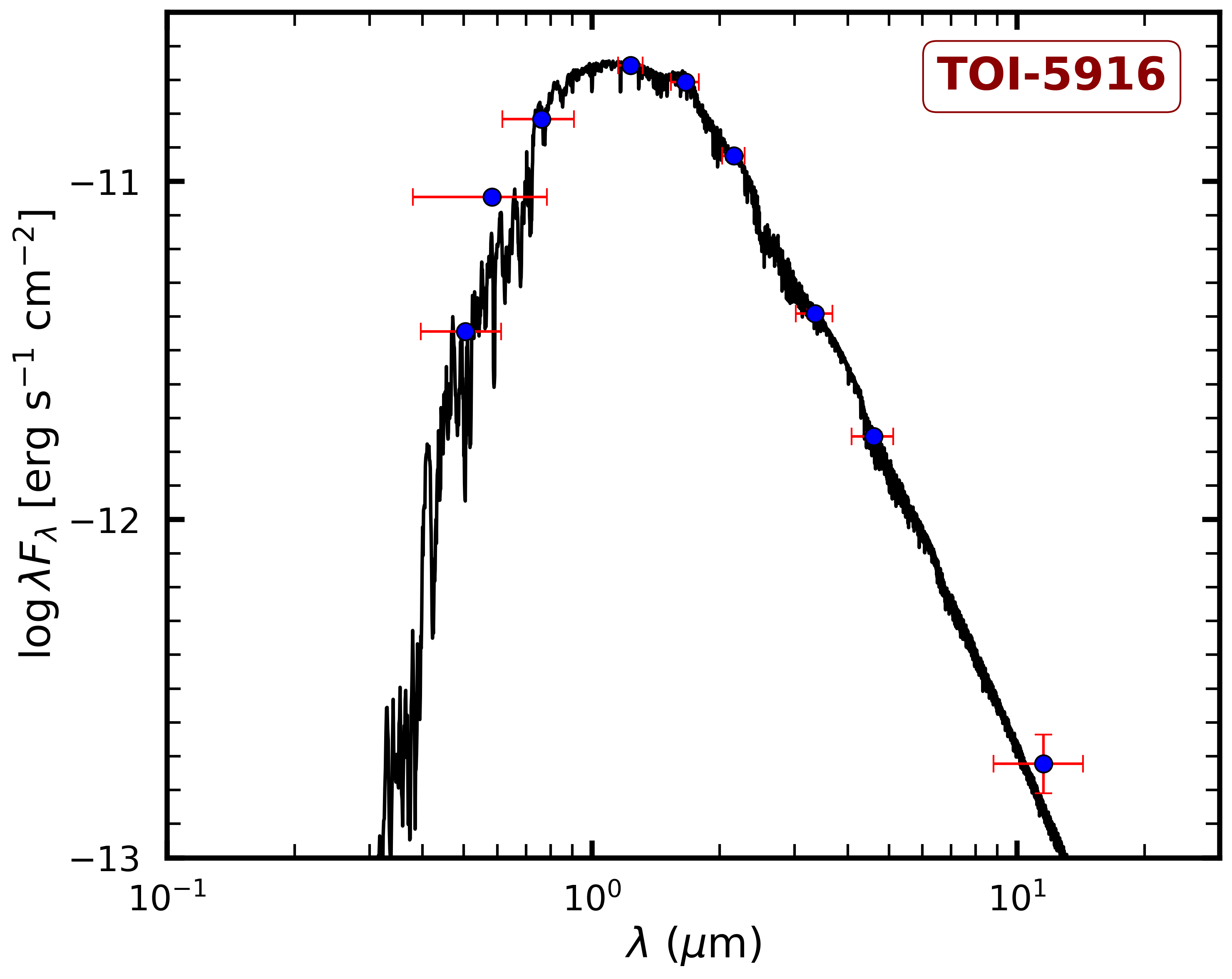}
\caption{Spectral Energy Distributions (SED) for TOI-3288~A, TOI-4666, TOI-5007, TOI-5292~A,  and TOI-5916 . The red symbols represent the observed photometric measurements, and the horizontal bars represent the effective width of the passband.}
\label{fig:sed}
\end{figure*}

\section{RV datasets}
\label{ap:RV_data}

We present the radial velocity points collected with ESPRESSO and PFS for the confirmation of the planets in this work. 

\begin{table}
    \centering
    \begin{tabular}{ccc}
    \midrule \midrule
\textbf{Time (BJD)} & \textbf{RV ($\mathrm{km.s}^{-1}$)} & \textbf{Error ($\mathrm{km.s}^{-1}$)}\\
\midrule
2460588.59382091 & 29.642753 & 0.028918\\
2460589.58301817 & 28.930989 & 0.020481\\
2460591.53703018 & 29.710953 & 0.049006\\
2460596.55754534 & 28.877204 & 0.020749\\
2460764.88300042 & 29.537801 & 0.021033\\
2460927.5789118 & 29.1194043 & 0.022688\\
    \end{tabular}
    \caption{Radial velocity measurements obtained with ESPRESSO for TOI-3288~A.}
    \label{tab:RV_TOI-3288}
\end{table}

\begin{table}
    \centering
    \begin{tabular}{ccc}
    \midrule \midrule
\textbf{Time (BJD)} & \textbf{RV ($\mathrm{km.s}^{-1}$)} & \textbf{Error ($\mathrm{km.s}^{-1}$)}\\
\midrule
2460296.54909425 & 7.873443 & 0.018208 \\
2460611.60882457 & 7.715562 & 0.016224 \\
2460612.75344893 & 7.801105 & 0.013576 \\
2460613.62217581 & 7.895610 & 0.028190 \\
2460614.60423935 & 7.705984 & 0.037956 \\
    \end{tabular}
    \caption{Radial velocity measurements obtained with ESPRESSO for TOI-4666.}
    \label{tab:RV_TOI-4666}
\end{table}

\begin{table}
    \centering
    \begin{tabular}{ccc}
    \midrule \midrule
\textbf{Time (BJD)} & \textbf{RV ($\mathrm{km.s}^{-1}$)} & \textbf{Error ($\mathrm{km.s}^{-1}$)}\\
\midrule
2460819.88751442 & -8.543389 & 0.016603 \\
2460837.82802106 & -8.389539 & 0.017020 \\
2460850.7915149 & -8.9525730 & 0.025503 \\
2460867.8634546 & -8.4141739 & 0.013828 \\
2460880.89505529 & -8.853235 & 0.014809 \\
2460895.72150212 & -8.821266 & 0.019857 \\
    \end{tabular}
    \caption{Radial velocity measurements obtained with ESPRESSO for TOI-5292~A.}
    \label{tab:RV_TOI-5292}
\end{table}

\begin{table}
    \centering
    \begin{tabular}{ccc}
    \midrule \midrule
\textbf{Time (BJD)} & \textbf{RV ($\mathrm{km.s}^{-1}$)} & \textbf{Error ($\mathrm{km.s}^{-1}$)}\\
\midrule
2460587.64343877 & -2.227842 & 0.022585 \\
2460588.5093216 &  -1.919660 & 0.024024 \\
2460589.57345587 & -2.145545 & 0.020670 \\
2460594.60454244 & -2.169590 & 0.038107 \\
2460595.52760945 & -1.939142 & 0.017934 \\
    \end{tabular}
    \caption{Radial velocity measurements obtained with ESPRESSO for TOI-5916.}
    \label{tab:RV_TOI-5916}
\end{table}

\begin{table}
    \centering
    \begin{tabular}{ccc}
    \midrule \midrule
\textbf{Time (BJD)} & \textbf{RV ($\mathrm{km.s}^{-1}$)} & \textbf{Error ($\mathrm{km.s}^{-1}$)}\\
\midrule
2460728.82822461 & 4.506280 & 0.017591 \\
2460731.84552939 & 4.617750 & 0.025800 \\
2460735.79607892 & 4.365100 & 0.016645 \\
2460736.85480744 & 4.604084 & 0.033525 \\ \midrule
2460042.84358 & 0.15931 & 0.01152\\
2460042.85728 & 0.16258 & 0.01183\\
2460042.87098 & 0.15394 & 0.01051\\
2460043.82899 & -0.13012 & 0.01167\\
2460043.84350 & -0.13319 & 0.01128\\
2460043.85775 & -0.08936 & 0.01199\\
2460044.82447 & 0.06492 & 0.01176\\
2460044.83907 & 0.05545 & 0.01314\\
2460044.85367 & 0.03082 & 0.01207\\
2460063.75571 & -0.01956 & 0.01572\\
2460063.77096 & -0.01524 & 0.01331\\
2460063.78556 & -0.0178 & 0.01180\\
2460067.76870 & 0.06425 & 0.01436\\
2460067.78289 & 0.06838 & 0.01499\\
2460067.79737 & 0.10587 & 0.01439\\
2460509.52196 & -0.08646 & 0.01189\\
2460509.53575 & -0.09951 & 0.01101\\
2460509.55012 & -0.11911 & 0.01171\\
2460517.50560 & 0.00900 & 0.0174\\
2460517.52053 & -0.04869 & 0.0181\\
2460517.53594 & 0.0000 & 0.01712\\
    \end{tabular}
    \caption{Radial velocity measurements obtained with ESPRESSO (top) and PFS (bottom) for TOI-5007.}
    \label{tab:RV_TOI-5007}
\end{table}

\section{Additional fitted parameters}

In these Tables we present additional fitted and derived parameters (dilutions, radial velocity jitter, and quadratic limb darkening coefficients) for the planets in our sample. 

\begin{table}
    \centering
    \begin{tabular}{lc}
    \midrule \midrule
    \multicolumn{1}{c}{\textbf{Parameter}} & \textbf{Value} \\ \midrule \vspace{0.1cm}
        Dilution TS \textit{I+z} &  $0.006 ^{+0.005}_{-0.008}$\\ \vspace{0.1cm}
        Dilution SSO/Europa \textit{I+z} &  $0.013 ^{+0.009}_{-0.013}$\\ \vspace{0.1cm}
        Dilution SSO/Io \textit{I+z} &  $0.079 ^{+0.020}_{-0.014}$\\ \vspace{0.1cm}
        Dilution SSO/Callisto \textit{Sloan-z'} & $0.044 ^{+0.025}_{-0.027}$\\ \vspace{0.1cm}
        Dilution SSO/Europa \textit{Sloan-r'} & $0.004 ^{+0.003}_{-0.005}$\\ \vspace{0.1cm}
        Dilution SSO/Ganymede \textit{Sloan-r'} & $0.007 ^{+0.005}_{-0.008}$\\ \vspace{0.1cm}
        Dilution SSO/Ganymede \textit{Sloan-i'} & $0.065 ^{+0.022}_{-0.019}$\\ \vspace{0.1cm}
        Dilution SSO/Callisto/SPIRIT1 \textit{zYJ} & $0.088 ^{+0.011}_{-0.008}$\\ \vspace{0.1cm}
        Dilution SSO/Callisto/SPIRIT2 \textit{zYJ} & $0.097 ^{+0.004}_{-0.002}$\\ \vspace{0.1cm}
Quadratic limb-darkening coefficient $u_{1,i'}$ & 0.3863 $_{-0.0477}^{0.0433}$ \\ \vspace{0.1cm}
Quadratic limb-darkening coefficient $u_{2,i'}$ & 0.1715 $_{-0.0474}^{0.0552}$ \\ \vspace{0.1cm}
Quadratic limb-darkening coefficient $u_{1,z'}$ & 0.2589 $_{-0.0411}^{0.0421}$ \\ \vspace{0.1cm}
Quadratic limb-darkening coefficient $u_{2,z'}$ & 0.1990 $_{-0.0457}^{+0.0499}$ \\ \vspace{0.1cm}
Quadratic limb-darkening coefficient $u_{1,r'}$ & 0.6896 $_{-0.0390}^{+0.0417}$ \\ \vspace{0.1cm}
Quadratic limb-darkening coefficient $u_{2,r'}$ & 0.0094 $_{-0.0491}^{+0.0492}$ \\ \vspace{0.1cm}
Quadratic limb-darkening coefficient $u_{1,I+z}$ & 0.3824 $_{-0.0388}^{+0.0419}$ \\ \vspace{0.1cm}
Quadratic limb-darkening coefficient $u_{2,I+z}$ & 0.1671 $_{-0.0491}^{+0.0498}$ \\ \vspace{0.1cm}
Quadratic limb-darkening coefficient $u_{1,g'}$ & 0.6725 $_{-0.0513}^{+0.0566}$ \\ \vspace{0.1cm}
Quadratic limb-darkening coefficient $u_{2,g'}$ & 0.0790 $_{-0.0547}^{+0.0545}$ \\ \vspace{0.1cm}
Quadratic limb-darkening coefficient $u_{1,zYJ}$ & 0.2623 $_{-0.0355}^{0.0362}$ \\ \vspace{0.1cm}
Quadratic limb-darkening coefficient $u_{2,zYJ}$ & 0.1523 $_{-0.0403}^{0.0400}$\\ \vspace{0.1cm}
$\ln{\sigma_\mathrm{jitter; ESPRESSO}}$ (km\,s\,$^{-1}$)  & $-9.0_{-3.5}^{+3.2}$  \\ 
    \end{tabular}
    \caption{Fitted dilution coefficients for the contamination of TOI-3288~A by its bound companion TOI-3288~B, as well as RV jitter and derived quadratic limb darkening coefficients for TOI-3288~Ab.}
    \label{tab:ldcs_3288Ab} 
\end{table}

\begin{table}
    \centering
    \begin{tabular}{lc}
    \midrule \midrule
    \multicolumn{1}{c}{\textbf{Parameter}} & \textbf{Value} \\ \midrule \vspace{0.1cm}
Quadratic limb-darkening coefficient $u_{1,i'}$ & 0.4212 $_{-0.0367}^{0.0390}$ \\ \vspace{0.1cm}
Quadratic limb-darkening coefficient $u_{2,i'}$ & 0.1744 $_{0.0501}^{0.0498}$ \\ \vspace{0.1cm}
Quadratic limb-darkening coefficient $u_{1,z'}$ & 0.3333 $_{0.0351}^{0.0365}$ \\ \vspace{0.1cm}
Quadratic limb-darkening coefficient $u_{2,z'}$ & 0.1602 $_{-0.0445}^{+0.0442}$ \\ \vspace{0.1cm}
Quadratic limb-darkening coefficient $u_{1,r'}$ & 0.5882 $_{-0.0420}^{+0.0432}$ \\ \vspace{0.1cm}
Quadratic limb-darkening coefficient $u_{2,r'}$ & 0.1268 $_{-0.0568}^{+0.0564}$ \\ \vspace{0.1cm}
Quadratic limb-darkening coefficient $u_{1,I+z}$ & 0.3930 $_{-0.0284}^{+0.0300}$ \\ \vspace{0.1cm}
Quadratic limb-darkening coefficient $u_{2,I+z}$ & 0.1323 $_{-0.0423}^{+0.0441}$ \\ \vspace{0.1cm}
Quadratic limb-darkening coefficient $u_{1,g'}$ & 0.5857 $_{-0.0494}^{+0.0493}$ \\ \vspace{0.1cm}
Quadratic limb-darkening coefficient $u_{2,g'}$ & 0.1690 $_{-0.0577}^{+0.0582}$ \\ \vspace{0.1cm}
Quadratic limb-darkening coefficient $u_{1,zYJ}$ & 0.2290 $_{-0.0287}^{+0.0280}$ \\ \vspace{0.1cm}
Quadratic limb-darkening coefficient $u_{2,zYJ}$ & 0.1959 $_{-0.0427}^{+0.0451}$\\ \vspace{0.1cm}
$\ln{\sigma_\mathrm{jitter; ESPRESSO}}$ (km\,s\,$^{-1}$) & $-9.5\pm3.6$\\
    \end{tabular}
    \caption{Fitted limb darkening coefficients and RV jitter for TOI-4666 b.}
    \label{tab:ldcs_4666b} 
\end{table}

\begin{table}
    \centering
    \begin{tabular}{lc}
    \midrule \midrule
    \multicolumn{1}{c}{\textbf{Parameter}} & \textbf{Value} \\ \midrule \vspace{0.1cm}
Quadratic limb-darkening coefficient $u_{1,TESS}$ & 0.3242 $_{-0.0538}^{0.0522}$ \\ \vspace{0.1cm}
Quadratic limb-darkening coefficient $u_{2,TESS}$ & 0.1904 $_{0.0489}^{0.0502}$ \\ \vspace{0.1cm}
Quadratic limb-darkening coefficient $u_{1,i'}$ & 0.4138 $_{-0.0632}^{0.0651}$ \\ \vspace{0.1cm}
Quadratic limb-darkening coefficient $u_{2,i'}$ & 0.1815 $_{0.0562}^{0.0531}$ \\ \vspace{0.1cm}
Quadratic limb-darkening coefficient $u_{1,z'}$ & 0.2455 $_{0.0476}^{0.0503}$ \\ \vspace{0.1cm}
Quadratic limb-darkening coefficient $u_{2,z'}$ & 0.1598 $_{-0.0443}^{+0.0469}$ \\ \vspace{0.1cm}
Quadratic limb-darkening coefficient $u_{1,r'}$ & 0.5537 $_{-0.0713}^{+0.0696}$ \\ \vspace{0.1cm}
Quadratic limb-darkening coefficient $u_{2,r'}$ & 0.1500 $_{-0.0616}^{+0.0645}$ \\ \vspace{0.1cm}
Quadratic limb-darkening coefficient $u_{1,g'}$ & 0.6561 $_{-0.0714}^{+0.0790}$ \\ \vspace{0.1cm}
Quadratic limb-darkening coefficient $u_{2,g'}$ & 0.1324 $_{-0.0769}^{+0.0668}$ \\  \vspace{0.1cm}
Quadratic limb-darkening coefficient $u_{1,ExTrA}$ & 0.2047 $_{-0.0462}^{+0.0449}$ \\ \vspace{0.1cm}
Quadratic limb-darkening coefficient $u_{2,ExTrA}$ & 0.1451 $_{-0.0400}^{+0.0434}$ \\ \vspace{0.1cm}
Quadratic limb-darkening coefficient $u_{1,zYJ}$ & 0.2633 $_{-0.0933}^{+0.1074}$ \\ \vspace{0.1cm}
Quadratic limb-darkening coefficient $u_{2,zYJ}$ & 0.1591 $_{-0.0807}^{+0.0954}$ \\ \vspace{0.1cm}
$\ln{\sigma_\mathrm{jitter; ESPRESSO}}$ (km\,s\,$^{-1}$) & $-9.6\pm3.5$  \\ \vspace{0.1cm}
$\ln{\sigma_\mathrm{jitter; PFS}}$  (km\,s\,$^{-1}$) & $-4.21_{-0.36}^{+0.31}$
    \end{tabular}
    \caption{Fitted limb darkening coefficients and RV jitter for TOI-5007 b.}
    \label{tab:ldcs_5007b} 
\end{table}

\begin{table}
    \centering
    \begin{tabular}{lc}
    \midrule \midrule
    \multicolumn{1}{c}{\textbf{Parameter}} & \textbf{Value} \\ \midrule \vspace{0.1cm}
Quadratic limb-darkening coefficient $u_{1,TESS}$ & 0.3639 $_{-0.0563}^{0.0620}$ \\ \vspace{0.1cm}
Quadratic limb-darkening coefficient $u_{2,TESS}$ & 0.1801 $_{0.0546}^{0.0566}$ \\ \vspace{0.1cm}
Quadratic limb-darkening coefficient $u_{1,I+z}$ & 0.2871 $_{-0.0514}^{0.0552}$ \\ \vspace{0.1cm}
Quadratic limb-darkening coefficient $u_{2,I+z}$ & 0.1726 $_{0.0488}^{0.0522}$ \\ \vspace{0.1cm}
Quadratic limb-darkening coefficient $u_{1,z'}$ & 0.2911 $_{0.0485}^{0.0511}$ \\ \vspace{0.1cm}
Quadratic limb-darkening coefficient $u_{2,z'}$ & 0.1738 $_{-0.0477}^{+0.0505}$ \\ \vspace{0.1cm}
Quadratic limb-darkening coefficient $u_{1,g'}$ & 0.5610 $_{-0.0707}^{+0.0708}$ \\ \vspace{0.1cm}
Quadratic limb-darkening coefficient $u_{2,g'}$ & 0.1771 $_{-0.0672}^{+0.0692}$ \\  \vspace{0.1cm}
Quadratic limb-darkening coefficient $u_{1,i'}$ & 0.3762 $_{-0.0567}^{+0.0579}$ \\ \vspace{0.1cm}
Quadratic limb-darkening coefficient $u_{2,i'}$ & 0.1816 $_{-0.0546}^{+0.0569}$ \\  \vspace{0.1cm}
$\ln{\sigma_\mathrm{jitter; ESPRESSO}}$  (km\,s\,$^{-1}$) & $-9.2_{-3.8}^{+3.5}$
    \end{tabular}
    \caption{Fitted limb darkening coefficients and RV jitter for TOI-5292~Ab.}
    \label{tab:ldcs_5292b} 
\end{table}

\begin{table}
    \centering
    \begin{tabular}{lc}
    \midrule \midrule
    \multicolumn{1}{c}{\textbf{Parameter}} & \textbf{Value} \\ \midrule \vspace{0.1cm}
Quadratic limb-darkening coefficient $u_{1,TESS}$ & 0.2804 $_{-0.0469}^{0.0494}$ \\ \vspace{0.1cm}
Quadratic limb-darkening coefficient $u_{2,TESS}$ & 0.2062 $_{0.0440}^{0.0490}$ \\ \vspace{0.1cm}
Quadratic limb-darkening coefficient $u_{1,g'}$ & 0.5287 $_{-0.0610}^{0.0597}$ \\ \vspace{0.1cm}
Quadratic limb-darkening coefficient $u_{2,g'}$ & 0.2304 $_{0.0586}^{0.0617}$ \\ \vspace{0.1cm}
Quadratic limb-darkening coefficient $u_{1,I+z}$ & 0.2657 $_{0.0444}^{0.0463}$ \\ \vspace{0.1cm}
Quadratic limb-darkening coefficient $u_{2,I+z}$ & 0.1904 $_{-0.0420}^{+0.0432}$ \\ \vspace{0.1cm}
Quadratic limb-darkening coefficient $u_{1,r'}$ & 0.5381 $_{-0.0614}^{+0.0663}$ \\ \vspace{0.1cm}
Quadratic limb-darkening coefficient $u_{2,r'}$ & 0.1824 $_{-0.0634}^{+0.0651}$ \\  \vspace{0.1cm}
Quadratic limb-darkening coefficient $u_{1,z'}$ & 0.2514 $_{-0.0466}^{+0.0471}$ \\ \vspace{0.1cm}
Quadratic limb-darkening coefficient $u_{2,z'}$ & 0.1651 $_{-0.0395}^{+0.0421}$ \\  \vspace{0.1cm}
Quadratic limb-darkening coefficient $u_{1,zs}$ & 0.2625 $_{-0.0461}^{+0.0507}$ \\ \vspace{0.1cm}
Quadratic limb-darkening coefficient $u_{2,zs}$ & 0.2037 $_{-0.0457}^{+0.0466}$ \\  \vspace{0.1cm}
Quadratic limb-darkening coefficient $u_{1,i'}$ & 0.3384 $_{-0.0525}^{+0.0501}$ \\ \vspace{0.1cm}
Quadratic limb-darkening coefficient $u_{2,i'}$ & 0.1977 $_{-0.0470}^{+0.0505}$ \\  \vspace{0.1cm}
$\ln{\sigma_\mathrm{jitter; ESPRESSO}}$ (km\,s\,$^{-1}$) & $-6.6_{-3.1}^{+2.3}$
    \end{tabular}
    \caption{Fitted limb darkening coefficients and RV jitter for TOI-5916 b.}
    \label{tab:ldcs_5916b} 
\end{table}




\bsp	
\label{lastpage}
\end{document}